\documentclass[a4paper,twoside]{ifj-thesis}
\usepackage{amssymb}

\usepackage{epsfig,mytab}

\usepackage{graphicx}
\usepackage{amsmath}

\def\vec#1{\mathchoice{\mbox{\boldmath$\displaystyle#1$}}
{\mbox{\boldmath$\textstyle#1$}}
{\mbox{\boldmath$\scriptstyle#1$}}
{\mbox{\boldmath$\scriptscriptstyle#1$}}}
\makeatletter
\newcommand\erfc{\mathop{\operator@font erfc}\nolimits}
\def\slashchar#1{\setbox0=\hbox{$#1$}
   \dimen0=\wd0 \setbox1=\hbox{/} \dimen1=\wd1
   \ifdim\dimen0>\dimen1 \rlap{\hbox to \dimen0{\hfil/\hfil}} #1
   \else  \rlap{\hbox to \dimen1{\hfil$#1$\hfil}} / \fi}

\makeatother

\centerline{\large\bf MODIFICATION OF THE \bf{$\pi \omega \rho $} VERTEX}
\centerline{\large\bf IN NUCLEAR MEDIUM AND ITS INFLUENCE}
\centerline{\large\bf ON THE DILEPTON PRODUCTION RATE}
\centerline{\large\bf IN RELATIVISTIC HEAVY-ION COLLISIONS}
\centerline{\Large Agnieszka Bieniek}
\centerline{\large PhD Thesis}
\bigskip
\bigskip
\centerline{\large Department of Theoretical Physics}
\centerline{\large The Henryk Niewodnicza\'nski}
\centerline{\large Institute of Nuclear Physics}
\centerline{\large Polish Academy of Sciences}
\centerline{\large Krak\'ow}
\centerline{\large SUPERVISOR: Assoc. Prof. Wojciech Broniowski}

\linespacing{1.33}

\begin{document}
\newpage
.
\eject
\newpage
{\small \bf This PhD thesis is supported by the Polish State Committee for Science Research, grant 2P03B 13324.}
\newpage
.
\eject
\newpage
\centerline{\large\bf Abstract}
\linespacing{1.2} 
\vskip 1.cm
In this PhD thesis we analyze the medium modification of the $\pi\omega\rho$ coupling and subsequently discuss 
its possible 
influence on the dilepton production rate in relativistic heavy-ion collisions. 

The first part of the thesis is devoted to the medium modifications  
of the $\omega \rightarrow \pi ^{0}\gamma ^{\ast }$, $\rho ^{a}\rightarrow \pi
^{a}\gamma ^{\ast }$ and $\pi ^{0}\rightarrow \gamma \gamma ^{\ast }$ decays in nuclear matter. 
We use the relativistic field theory formalism describing modifications of the coupling  by 
one-loop Feynman diagrams  
with nucleons and $\Delta$(1232) isobars. For simplicity we work in the leading baryon density approximation 
and at zero 
temperature. We consider cases when the decaying particle is at rest or at motion with respect 
to the medium. The kinematics is considered in the rest frame of the medium.
In the medium, the structure of the $\pi\omega\rho$ amplitude is more complex than in the 
vacuum, which is due to the presence of the additional four-vector describing the velocity of the medium. 
We investigate the dependence of medium effects on the dilepton invariant mass and observe a sizeable increase of 
the effective coupling constants compared to their vacuum values. The $\pi\omega\rho$ 
coupling constant increases by about a factor of 2 for the $\omega\to\pi^{0}\gamma^{*}$ decay and by 
about a factor of 5 for the $\rho^{a}\to\pi^{a}\gamma^{*}$ decay. The effect grows with  
density. Similar conclusions are obtained for a particle moving with respect to the medium.

In the second part of the thesis we apply our model to evaluate the influence of the medium effects on 
the dilepton production from the Dalitz decays. 
In the calculations of the dilepton spectrum we use the model of the hydrodynamic expansion of the 
fire cylinder, including the longitudinal and transverse expansion. 
In our comparison to the CERES dilepton data 
we take into account the experimental cuts, which is very important in the detailed numerical analysis. 
The medium modifications enhance the effect of the Dalitz decays, however, the numbers are still below 
the experimental data. 

It is worth to point out that in the medium the $\rho$ decay due to 
the isospin degeneracy becomes as important as the $\omega$ decay. 
The main new and original results of the thesis are:
\begin{itemize}
\item[
1.] Significant in-medium increase of the $\pi\omega\rho$ coupling constant. 
\item[2.] Large enhancement of the Dalitz
$\omega \rightarrow \pi ^{0}\gamma ^{\ast }$ and $\rho ^{a}\rightarrow \pi
^{a}\gamma ^{\ast }$ decays in the range $0.2-0.6~{\rm~GeV}$ of the dilepton invariant mass.
\end{itemize}
\newpage
.
\eject

\newpage
\linespacing{1.33} 

\begin{acknowledgements}
\vskip 2 cm

First and foremost I have the pleasure to thank my supervisor Dr. Wojciech Broniowski for guiding me through 
my PhD.

I am also very grateful to Prof. Wojciech Florkowski for many valuable comments. 

I would like to thank Prof. Edward Kapu\'scik, the head of International PhD studies at IFJ PAN for help 
and for many interesting discussions.

Many warm thanks to the staff and colleagues of Theory Department in the Institute of Nuclear 
Physics for numerous discussions during my PhD studies.

I warmly thank my sister Marta and Steven Steinke from University of Arizona for linguistic help.

Finally I would like to thank my family for support, and all people who in some way have helped 
and encouraged me.

Special thanks to Piotr Czerski.

\end{acknowledgements}

\newpage

\tableofcontents
\pagestyle{plain}

\chapter*{Introduction}

\addcontentsline{toc}{chapter}{Introduction}

Physics of ultra-relativistic heavy-ion collisions is a quickly developing branch of science which 
allow us to investigate the properties of nuclear matter in terms of elementary interactions. 
The production of dileptons ($e^{+}e^{-}$, $\mu^{+}\mu^{-}$) 
is one of the most important probes in studying the dynamical evolution of the nuclear collision processes.
Since they do not interact strongly, the dileptons 
escape unthermalized from the hot and dense matter during all stages of the evolution. 
A major source of dileptons comes from the  
direct and the Dalitz decays of meson resonances in particular the  $\omega$, $\rho$, $\eta$, $\eta'$ or $\phi$. 
The properties of vector  
mesons such as  masses or widths, are known to be significantly modified in nuclear 
matter, \cite{brscale, chin, hatsuda}. 
These in-medium properties play an important role in the low mass dilepton production region, namely 
between $0.2-0.6~\rm{GeV}$ of invariant masses, covered recent heavy-ion collision experiments, \cite{ceres, helios, newceres}. 
The dilepton production enhancement   
in central nucleus-nucleus collisions in the low-mass region has been a topic 
of great interest in the last years, \cite{celenza,herrmann1,herrmann2,Mishra,renkMish}.  
This phenomenon is frequently explained by assuming that the 
vector meson spectral functions undergo substantial modifications through the strong 
interactions with the nucleons in the medium. Many research groups search for 
a theoretical description of the low-mass dilepton enhancement however, a convincing explanation has not been 
reached and the problem still remains a major puzzle 
in relativistic heavy-ion physics. 

Trying to explain this relevant problem, in this thesis 
we investigate modifications of the $\pi\omega\rho$ coupling constant in the context of the $\omega$ and $\rho$ 
Dalitz decays. We analyze these decays in nuclear matter applying a relativistic hadronic framework 
incorporating nucleons and $\Delta$(1232) isobars. We find that the $\pi\omega\rho$ coupling constant 
for $\omega\to\pi^0 e^{+}e^{-}$ and $\rho^{a} \to \pi^{a} e^{+}e^{-}$ decays are 
considerably enhanced by the medium as compared to the vacuum values. We observe an enhancement by 
about a factor of 2 for the $\omega$ meson and by about a factor of 5 for the $\rho$ meson. We have to 
stress that the Dalitz decay of the $\rho$ meson is found to be equally important to the omega decay, \cite{ab1}.
The in-medium-modified $\pi\omega\rho$ coupling constant increases the corresponding widths of considered 
decays. 
In order to estimate the dilepton yields we use the vector dominance model (VDM) and include the effect 
of the expansion of the medium. Next, we compare our model results to the dilepton yield 
from the CERES Pb+Au experiment in the low-mass region carefully including the experimental 
kinematic cuts. Finally, our results show that medium effects from the $\omega$ and $\rho$ Dalitz 
decays are significant and about two times larger compared to the vacuum.

\subsubsection{Highlights of the experimental situation}
The physics of ultra-relativistic heavy ion collisions, \emph{i.e.}, collisions of 
atomic nuclei in which the center-of-mass energy per nucleon is much larger than the nucleon rest mass, has been  
studied experimentally with immense intensity. 
First experiments have been performed at the Alternating Gradient Synchrotron (AGS) in the 
Brookhaven National Laboratory (BNL) and at the Super Proton Synchrotron (SPS) at CERN, where 
ion beams were accelerated with the center of mass energies around $\sqrt{s} \sim 5~\rm{A~GeV}$ and 
$\sqrt{s} \sim 20~\rm{A~GeV}$, respectively.  

In 2001 the first data were collected from the Relativistic Heavy Ion Collider (RHIC) at BNL at the energy of 
$\sqrt{s} \sim 200~\rm{A~GeV}$. 
Between the years 2001-2004 the next three runs took place. 
In the near future heavy ions will also be injected into the CERN Large Hadron Collider (LHC) reaching 
the energy of $\sqrt{s}=10~A~\rm{TeV}$.
The principal goal of experimental and theoretical initiative in ultra-relativistic heavy-ion collisions 
is the observation and understanding the physics of hot and dense medium. 

Many probes have been proposed to analyze the behavior of hot and dense hadronic matter. 
Among these probes the dileptons stand out, because they 
couple directly to vector mesons and interact only electromagnetically. Therefore they are likely to 
bring information about the hot and dense matter formed in the early stages of nuclear collisions. 
\begin{figure}[h]
\centerline{
\epsfysize = 10 cm \centerline{\epsfbox{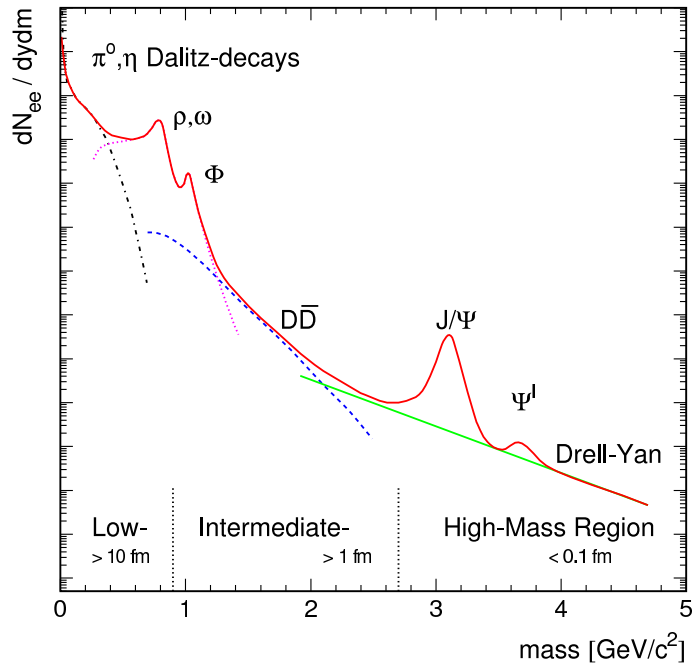}} \vspace{0mm}} 
\caption{\label{all} Major dilepton sources as a function of invariant mass in 
ultrarelativistic heavy-ion collisions. In the high-mass region above invariant masses $M_{e+e-}=3~\rm{GeV}$, 
dileptons come mostly from hard Drell-Yan annihilation processes. Here strong $J/\Psi$ suppression 
and $\Psi'$ abundance are interpreted as a sign of color deconfinement. In the intermediate-mass 
region between $1-2.5~\rm{GeV}$, the thermal signal from QGP 
could be related to the observation of associated $D \bar{D}$ production. In the low-mass region below 
$1~\rm{GeV}$, dileptons originate from Dalitz decays of neutral mesons 
such as $\pi^{0}$, $\eta$, $\eta' \to e^{+}e^{-}\gamma$, 
$\omega \to \pi^{0} e^{+}e^{-}$, and from direct decays, $\omega$, $\rho$, $\phi \to e^{+}e^{-}$. This region is 
very sensitive to hadron modifications in the nuclear matter.}
\end{figure}
A schematic view of characteristic dilepton sources in ultrarelativistic heavy-ion collisions is given 
in Fig.~\ref{all}. One can distinguish three basic regions with respect to the invariant mass of the 
dilepton pair.

In the high-mass region, $M_{e+e-}=3~\rm{GeV}$, dileptons come dominantly from 
hard processes such as the Drell-Yan annihilation 
occurring in the early stage, where the excited hadronic system is far from thermal equilibrium. 
After rapid thermalization the QGP phase may be established, where the dilepton production proceeds predominantly 
via the quark-antiquark annihilation. Upon expansion and cooling, the QGP is converted into a hot hadron gas. 
At this stage dileptons are radiated from pion and kaon annihilation processes as well as other collisions 
between various hadrons. In contrast to the light 
vector mesons ($\omega$, $\rho$), the lifetime for the heavy quarkonium states such as $J/\Psi$ and 
$\Upsilon$ is 
substantially longer than the typical life-time in the hadronic fireball. Therefore these mesons 
will mostly decay after freezeout, and hence will not feel the medium. 

The intermediate mass region, above $\sim 1.5~\rm{GeV}$,
might be proper to observe a thermal signal from the plasma radiation, because QGP can only be formed 
at higher temperatures than a hadronic gas. This signal could be revealed by 
the observation of the associated $D\bar{D}$ production in this intermediate region. 

Next, when the freezeout stage is reached, the dominant sources are hadronic resonances as well as Dalitz decays, 
mostly from $\pi^{0}$, $\eta$, $\omega$, and $\rho$ mesons. 
The two-body annihilation processes are dynamically enhanced through the formation 
of light vector meson resonances  such as the $\rho$, $\omega$, and $\phi$ mesons, which directly couple 
to lepton pairs. Thus, the invariant mass distribution of the lepton pairs reflects the mass distribution of the 
vector mesons at the moment of decay. 
Therefore, vector mesons, possibly modified by medium effects, play an important role 
in the measurement of dilepton rates in relativistic heavy-ion reactions.
All these above-mentioned dominant sources feed into the low mass region, $M_{l+l-}\leq 1~\rm{GeV}$, which  
is particularly sensitive to in-medium modifications of the light hadrons.

The experimental measurements of the dilepton spectra in relativistic heavy-ion collisions have mainly been 
carried out at the CERN SPS by the CERES/NA45 (low-mass region)\cite{ceres,ceres2,ceres3}, 
HELIOS-3 (low- and intermediate-mass region) \cite{helios} 
and NA38/NA50 (intermediate- and high-mass region) \cite{na38a,na38b,na50}. In the near future, at RHIC and CERN-SPS, 
dilepton spectra will be measured by the PHENIX and NA60 \cite{na60} collaborations, respectively. 
At much lower bombarding 
energies, dilepton data have also been taken by the DLS collaboration at SIS (GSI) \cite{bevalac1,bevalac2}, 
where only the low-mass region up to the kinematical limit of around $1~\rm{GeV}$ may accessible. 
These experiments focused on the role of high baryon density.

Recently, the CERES \cite{ceres} and HELIOS \cite{helios} collaborations have observed that central 
nucleus-nucleus collisions exhibit a strong enhancement of low-mass dilepton production as compared 
to proton-nucleus reactions. The data analysis is based on 
the so-called \emph{hadronic cocktail model} which works very well to describe the proton-nucleus data. 
On the other hand, the hadronic cocktail strongly underestimates the dilepton yield from nucleus-nucleus 
collisions. The model is based on data from proton-induced collisions which are scaled with the 
multiplicity of charged particles.
A lot of efforts have been undertaken in order to understand phenomenon of the low-mass enhancement.

In theoretical attempts to address the problem, the increased dilepton yield in nucleus-nucleus collisions 
was attributed to enhanced $\rho$-meson production via $\pi^{+}\pi^{-}$ annihilation. 
Thus, many theoretical groups, \cite{li,LiKoBrown1,CassingDil}, have included 
the $\rho \to \pi^{+}\pi^{-}$
process within different models for the space-time evolution of nucleus-nucleus reactions. Their 
calculations are based on the vacuum meson properties, and all results are in disagreement with the data. This 
has led to the suggestion of various medium effects, such as the reduction of the vector 
meson masses, that might be responsible for the observed enhancement. 
Brown and Rho suggested the hypothesis that the vector meson masses drop in the medium 
according to a simple scaling law, \cite{brscale, celenza}. 

There are also many approaches based on QCD sum rules \cite{hatlee, leupold}, which confirm the scaling hypothesis.

The conclusion from the existing studies is that although conventional mechanisms are sufficient to account for 
the dilepton spectra from proton-induced reactions, medium effects are needed to explain the low-mass dilepton 
enhancement observed in nucleus-nucleus reactions. Thus, in this thesis we examine the in-medium 
$\pi\omega\rho$ vertex and its large influence on the dilepton production rate in relativistic 
heavy-ion collisions. We believe that our model together with other medium effects may 
resolve the dilepton puzzle. This is the main reason of our study. 

\subsubsection{Organization of the Thesis}

The first part of this dissertation is devoted to the modification of hadron properties, in particular the 
$\pi\omega\rho$ vertex, in the presence of nuclear medium. 

As the introduction to the subject, in Chapter 2 we first review some theoretical approaches 
describing the medium modifications of hadron properties. Next, we present our original analysis, 
describing our model (section 2.1). The model is  
based on fully relativistic hadronic theory where mesons and nucleons interact with each 
other and with the $\Delta$ isobars. The presentation concentrates on Dalitz decays such as 
$\pi^{0}\to\gamma\gamma^*$, $\omega\to\pi^{0}\gamma^*$ and $\rho^{a}\to\pi^{a}\gamma^*$.
In such processes the $\pi\omega\rho$ coupling constant appears. 
We present propagator, vertices for the interactions with 
and without $\Delta$ resonance, and the in-medium structure of the $\pi\omega\rho$ amplitude. This structure is 
more complicated because of the additional four-vector, the four-velocity of the medium.
We analyze the Dalitz decays at rest, section 2.2, as well at motion, section 2.3, 
with respect to the medium. In the case when the decaying particles are moving with respect to the medium we 
consider the kinematics in a convenient and non-textbook way. The kinematics is usually analyzed in the rest frame of the decaying 
particle, whereas we consider it in the rest frame of the nuclear medium (with the four-velocity  
of the medium $u=(1,0,0,0)$). 
We investigate the dependence of medium effects in the dilepton invariant mass and 
observe their influence on the effective $\pi\omega\rho$ coupling constant compared to its vacuum value. 
We point out that 
the $\pi\omega\rho$ coupling constant is considerably enhanced in the presence of the nuclear medium 
for both the $\omega$ and $\rho$ Dalitz decays. The $\omega$ and $\rho$ can have two polarizations.
When the decaying particle moves with respect to the medium each polarization behaves differently, thus 
the properties of decaying particle are different. 
We calculate the dependence of the width for both the transverse and longitudinal 
polarizations as a function of the invariant mass of the dileptons, and show significant medium effects.
Finally, in Chapter 3 we discuss how nuclear matter influences the   
$\pi\omega\rho$ coupling constant in the previously mentioned Dalitz processes.

In the second part of the thesis, we focus on applying our model to the production of 
dileptons in relativistic heavy-ion collisions. We begin this part with a review of major dilepton experiments and 
some theoretical models connected to the low-mass region. This part 
starts out by describing the experimental measurements of the dilepton yield. 
Here we concentrate on the experimental background and discuss relevant dilepton experiments, mainly at CERN SPS 
in the proton-nucleus and nucleus-nucleus reactions. 
We discuss results at high bombarding energies of $158~\rm{A~GeV}$ or $200~\rm{A~GeV}$ and at the lower energies 
of $40~\rm{A~GeV}$ and $1-5~\rm{A~GeV}$. We focus on the low-mass region, because here in particular 
the experimental results are in disagreement with many theoretical models. 
The experimental data are usually compared to the hadronic 
cocktail model, therefore we present a short description of this approach. 
In section 4.2 of Chapter 4, we describe the most popular theoretical explanations related to the 
low-mass dilepton enhancement, based on the in-medium modifications of vector mesons. 

In order to estimate the dilepton production rate from the Dalitz decays of vector mesons we use the 
vector dominance model (VDM) reviewed in section 5.1 of Chapter 5. Section 5.2 
discusses in detail the structure of the three-body Dalitz decays. In section 5.3 we describe briefly the 
model considered for the hydrodynamic evolution of the fire cylinder from Ref.~\cite{Rappevol, RappShur}.  
Next, in section 5.4, we present details of our formalism used 
to calculate the dilepton production 
from the Dalitz and direct decays of the $\omega$ and $\rho$ mesons. 
We use the experimental acceptance cuts in our analysis.  
It is quite nontrivial how to construct and include these cuts in the calculations, therefore 
in many other theoretical works the acceptance cuts are frequently not taken into account.

In section 5.5 we present our numerical results, starting from calculations without expansion of the fireball 
and without the CERES acceptance function. Our plots are done for two typical temperatures of the fireball. 
Next, we present the dilepton yield from direct and Dalitz decays with 
expansion and acceptance cuts. We compare our numerical results to the CERES experimental data also 
for two different temperatures. 

In the medium we observe a significant enhancement for the Dalitz decays of the 
$\omega$ and $\rho$ mesons in the dilepton mass region $0.2-0.6~\rm{GeV}$ compared to the vacuum case.

Although the modification of the $\pi\omega\rho$ vertex increases the contribution from the Dalitz decays, 
the obtained yields alone fall short of describing the data. Yet, they help by contributing more to the 
theoretical curves, and together with other effects not considered here 
(such as dropping masses, broadening widths, 
or Dalitz decays of $\eta$ and $\eta'$ mesons), may help to explain the long standing 
problem of the low-mass dilepton enhancement seen in relativistic heavy-ion collisions. 

The new and original material of this thesis is contained in Chapter 2, section 5.4 and Chapter 6, 
while Chapter 1, Chapter 4 and sections 5.1, 5.2, 5.3  
contain useful reviews necessary for the integrity of the text.

\part{Modification of the  $\pi \omega \rho$ coupling in nuclear medium}
\pagestyle{headings}

\chapter{Modification of Hadrons in medium}

As mentioned in the Introduction, in heavy-ion collisions the particle production is related
to the evolution of hot and dense matter. 
Heavy-ion collisions provide the only way to compress and heat up 
nuclear matter in the laboratory conditions. 
From these heavy-ion data we obtain important information useful in 
construction of models for the early universe, supernova explosions or neutron stars. 
In the so-called \emph{relativistic regime}, 
at energies above 1 GeV per nucleon, heavy-ion collisions  
are likely to yield particle densities of the order of a few times the normal nuclear matter density. 
In such system the particles cannot propagate completely freely (the Compton wavelength is comparable with  
mean free paths of particles). In this situation we expect that in hot and dense 
environment the particle properties, \emph{i.e.} their masses, widths, or coupling constants, 
are changed and these \emph{in-medium modifications} may lead to experimentally observed phenomena.

The problem of how the properties of hadrons change in hadronic or nuclear matter 
in comparison to their free values has attracted a lot of attention \cite{chin,hatsuda,herrmann1,herrmann2,
hatlee,leupold,hadrons,tsuk,torino,jean,herrmann3,pirner,Urban0,urban,rapp,Post,Peters,LeupoldRev,
Gao,serot,lee2000}.  
Till a few years ago there were only theoretical predictions \cite{hadrons,tsuk,torino}, 
for the in-medium modifications. At present we also have some indirect 
experimental evidence from the dilepton production in hadronic collisions, (which we discuss in more  
detail in the Part II, Chapter 4) \cite{ceres, helios} 
or from data on the $\pi^{+}\pi^{-}$ correlations from STAR, where the   
$\rho$ meson peak moves to lower masses \cite{ fachini}.
The existing calculations are based on simple scaling of masses \cite{brscale,celenza}, 
numerous hadronic models \cite{chin,hatsuda,herrmann1,herrmann2,jean,herrmann3,pirner,
Urban0,urban,rapp,Post,Peters,LeupoldRev,Gao,serot}, or
QCD sum rule techniques \cite{hatlee,leupold,lee2000}. More fundamental 
predictions are based on low-density expansion and dispersion relations \cite{eletsky, friman2,Lutz99}. 
In all these calculations properties of hadrons are changed by the presence of the medium 
through strong interactions with nucleons from the medium. 

Below we will briefly review major theoretical approaches for in-medium properties of 
vector mesons, namely, $\rho$, $\omega$, and $\phi$ mesons.

\section{The Walecka model}
 
A simple and very illustrative approach showing the effects of medium modifications is  
Quantum Hydrodynamics (QHD) \cite{chin, serot} in its simplest form known as the 
Walecka model. It is based on the mean-field theory of hadrons, where the scalar and vector mean fields  
generated by the nucleon sources are themselves self-consistently responsible for the modification 
of the nucleon mass. The Lagrangian density for the Walecka model is
\begin{eqnarray}
{\cal L}=&\bar{\Psi}&[\gamma_{\mu}(i \partial^{\mu} -g_{v}V^{\mu})-(M-g_{s}\phi)]\Psi+\frac{1}{2}
(\partial_{\mu} \phi\partial^{\mu} \phi-m_{s}^{2}\phi^2) \nonumber\\
&-&\frac{1}{4}F_{\mu\nu}F^{\mu\nu}+\frac{1}{2}m_{v}^2 
V_{\mu}V^{\nu}+\delta{\cal L},
\end{eqnarray}
where $\Psi$ is the baryon field of mass $M$, $\phi$ is the neutral scalar-meson field with mass $m_{s}$, 
$V^{\mu}$ is the neutral vector meson field of mass $m_{v}$, and $F^{\mu\nu}\equiv\partial^{\mu}V^{\nu}-
\partial^{\nu}V^{\mu}$. The term $\delta{\cal L}$ contains renormalization counterterms. The nucleons 
interact via the exchange of isoscalar mesons with the coupling of the scalar field $\phi$ 
to the baryon scalar density $\bar{\Psi}\Psi$, and the vector field $V^{\mu}$ to the conserved baryon current 
$\bar{\Psi}\gamma_{\mu}\Psi$, obtained through minimal substitution. Because the exact solutions to the 
field equations are very complicated, a mean-field approximation is used. In a mean-field approximation, 
the meson field operators are replaced by their expectations values, which are classical fields:
\begin{eqnarray}
\phi \rightarrow \langle \phi \rangle \equiv \phi_{0}, \nonumber \\
V_{\mu} \rightarrow \langle V_{\mu} \rangle \equiv \delta_{\mu0} V_{0}.
\end{eqnarray}
Now the mean-field equations can be solved exactly with the solution becoming increasingly 
valid with increasing baryon density. Usually, the mean-field equations are solved in two approximations. 
In the mean-field theory one calculates a baryon self-energy which is generated by the presence 
of all the nucleons in the occupied Fermi sea, and the effect of the Dirac sea is neglected. In contrast, 
in the relativistic Hartree approximation one includes the contribution to the baryon self-energy 
arising from the occupied Fermi sea as well as from the full Dirac sea. In consequence, the baryon self-energy 
diverges in the relativistic Hartree approximation and must be renormalized. 
In both approximations, the mean scalar meson field $\phi_{0}$ 
is responsible for a shift of the nucleon mass $M^{\ast}$ in the nuclear medium relative to its vacuum 
value $M$. In contrast to the ground-state expectation value of the vector field, which is determined by the 
conserved baryon density, $\rho_{B}$:
\begin{eqnarray}
V^{0}=\frac{g_{v}}{m_{v}^2}\langle\bar{\Psi}\gamma^{0}\Psi\rangle=\frac{g_{v}}{m_{v}^2}\rho_{B}.
\end{eqnarray}
The expectation value of the scalar field, and consequently the effective mass of the nucleon, is a dynamical 
quantity that must be determined self-consistently from the equations of motion
\begin{eqnarray}
M^{\ast}=M-g_{s}\phi_{0}=M-\frac{g_{s}^2}{m_{s}^2}\langle\bar{\Psi}\Psi\rangle=
M-\frac{g_{s}^2}{m_{s}^2}\rho_{s},
\end{eqnarray}
where $\rho_{s}=\langle\bar{\Psi}\Psi\rangle$ is the scalar density of the nucleons. Since $\rho_{s}>0$, 
from the above equation we see that indeed nucleon mass, $M^{\ast}$, is lower in medium than in vacuum.

The model has already been used extensively in calculations of nuclear matter and finite nuclei. The 
saturation of nuclear matter and the strong spin-orbit splitting observed in finite nuclei were among 
the first successes of the model \cite{serot, walecka, serot2, serot3}. The model has also been used 
with considerable success to analyze such diverse topics as collective modes in nuclear matter, 
isoscalar magnetic moments, and electroweak and hadronic responses from finite nuclei.

\section{Quark condensate in medium}

The authors of Ref.~\cite{griegel} have studied quark and gluon condensates in nuclear matter. 
These condensates are expectation values of local composite operators 
such as $\bar{q}q$ and $G_{\mu\nu}^{a}G^{a\mu\nu}$, where $q$ is an up or down quark field and $G_{\mu\nu}^{a}$ 
is the gluon field-strength tensor. 
The authors of Ref.~\cite{griegel} focused on describing the properties of hadrons 
in nuclear matter in terms of \emph{in-medium} quark and gluon condensates, which are shifted from their vacuum 
values. The condensates are analyzed in the ground state of nuclear matter. 
Simple expressions for these quantities, which are model independent to first order in the nucleon density, 
have been developed. 

As a result, the ratio of the in-medium quark condensate to its vacuum value is given by 
\begin{eqnarray}
\langle\bar{q}q\rangle_{\rho_{N}}=\langle\bar{q}q\rangle_{vac} 
(1-\frac{\sigma_{N} \rho_{N}}{m_{\pi}^{2}f_{\pi}^2}),
\label{qquark}
\end{eqnarray}
where $\langle\bar{q}q\rangle_{\rho_{N}}$ and $\langle\bar{q}q\rangle_{vac}$ are the medium and vacuum values 
of the quark condensate, $m_{\pi}$ is the pion mass and $f_{\pi}$ is the pion decay constant. 
The quark condensate at low densities is related to the nucleon $\sigma$ term, $\sigma_{N}$, which can be 
expressed as $\sigma_{N}=\langle N |m \bar{q}q| N \rangle$, where $m$ is the current mass of light quarks.
Eq.~\ref{qquark} shows the in-medium quark condensate to first order in the nucleon density. There are 
higher-order corrections, see Ref.~\cite{griegel}; 
therefore, the model-independent results are valid at sufficiently low nucleon densities.
To first order in the nucleon density, the relationship of the in-medium quark condensate to its vacuum value 
depends on the values of the pion mass ($138~\rm{MeV}$), the pion decay constant ($93~\rm{MeV}$) 
and the $\sigma$ term (about $45~\rm{MeV}$). The value of nucleon density at saturation 
is $\rho_{N}=0.17~\rm{fm^{-3}}$. 
Depending on the precise value of the $\sigma$ term, the quark condensate is reduced considerably at nuclear 
matter saturation density - it is roughly $25-50 \%$ smaller than the vacuum value,
thus the shift in the quark condensate is large. 
This description is complementary to the Walecka model from QCD ($M \sim \langle \bar{q}q\rangle$).
On the other hand, the finite-density effects modify the gluon condensate to a 
much smaller extent than they do for the quark condensate. The decrease in the gluon condensate at nuclear matter 
saturation density is around $5 \%$. Hence, it does not affect the particle properties as much.

\section{Brown-Rho scaling}
Brown and Rho (BR) have suggested the hypothesis that the hadron masses drop in the medium 
according to a simple scaling law \cite{brscale}, given by 

\begin{eqnarray}
\frac{m_{N}^{\ast}}{m_{N}}=\frac{m_{\rho}^{\ast}}{m_{\rho}}=
\frac{m_{\omega}^{\ast}}{m_{\omega}}=\frac{f_{\pi}^{\ast}}{f_{\pi}},
\end{eqnarray}
where $f_{\pi}$ is a pion decay constant and the medium-modified parameters are indicated by asterisks. 
This hypothesis is based on phenomenological implementation
of the restoration of chiral symmetry in the framework of an effective theory. 

At least qualitatively, this scaling conjecture is supported by studies based on the QCD sum rule approach,
as well as by other hadronic models \cite{chin,hatsuda,jean,herrmann1,herrmann2,herrmann3,pirner,%
Urban0,urban,rapp,Post,Peters,LeupoldRev,Gao,serot}.

\section{QCD sum rules}
A natural and direct use for in-medium condensates is made in QCD sum rule calculations of hadronic properties 
in nuclear matter. This approach can be used to predict in-medium spectral properties (effective masses or 
self-energies) of baryons and mesons.

The QCD sum rule approach aims to understand the physical current-current correlation 
functions in terms of QCD. It is done by relating the observed hadron spectrum to the 
non-perturbative vacuum structure of QCD. 
Through the use of this technique 
Hatsuda and Lee, \cite{ hatlee} have extracted the medium dependence of the 
non-strange vector-meson masses. They have obtained the following result
\begin{eqnarray}
\frac{m_{\rho,\omega}^{\ast}}{m_{\rho,\omega}} \approx 1-0.18(\frac{\rho_{N}}{\rho_{0}}),
\end{eqnarray}
and
\begin{eqnarray}
\frac{m_{\phi}^{\ast}}{m_{\phi}} \approx 1-0.15a(\frac{\rho_{N}}{\rho_{0}})
\end{eqnarray}
where $a$ is the nucleon strangeness content, $a=\langle N| \bar{s}s| N \rangle/\langle N |\bar{q}q|N \rangle$, 
taken to be $a=0.17$. The $\rho$ and $\omega$ meson masses 
decrease significantly with density due to the strong density dependence of the light quark condensate. Since 
the strange quark condensate does not change much in nuclear media as a result of the small nucleon strangeness 
content, the $\phi$ meson mass shows a weaker density dependence.
The fact that these masses decrease as density increases 
has been taken as a confirmation of the Brown-Rho scaling \cite{brscale}. 

The QCD sum rules for vector mesons were reanalyzed in Ref.~\cite{Jin}. 
The uncertainties in various in-medium quark condensates were assessed using the Monte-Carlo 
method analysis. It was found that at normal nuclear matter density $m_{\rho}^{\ast}/m_{\rho}=0.78\pm0.08$, 
in nice agreement with the results of Refs.~\cite{ hatlee}. In Ref.~\cite{asakawa} a decrease of the $\rho$ 
meson mass with increasing density like that of Ref.~\cite{ hatlee} has also been obtained. The predictive 
power of the QCD sum rule approach was addressed in Ref.~\cite{leupold} taking into account the 
possible modifications of the $\rho$ meson mass and width. In this approach the $\rho$ meson spectral function 
was parameterized by a simple Breit-Wigner form containing two fitting parameters, the pole mass and the width, 
and the values for mass and width are determined by requiring that the QCD sum rules are satisfied. It was found 
that in the case of vanishing width, the $\rho$ meson mass indeed decreases in nuclear matter. On the other hand 
if the width increases it is then possible to satisfy the QCD sum rules with a constant or even an increasing 
$\rho$ meson mass. The predictive power of the QCD sum rule approach thus depends on how well one can constrain 
the phenomenological $\rho$ meson spectral function in the nuclear medium.

\section{Predictions from dispersion relations}
 
The in-medium mass shift and broadening of the width  
are due to interactions of the hadron with the medium. Thus, one can use phenomenological information on 
this interaction to calculate the mass shifts. The authors of \cite{ eletsky2} 
have used the result that the mass shift 
of a particle in medium is related to the forward scattering amplitude, $f_{sc}$, of this particle on the 
constituents of the medium.  

This gives the mass shift 
\begin{eqnarray}
\Delta m=-2\pi \frac{\rho}{m}Re f_{sc}
\label{delta} 
\end{eqnarray}
where $m$ is the vacuum mass of the particle and $\rho$ denotes the density of the medium. 
The width broadening is given by 
\begin{eqnarray}
\Delta\Gamma=\frac{\rho}{m} k \sigma,
\label{deltagamma} 
\end{eqnarray}
where $k$ is the particle momentum and $k\sigma=4\pi Im f_{sc}$. Equations \ref{delta} and \ref{deltagamma} 
are correct when the particle's momentum $k$ is larger than few hundred $\rm{MeV}$, and $Re f_{sc}$ 
fullfils the requirement $|Re f_{sc}|< d$, (where $d$ is the distance between medium constituents). 
Additionally the main part of the scattering should proceed through small angles, less than 1, 
and the medium constituents should have some momentum distributions, 
such as Fermi-Dirac or Bose-Einstein distributions, for finite temperatures and chemical potentials. 
In such cases, in the right-hand sides of Eqs.~\ref{delta} and, \ref{deltagamma} an averaging over the momentum 
distribution of constituents must be performed.
Thanks to this approach one may formulates explicitly the applicability conditions discussed above.
When gas in thermal equilibrium plays the role of the medium, the equivalents of Eqs.~\ref{delta} 
and, \ref{deltagamma}  can be derived in the framework of the thermal field theory.

Using this approach, authors of Ref. \cite{eletsky2} have estimated $\rho$ meson mass shift and width broadening 
in the case of $\rho$-mesons produced in heavy-ion collisions, with following conclusions: 
For $\rho$-mesons produced in high energy heavy ion collisions the mass shift is 
small, but the width broadening is so large that one can hardly observe a $\rho$-peak in $e^{+}e^{-}$ 
or $\mu^{+}\mu^{-}$ mass distributions. In low energy heavy-ion collisions a $\rho$-peak may be observed 
in $e^{+}e^{-}$ or $\mu^{+}\mu^{-}$ mass distributions as 
a broad enhancement approximately at the position of $\rho$-mass.

\section{Lowering of the position of $\rho^{0}$ peak}
\begin{figure}[t]
\begin{center}
\includegraphics[width=0.7 \textwidth]{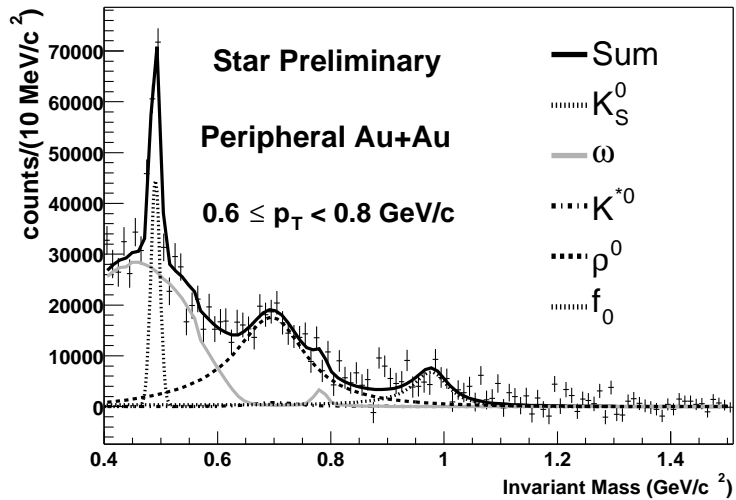}
\caption{\label{auau} The $\pi^{+}\pi^{-}$ invariant mass distributions for peripheral Au+Au 
interactions. The $\rho$ peak is around $700~\rm{MeV}$, much lower than the vacuum value.}
\end{center}
\end{figure}
Experimental indications for in-medium physics also come from RHIC \cite{ fachini}, where the data obtained 
by the STAR collaboration have been interpreted as the lowering of the 
position of the $\rho^{0}$ meson peak, see Fig. \ref{auau}. 
This is the first direct measurement of $\rho(770)^{0} \to \pi^{+} 
\pi^{-}$ in heavy-ion collisions. The important observation is that the $\rho^{0}$ mass is lower in 
peripheral Au+Au collisions than in p+p collisions. 
Apart from the dynamical interactions with the surrounding matter, 
interference between various $\pi^{+}\pi^{-}$ scattering channels, phase space distortions due to the 
rescattering of pions forming $\rho^{0}$, and Bose-Einstein correlations between $\rho^{0}$ decay 
daughters and pions in the surrounding matter are among possible explanations for the apparent change of 
the $\rho^{0}$ meson. Hopefully, more detailed information will be provided by future correlation experiments.

\section{Modification of coupling constants}
 
Meson properties are changed due to strong interactions 
with nucleons of the medium. 
If the mass and widths of hadrons can be significantly
modified, one can expect that also the coupling constants are altered. There are numerous 
studies of mesonic two-point functions in the literature, Ref.~\cite {galeLich}, 
which in general contained medium modified mass $M^{\ast}$ and medium modified width $\Gamma^{\ast}$, 
but only a few devoted to meson three-point functions. In their studies of the $\rho$ 
meson in-medium spectral function
in Ref. \cite{herrmann1}, Herrmann, Friman and Norenberg have analyzed the 
$\rho\pi\pi$ vertex for the $\rho$ at rest with respect to the medium. 
They have considered the above vertex on the 
basis of a hadronic model with the $\Delta$ isobar and with nonrelativistic couplings. 
Song and Koch in Ref. \cite{Song2} have taken into account the temperature effects on the $\rho \pi \pi$ 
interaction. Krippa in Ref. \cite{Krippa} has computed the effects of density on chiral mixing of meson 
three-point functions. The authors of Ref.\cite {bfh,BFH2,BFH3,rhopipi} have analyzed the 
$\omega \rightarrow \pi \pi$ and $\rho \rightarrow \pi \pi$ decays in nuclear medium. 
They have studied the $\rho\pi\pi$ vertex in the fully relativistic framework with relativistic interactions, 
incorporating nucleons and $\Delta$(1232) isobars, and 
found that the medium effects on the $\rho\pi\pi$ constant are large, and
dominantly come from the processes where the $\Delta$ is excited in the intermediate state. This has 
immediate consequences for the direct decay of the $\rho$ meson into dileptons (such as broadening of width).

An important factor brought in by the presence of the medium is that processes 
which are forbidden in the vacuum by symmetry principles are now made possible. In this work we concentrate on 
the $\pi\omega\rho$ coupling constant and analyze its significance for the Dalitz decays of vector mesons.

\chapter{$\protect\pi \protect\omega \protect\rho $ vertex in nuclear medium}

\begin{figure}[h]
\includegraphics[width=1.0\textwidth]{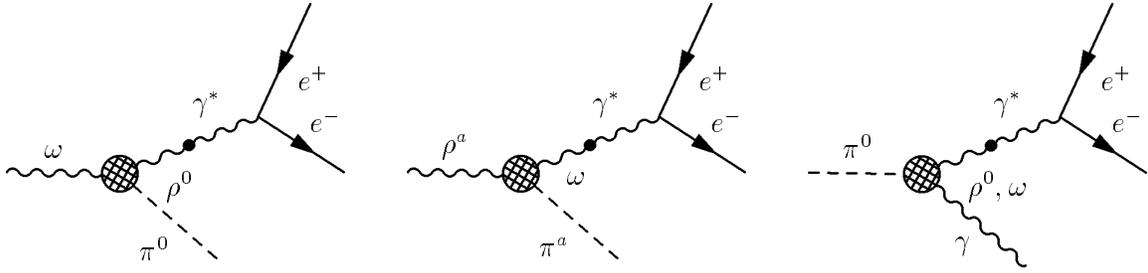}
\caption{\label{dalitz} The physical processes where the $\pi \omega \rho$ vertex appears. 
Wavy lines indicate the $\rho$ or $\omega$, 
dashed lines the pions, the hatched blob indicates the medium-modified vertex, 
the black dot is equal to $\frac{em_{\omega }^{2}}{g_{\omega }}$ 
or $\frac{em_{\rho }^{2}}{g_{\rho }}$ as follows from the Vector Dominance Model (VDM).}
\end{figure}

So far we have discussed major theoretical approaches related to the medium modifications of particle properties. 
In this chapter we begin the original material of the thesis. We focus on the $\pi\omega\rho$ coupling constant and  
its medium modifications. The $\pi\omega\rho$ vertex appears in such processes as: 
$\omega \rightarrow \pi ^{0}\gamma ^{\ast }$, $\rho ^{a}\rightarrow \pi
^{a}\gamma ^{\ast }$, and $\pi ^{0}\rightarrow \gamma \gamma ^{\ast }$, see Fig~~\ref{dalitz}. 
We expect that medium effects may have a significant influence and may affect the existing 
calculations based solely on the vacuum values.  
Our calculation of the in-medium $\pi \omega \rho$ vertex is made in the framework of a fully 
relativistic hadronic theory, where mesons interact with nucleons and $\Delta$ isobars. 
Among the baryon resonances, the $\Delta$ is most important due to the small $\Delta-N$ 
mass splitting and a large value of the $\pi N \Delta$ coupling constant. 
In our investigation we work at zero temperature to obtain the vertices of Fig.~\ref{dalitz}, at least as a first 
approximation.
We use the \emph{leading-density} approximation, which makes the calculation simpler, as no integration over 
the nucleon momenta is necessary. 
This Chapter is  divided into three sections.
In section 2.1 we present our model with all the needed diagrams, propagators and vertices for interactions 
of mesons with nucleons and with $\Delta$ resonances which we use in our calculations. 
Also, the tensor structure of the $\pi\omega\rho$ vertex is presented. 
Next, we look at how the effective coupling constant is changed compared to 
the vacuum value. First, in section 2.2 we analyze our decays in the rest 
frame with respect to the medium. Later, in section 2.3, keeping non-zero 
three momenta of the $\omega$ and $\rho$ we look separately at the longitudinal and transverse 
polarizations. In both cases we observe that the effective coupling constant is significantly 
enhanced in the presence of nuclear matter.

\section{Our model}

\begin{figure}[h]
\includegraphics[width=1.0\textwidth]{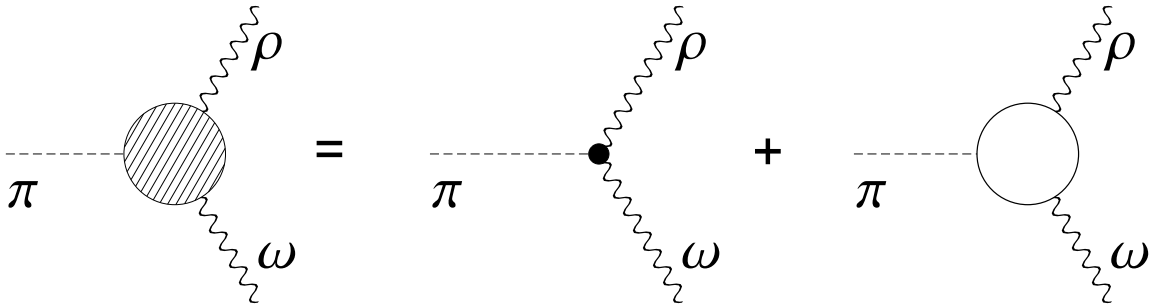}
\caption{\label{vacmed} The full $\pi \omega \rho$ vertex consisting of two parts, 
the vacuum part and the medium part, respectively.}
\end{figure}

In our analysis we apply a conventional hadronic model with meson, nucleon
and $\Delta $ degrees of freedom. This model is similar to the Walecka Model, which we have discussed in Chapter I. 
In Fig.~~\ref{vacmed} the considered diagram for the $\pi\omega\rho$ coupling consists of two parts, 
the vacuum one and the induced by the medium. 

\begin{figure}[h]
\includegraphics[width=1.0\textwidth]{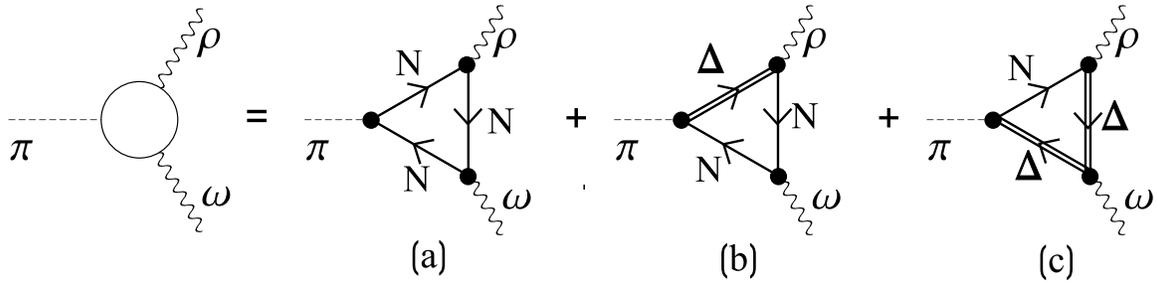}
\caption{\label{meddiag} In-medium diagrams included in the present calculation 
(crossed diagrams not shown). Solid lines
indicate the in-medium nucleon propagator, double lines the $\Delta$ propagator.} 
\end{figure}

There are three types of in-medium diagrams, see Fig.~~\ref{meddiag}:
diagrams with three nucleons, Fig.~~\ref{meddiag} (a), two nucleons and one $\Delta $ isobar, 
Fig.~~\ref{meddiag} (b), and with one nucleon and two $\Delta $\ isobars, Fig.~~\ref{meddiag} (c). 

\bigskip
The diagrams of interest are those
which include at least one nucleon, because only these depend on the baryon
density, carried by the Fermi sea. Since the $\Delta $ isobar has a large 
coupling constant it has a large contribution to the considered processes. 
The diagram with three $\Delta $ is not considered, because 
such a diagram corresponds to vacuum polarization effects, which are not considered. 

\subsubsection{The in-medium nucleon propagator}
Next, we introduce the in-medium nucleon propagator and the Rarita-Schwinger $\Delta$ propagator which we use 
in our calculations. We show also all the needed vertices, and the in-medium tensor structure 
of the in-medium $\pi\omega\rho$ coupling constant. 

Solid lines in Fig.~~\ref{meddiag} denote 
the in-medium nucleon propagator, which can be conveniently decomposed into the 
\emph{free} and \emph{density} parts \cite{chin}: 
\begin{eqnarray}
iS(k) &=&iS_{F}(k)+iS_{D}(k)=i(\gamma ^{\mu }k_{\mu }+m_{N})[\frac{1}{%
k^{2}-m_{N}^{2}+i\varepsilon }+  \notag \\
&&\frac{i\pi }{E_{k}}\delta (k_{0}-E_{k})\theta (k_{F}-|k|)],  \label{S}
\end{eqnarray}
where $m_{N}$ denotes the nucleon mass, $E_{k}=\sqrt{m_{N}^{2}+k^{2}}$, and $%
k_{F}$ is the Fermi momentum. 
The density effects are produced when exactly one of the nucleon lines in each of the 
diagrams of Fig.~~\ref{meddiag} involves the nucleon density propagator, $S_{D}$. 
The diagrams with more than one $S_{D}$ vanish for kinematic reasons, see Appendix D. 
On the other hand, diagrams with only $S_{F}$ propagators do not contain Fermi-sea effects. 
Therefore, these vacuum diagrams are not considered in the present study. Finally, we 
have only one possibility of having exactly one nucleon density propagator in each 
of the diagrams of Fig.~\ref{meddiag}.

\subsubsection{The Rarita-Schwinger propagator for the $\Delta$ baryon}

In relativistic field theory 
the covariant description of spin $3/2$ particles is based on the Rarita-Schwinger formalism 
\cite{rarita}, where the fundamental object is the spin-vector field $\Psi^{\mu}$. 
The most general Lagrangian for the massive spin $3/2$ field has the following form:
\begin{eqnarray}
{\cal L}=\bar{\Psi}^{\alpha} \Lambda_{\alpha\beta} \Psi^{\beta},
\label{lag1}
\end{eqnarray}
with
\begin{eqnarray}
\Lambda_{\alpha\beta}=-[(-i \partial_{\mu}\gamma^{\mu}+M_{\Delta})g_{\alpha\beta}&-&iA(\gamma_{\alpha}\partial_{\beta}+
\gamma_{\beta}\partial_{\alpha})-\frac{i}{2}(3A^{2}+2A+1)\gamma_{\alpha}\partial^{\mu}
\gamma_{\mu}\gamma_{\beta}\nonumber\\
&-& M_{\Delta}(3A^{2}+3A+1)\gamma_{\alpha}\gamma_{\beta}],
\end{eqnarray}
where $M_{\Delta}$ is the mass of the spin $\frac{3}{2}$ particle ($\Delta$(1232) is the most familiar 
example), and $A$ is an arbitrary parameter ($A\not= -\frac{1}{2}$). 
This Lagrangian is invariant under the transformation
\begin{eqnarray}
\Psi^{\mu}&\rightarrow& \Psi^{\mu}+\alpha\gamma^{\mu}\gamma^{\nu}\Psi_{\nu},\nonumber\\
A&\rightarrow& \frac{A-2\alpha}{1+4\alpha},
\end{eqnarray}
where $\alpha\not= -\frac{1}{4}$, but is otherwise arbitrary. The Lagrangian \ref{lag1} leads to the equations 
of motion
\begin{eqnarray}
\Lambda_{\alpha\beta}\Psi^{\beta}=0.
\end{eqnarray}
The propagator for the massive spin $3/2$ particle satisfies the following equation in momentum space
\begin{eqnarray}
\Lambda_{\alpha\beta}(k)G_{\delta}^{\beta}(k)=g_{\alpha\delta},
\end{eqnarray}
where $g_{\alpha\delta}$ is the metric tensor, and 
\begin{eqnarray}
\Lambda_{\alpha\beta}= -[(-\gamma\cdot k+M_{\Delta})g_{\alpha\beta}&-&A(\gamma_{\alpha}k_{\beta}+
\gamma_{\beta}k_{\alpha})-\frac{1}{2}(3A^{2}+2A+1)\gamma_{\alpha}\gamma\cdot k\gamma_{\beta}\nonumber\\
&-& M_{\Delta}(3A^2 + 3A+1)\gamma_{\alpha}\gamma_{\beta}]\label{lag}.
\end{eqnarray}
Solving for $G$, one gets
\begin{eqnarray}
G_{\alpha\beta}(k)&=&\frac{\gamma\cdot k+M_{\Delta}}{k^2-M_{\Delta}^2}\left[g_{\alpha\beta}-
\frac{1}{3}\gamma_{\alpha}\gamma_{\beta}
-\frac{1}{3M_{\Delta}}(\gamma_{\alpha}k_{\beta}-\gamma_{\beta}k_{\alpha})-
\frac{2}{3M_{\Delta}^2}k_{\alpha}k_{\beta}\right]-\nonumber\\
&-&\frac{1}{3M_{\Delta}^2}\frac{A+1}{2A+1} 
\left[\gamma_{\alpha}k_{\beta}+\frac{A}{2A+1}\gamma_{\alpha}k_{\beta}+\left[\frac{1}{2}\frac{A-1}
{2A+1}\gamma\cdot k -\frac{A M_{\Delta}}{2A+1}\right]\gamma_{\alpha}\gamma_{\beta}\right].\nonumber \\
\end{eqnarray}
The physical properties of the free field are independent of the parameter $A$, so finally, with 
the parameter choice $A=-1$, 
we obtain the Rarita-Schwinger \cite{Ben,rarita} propagator for the spin $\Delta$ particle, 

\begin{equation}
iS_{\Delta }^{\mu \nu }(k)=i\frac{\gamma ^{\mu }k_{\mu }+M_{\Delta }}{%
k^{2}-M_{\Delta }^{2}}(-g^{\mu \nu }+\frac{1}{3}\gamma ^{\mu }\gamma ^{\nu }+%
\frac{2k^{\mu }k^{\nu }}{3M_{\Delta }^{2}}+\frac{\gamma ^{\mu }k^{\nu
}-\gamma ^{\nu }k^{\mu }}{3M_{\Delta }})  \label{rarita}.
\end{equation}
This propagator, often found in the literature, is denoted by the double lines in the diagrams 
of Fig.~~\ref{meddiag}. 

We incorporate phenomenologically the effects of the non-zero width of the $
\Delta $ by the replacement $M_{\Delta }\rightarrow M_{\Delta }-i\Gamma_{\Delta} /2$.
This treatment of the finite width of the $\Delta $ is consistent with the
Ward-Takahashi identities for the $\pi \rho \omega $ vertex. This would not
be true if $\Gamma _{\Delta }$ were introduced in the denominator of Eq. (\ref{rarita}) only.

\subsubsection{Feynman rules}

\begin{figure}[h]
\centerline{\includegraphics[width=0.3\textwidth]{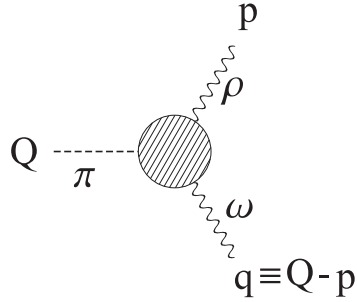}}
\caption{\label{not} The assignment of momenta in the $\pi \omega \rho$ vertex.} 
\end{figure}

In the vacuum, the $\pi \omega \rho $ vertex has the form 
\begin{equation}
-iV_{\pi \omega ^{\mu }\rho ^{\nu }}=i\frac{g_{\pi \omega \rho }}{F_{\pi }}%
\epsilon ^{\mu \nu pQ},  \label{vac}
\end{equation}
where $g_{\pi\omega\rho}$ is a $\pi\omega\rho$ coupling constant, and $F_{\pi}=93~{\rm MeV}$
is the pion coupling constant. We have also used a convenient short-hand notation 
\begin{equation}
\epsilon ^{\mu \nu
pQ}=\epsilon ^{\mu \nu \alpha \beta }p_{\alpha }Q_{\beta }. 
\end{equation}
We choose $Q$ as the incoming momentum of the $\pi $, $p$ as the outgoing momentum of
the $\rho $, and $q\equiv Q-p$ as the outgoing momentum of the $\omega $, see Fig.~~\ref{not}.

In the general case $\omega $ and $\rho $ are virtual particles. The value of $g_{\pi \omega \rho }$ can
be obtained with the help of the Vector Meson Dominance Model from the anomalous $\pi
\gamma \gamma $ decay. According to this model the electromagnetic interactions of 
hadrons are described by intermediate coupling to vector mesons, see Fig.~\ref{dalitz}. 
The vector meson is subsequently converted 
into a virtual photon and finally the photon decays into the lepton pair, $e^{+}e^{-}$. 

\subsubsection{The vacuum $\pi^{0} \rightarrow \gamma \gamma$ decay} 
\begin{figure}[h]
\centerline{\includegraphics[width=0.7\textwidth]{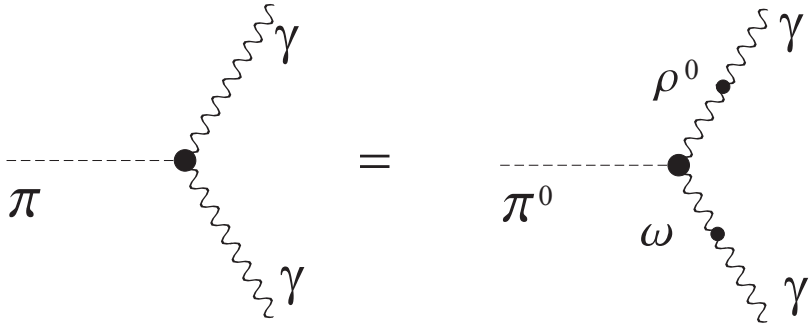}}
\caption{\label{comp} Comparison of two diagrams, $\pi \to \gamma\gamma$ and $\pi^{0}\to\omega\rho$. 
Black dot indicates $g_{\pi\gamma\gamma}=\frac{e^{2}}{4\pi^{2}F_{\pi}}$ and 
$\frac{g_{\pi\omega\rho}}{F_{\pi}^{2}}=\frac{g_{\omega}g_{\rho}}{F_{\pi}e^{2}}$ coupling constants. 
The smaller black dot is 
equal to $\frac{em_{\omega}^{2}}{g_{\omega}}$ or $\frac{em_{\rho}^{2}}{g_{\rho}}$, according to VDM.} 
\end{figure}

One can calculate the $g_{\pi\gamma\gamma}$ coupling constant from 
the width of the $\pi ^{0}\rightarrow \gamma \gamma$ decay in the vacuum.
The transition amplitude of $\pi ^{0}\rightarrow \gamma \gamma$ decay has the form
\begin {equation}
\mathcal{M}=g_{\pi\gamma\gamma}\epsilon ^{\mu \nu p Q} \epsilon^{\nu'}\epsilon^{\nu}, 
\end{equation}
where the value of $g_{\pi \gamma\gamma }=\frac{e^{2}}{4\pi ^{2}F_{\pi }}$  
is model independent and results from Adler-Bell-Jackiw anomaly. The formula for the decay width is
\begin{equation}
d{\Gamma}_{\pi\gamma\gamma}=\frac{1}{32\pi^{2}}\sum_{\epsilon \epsilon'}|\mathcal{M}|^{2}
\frac{|\mathbf{p}|}{m_{\pi}^{2}} d{\Omega}\times \frac{1}{2}, \label{width1} 
\end {equation}
where, ${|\mathbf{p}|}=m_{\pi}/2$ is momentum of one of the photons in the rest frame of the pion, 
and the factor of $1/2$ comes from the 
symmetrization over the indistinguishable bosons in the final state. Since the sum over 
all photon polarizations with momentum $p$ gives
\begin{equation}
\sum_{a} \epsilon_{(a)}^{\mu} \epsilon_{(a)}^{\nu^{\ast}} = g^{\mu \nu}-\frac{p^{\mu}p^{\nu}}{p^2}, \label{gmini} 
\end{equation}
and from kinematics $q^2=m_{\pi}^2$, $(Q-p)^2=0$, we find
\begin {equation}
\sum_{\epsilon \epsilon'}|\mathcal{M}|^{2}=g_{\pi\gamma\gamma}^2 \frac{m_{\pi}^4}{2}.
\end{equation}
Consequently, the width has a form
\begin{equation}
\Gamma_{\pi\gamma\gamma}=g_{\pi\gamma\gamma}^2 \frac{m_{\pi}^3}{64\pi},
\end{equation}
where $g_{\pi\gamma\gamma}$, applying the VDM (see Fig.~~\ref{dalitz}), can be expressed in 
terms of the $g_{\pi\omega\rho}$ coupling constant as (see Fig.~\ref{comp})
\begin{equation} 
g_{\pi\gamma\gamma}^{2}=g_{\pi\omega\rho}^{2} \frac{e^2}{F_{\pi}^2 g_{\rho}^2 g_{\omega}^2}.
\end{equation}
Finally, the width is 
\begin{equation}
\Gamma_{\pi\gamma\gamma}=g_{\pi\omega\rho}^2 \frac{e^2 }
{F_{\pi}^2 g_{\rho}^2 g_{\omega}^2}\frac {m_{\pi}^3}{64\pi}.
\end{equation}
Using the physical parameters 
which we collect in Eq.~\ref{par}, as well as $\Gamma_{\pi\gamma\gamma}=7.6\mathrm{eV}$, 
$m_{\pi}=135\mathrm{MeV}$, $e^2=4\pi\alpha$ (where $\alpha=\frac{1}{137}$), we obtain 
the value of coupling constant $g_{\pi\omega\rho}$
by comparison of the $\pi ^{0}\rightarrow \gamma \gamma$ and 
$\pi ^{0}\rightarrow \rho \omega $ diagrams, Fig.~\ref{comp}. 
It gives $g_{\pi\omega\rho }=-\frac{g_{\rho }g_{\omega }}{e^{2}}g_{\pi \gamma \gamma }=-1.36$ 
(in our convention the sign of the coupling is negative).

\subsubsection{Meson-NN and meson-N$\Delta$ vertices}
The meson-NN vertices needed for our calculation have the standard form
\begin{eqnarray}
-iV_{\omega ^{\mu }NN} &=&ig_{\omega }\gamma ^{\mu },  \notag \\
-iV_{\rho _{b}^{\mu }NN} &=&i\frac{g_{\rho }}{2}(\gamma ^{\mu }-\frac{%
i\kappa _{\rho }}{2m_{N}}\sigma ^{\mu \nu }p_{\nu })\tau ^{b}, \\
-iV_{\pi ^{a}NN} &=&\frac{g_{A}}{2F_{\pi }}\slashchar{Q}\gamma ^{5}\tau ^{a}, 
\notag
\end{eqnarray}
where $p$ is the outgoing momentum of the virtual $\rho $, $Q$ is the incoming
four-momentum of the $\pi $, and $a$, $b$ are the isospin indices of the 
$\pi $, and $\rho $ respectively, $g_{\omega}$ and $g_{\rho}$ are coupling constants 
for the $\omega$ and $\rho$ meson, $\kappa_{\rho}$ is tensor coupling of the $\rho$ meson, and 
$g_{A}$ denotes the axial coupling constant. 

For the interactions involving the $\Delta $ we use the same choice as in Ref. \cite{rhopipi}: 
\begin{eqnarray}
-iV_{N\Delta ^{\alpha }\pi ^{a}} &=&g_{N\Delta \pi }Q^{\alpha }T^{a},  
\notag \\
-iV_{N\Delta ^{\alpha }\rho _{b}^{\mu }} &=&ig_{N\Delta \rho }(\slashchar{p}
\gamma ^{5}g^{\alpha \mu }-\gamma ^{\mu }\gamma ^{5}p^{\alpha })T^{b}, \\
-iV_{\Delta ^{\alpha }\Delta ^{\beta }\omega ^{\mu }} &=&-ig_{\omega
}(\gamma ^{\alpha }\gamma ^{\mu }\gamma ^{\beta }-\gamma ^{\beta }g^{\alpha
\mu }-\gamma ^{\alpha }g^{\beta \mu }+\gamma ^{\mu }g^{\alpha \beta }), 
\notag
\end{eqnarray}
where $T^{a}$ are the isospin $\frac{1}{2}\rightarrow 
\frac{3}{2}$ transition matrices \cite{EW}. 
They are defined through the Clebsch-Gordan coefficients as follows: $\langle \frac{3%
}{2},I_{3}|T^{\mu }|\frac{1}{2},i_{3}\rangle =\langle \frac{1}{2}1i_{3}\mu |1%
\frac{1}{2}\frac{3}{2}I_{3}\rangle $, with $i_{3}$ and $I_{3}$ denoting the
isospin of the nucleon and $\Delta $, respectively. In the Cartesian basis the
explicit form reads 
\begin{equation}
T^{1}=\left( 
\begin{array}{cc}
-\frac{1}{\sqrt{2}} & 0 \\ 
0 & -\frac{1}{\sqrt{6}} \\ 
\frac{1}{\sqrt{6}} & 0 \\ 
0 & \frac{1}{\sqrt{2}}
\end{array}
\right) ,\qquad T^{2}=i\left( 
\begin{array}{cc}
\frac{1}{\sqrt{2}} & 0 \\ 
0 & \frac{1}{\sqrt{6}} \\ 
\frac{1}{\sqrt{6}} & 0 \\ 
0 & \frac{1}{\sqrt{2}}
\end{array}
\right) ,\qquad T^{3}=\left( 
\begin{array}{cc}
0 & 0 \\ 
\sqrt{\frac{2}{3}} & 0 \\ 
0 & \sqrt{\frac{2}{3}} \\ 
0 & 0
\end{array}
\right) ,  \label{Tmat}
\end{equation}
where the columns are labeled by $i_{3}=\frac{1}{2},-\frac{1}{2}$, left to
right, and the rows by $I_{3}=\frac{3}{2},\frac{1}{2},-\frac{1}{2},-\frac{3}{%
2}$, top to bottom. The following useful relation holds:
\begin{equation}
T^{a\dagger }T^{b}=\frac{2}{3}\delta ^{ab}-\frac{1}{3}\varepsilon ^{abc}\tau
_{c}.  \label{Talg}
\end{equation}
The values of the $N\Delta $ coupling constants follow from the
non-relativistic reduction of the vertices and comparison to the
non-relativistic values \cite{EW}. The $\Delta \Delta \omega $ coupling
incorporates the principle of universal coupling. It is obtained by replacing $k^{\mu}\to k^{\mu}+\omega^{\mu}$ in 
the Rarita-Schwinger Lagrangian, Eq.~\ref{lag}.

Note that the $\rho \Delta\Delta $ vertex does not appear in this study. Presence of $\rho\Delta\Delta$ 
coupling in diagram \ref{meddiag} implies that the $\omega N \Delta$ coupling should also appear. 
This is impossible from the Wigner-Eckart theorem, since $\omega$ is an isoscalar particle. 

The choice of physical parameters is taken from other works \cite{Gomez} and is as follows: 
\begin{eqnarray}
g_{\omega } &=&10.4,\quad g_{\rho }=5.2,\quad \quad g_{A}=1.26, \notag \\
g_{N\Delta \pi } &=&\frac{2.12}{m_{\pi }},\quad g_{N\Delta \rho }=\frac{2.12%
\sqrt{2}}{m_{\pi }}.  \label{par}
\end{eqnarray}
In the results presented
below we have chosen the tensor coupling of the $\rho$ meson, $\kappa _{\rho }=3.7$.
The principle of universal coupling  and the field current identity imply that the isovector 
part of the anomalous magnetic moment of the nucleon, $\kappa_{V}$, is equal to the tensor 
coupling of the $\rho$ meson,{\em \ i.e.} $\kappa _{\rho
}=\kappa _{V}=3.7$, \cite{ren}, $\kappa_{V}$ is the isovector anomalous magnetic moment of the nucleon. 
However, the empirical value of $\kappa_{\rho}/\kappa_{V}$ 
extracted by Hohler and Pietarinen \cite{hohler} 
using a dispersion relation analysis of the $\pi N$ scattering data is $1.78$.
Qualitatively, in our analysis similar conclusions follow when $\kappa
_{\rho }=6$ is used.

We have to admit here that the relativistic form of
couplings to the $\Delta $, as well as the values of the coupling constants,
are the subjects of an on-going discussion and research \cite
{Ben,Pascal,Hemmert,Haber}. We are interested in estimating the size of the medium effect 
on the $\pi\omega\rho$ vertex and we do not search for precise and accurate predictions of 
tensor couplings. 

In order to simplify the approach we do not include any form factors in the vertices. 
In general, a vertex may depend on outgoing meson momentum  squared, $p^{2}$, $q{^2}$, 
Fig.~\ref{not}, 
and the Lorentz scalars, $\bf{p} \cdot \bf{k}$ and $\bf{q} \cdot \bf{k}$, which enter 
the so-called sidewise form factors. Those form factors are not included in our 
calculations, because they lead to fundamental problems with gauge symmetry conservation. In particular the 
current conservation identities, Eq.~\ref{ward}, are violated.

\subsubsection{Tensor structure of the $\pi \omega \rho $ vertex in nuclear medium} 
In our calculation we evaluate the diagrams of Fig.~~\ref{meddiag} and obtain 
the in-medium $\pi \omega \rho $ vertex function 
\begin{eqnarray}
A^{\mu \nu }=\int \frac{d^{4}k%
}{(2\pi )^{4}}\frac{m_{N}}{E_{k}}\theta (k_{f}-\left| \vec{k}\right|) A'^{\mu\nu}\delta(k_{0}-E_{k}),
\label{ami1}
\end{eqnarray}
where
\begin{eqnarray}
A'^{\mu\nu}&=&A_{1}\varepsilon ^{\mu \nu pQ}+A_{2}\varepsilon ^{\mu \nu
kQ}+A_{3}\varepsilon ^{\mu \nu pk}+(A_{4}p^{\nu }+A_{5}Q^{\nu }+A_{6}k^{\nu})\varepsilon ^{\mu kpQ}\nonumber\\
\label{ami2}
&+&(A_{7}p^{\mu }+A_{8}Q^{\mu}+A_{9}k^{\mu})\varepsilon ^{\nu kpQ}.
\end{eqnarray}
Thus in medium we have 9 possible tensor structures, while in the vacuum there was only one, 
$\epsilon^{\mu\nu p Q}$, Eq~\ref{vac}.
All coefficients $A_{i}$ are scalar functions of the 
four-vectors \emph{Q}, \emph{p} and \emph{k}, with $k^{0}=E_{k}$. 
The occupation function is made explicitly Lorentz-invariant when we write $|{\bf k}|=\sqrt{%
(k \cdot u)^2 -m_N^2}$, where $u$ is the four-velocity of the medium. 

The term with $k^{\mu}$, upon the evaluation of the integral can be in general proportional to the three Lorentz 
vectors present in the problem, $p^{\mu}$, $Q^{\mu}$, and $u^{\mu}$, where $u^{\mu}$ is the four-velocity of the 
medium.
\begin{eqnarray}
\int \frac{d^{4}k
}{(2\pi )^{4}}\frac{m_{N}}{E_{k}}A k^{\mu} \theta (k_{f}-\left| \vec{k}\right|) \delta(k_{0}-E_{k})=
A_{p} p^{\mu} + A_{q} Q^{\mu} +A_{u} u^{\mu}, \label{term}
\end{eqnarray}
where $A_{p}$, $A_{q}$ and $A_{u}$ are scalar functions of $p^{2}$, $Q^{2}$, $p\cdot Q$, $p\cdot u$, 
$q\cdot u$, and $k_F$. Contracting Eq.~\ref{term} with $Q_{\mu}$, $p_{\mu}$ and $u_{\mu}$ we obtain a linear 
algebraic equations for $A_{p}$, $A_{q}$ and $A_{u}$ which may be solved.
To make this problem simpler we use the leading density approximation. 
This is equivalent to putting the three-vector ${\bf k}$ to zero in the 
functions $A_{2}-A_{9}$ in Eq.~\ref{ami1} in the rest frame of the medium.
Then we replace the loop momentum integration by 
\begin{eqnarray}
\int \frac{d^{4}k%
}{(2\pi )^{4}}\theta (k_{f}-\left| \vec{k}\right| )\delta
(k_{0}-E_{k})\rightarrow \frac{1}{8\pi }\rho _{B}, 
\end{eqnarray}
where $\rho _{B}$ is the baryon density.
Now, with ${\bf k}=0$, $k^{0}=m_{N}$, the 
contraction of Eq.~\ref{term} with $q_{\mu }$, $p_{\mu }$, and $u_{\mu }$
gives the set of equations 
\begin{eqnarray}
\frac{1}{8 \pi}\rho _{B}m_{N}p^{0} A({\bf k}=0) &=& A_{p}p^{2}+A_{q}Q\cdot
p+A_{u}p^{0},  \nonumber \\
\frac{1}{8 \pi}\rho _{B}m_{N}Q^{0} A({\bf k}=0) &=& A_{p}p\cdot
q+A_{q}Q^{2}+A_{u}q^{0},  \label{set} \\
\frac{1}{8 \pi}\rho _{B}m_{N}u^{0} A({\bf k}=0)  &=& A_{p}p\cdot u+A_{q}Q\cdot u+A_{u} u^{0},
\nonumber
\end{eqnarray}
where $u^{0}=1$ in the rest frame of the medium.

Since in the general case the 
vectors $p$, $Q$, and $u$ are linearly-independent, the solution of Eqs.~\ref{set} 
is $A_{p}=A_{q}=0$, $A_{u}=\frac{1}{8 \pi}\rho _{B}m_{N} A({\bf k}=0)$. 
Thus, only the term proportional to $u^{\mu }$ in Eq. (\ref{term}) is
present in the leading-density approximation. 
Summarizing, above  calculations, similarly as in 
\cite{rhopipi}, are equivalent to replacing $k^{\mu}$ by $m_{N}u^{\mu }$, 
what is valid for every occurrence of $k$ 
in Eq.~\ref{ami1}. For the terms in this equation involving second powers of $k$, the same prescription 
holds, as may be simply verified by an analysis analogous to the one presented above.

Finally, we obtain the most general tensor structure of the form  
\begin{eqnarray}
A^{\mu\nu}&=&A_{1}\varepsilon ^{\mu \nu pQ}+A_{2}\varepsilon ^{\mu \nu
uQ}+A_{3}\varepsilon ^{\mu \nu pu}+(A_{4}p^{\nu }+A_{5}Q^{\nu }+A_{6}u^{\nu})\varepsilon ^{\mu upQ}\nonumber\\
\label{struct}
&+&(A_{7}p^{\mu }+A_{8}Q^{\mu}+A_{9}u^{\mu})\varepsilon ^{\nu upQ}.
\end{eqnarray}

This structure, restricted by the Lorentz invariance and parity, is more
general than in the vacuum case Eq.~\ref{vac} due to the availability of the
four-velocity of the medium, $u$. The result of any dynamical calculation will
assume the form Eq~\ref{struct}. 
The vertex $A^{\mu \nu }$ satisfies the current conservation 
identities:
\begin{eqnarray}
Q_{\mu }A^{\mu \nu }=0, ~~~~\rm{and} ~~~~p_{\nu }A^{\mu \nu }=0,
\label{ward} 
\end{eqnarray}
which serves as a useful check of the algebra.

For the numerical analysis we have developed a program written in 'Mathematica'. In order to calculate traces from 
all diagrams of Fig.~\ref{meddiag}, we have to apply the Feynmann rules, use Lorentz index contractions, 
and compute traces of the  
Dirac matrices (up to as many as 14). It is rather complicated from the technical point of view, 
thus we use FeynCalc  \cite{feyncalc}. This is a package 
for algebraic calculation in elementary particle physics, which  
is very useful for many problems involving baryon loops with density-dependent nucleon propagators. 
All our calculations are covariant.

\section {Decays of particles at rest with respect to the medium}
We begin the presentation of our results for the case where the decaying particle 
is at rest with respect to the nuclear matter, ${\bf q}=0$.

\subsubsection{The in-medium $\pi ^{0}\rightarrow\gamma \gamma$ decay} 
First, numerical results are discussed for the process$\ \pi
^{0}\rightarrow \gamma \gamma $, because of its fundamental nature. This is
the decay into two real photons, where the pion is at rest with respect to
the medium. In this case the four-vectors 
$Q$ and $u$ are parallel, and out of the $9$ structures in
Eq. \ref{struct} only $A_{1}$ and $A_{3}$ survive: 
\begin{eqnarray}
A^{\mu \nu }=A_{1}\varepsilon ^{\mu
\nu pQ}+A_{3}\varepsilon ^{\mu \nu pu}. 
\end{eqnarray}
Thus in the rest frame of the
medium, where $Q=(m_{\pi },0,0,0),$ it is found that
\begin{eqnarray}
A^{\mu \nu}&=&A_{1}Q_{0}\varepsilon ^{\mu \nu p0}+A_{3}u_{0}\varepsilon ^{\mu \nu p0}
\label{vacuum}\\
&=&(m_{\pi }A_{1}+A_{3})\varepsilon ^{\mu \nu p0}=(A_{1}+A_{3}/m_{\pi
})\varepsilon ^{\mu \nu pQ}. \nonumber 
\end{eqnarray}
The following notation is introduced, which combines the vacuum and medium pieces: 
\begin{eqnarray}
A^{\mu \nu }=\frac{i}{F_{\pi }}\frac{e^{2}}{%
g_{\rho }g_{\omega }}\left( g_{\pi \rho \omega }+\rho _{B}B\right) \epsilon
^{\nu \mu pQ}. \label {amini}
\end{eqnarray}
All the medium effects are collected in the constant $B$. The
results of the medium-modified vertex at the nuclear-saturation density ($%
\rho_{B}=\rho _{0}=0.17~~\mathrm{fm}^{-3}$) are presented relative to the vacuum
value $g_{\mathrm{vac}}=g_{\pi \rho \omega }$. In this 
analysis the medium effects are linear in the baryon density. We define 
\begin{equation}
g_{\mathrm{eff}}=g_{\pi \rho \omega }+\rho _{B}B,  \label{geff}
\end{equation}
with $B$ denoting the medium contribution. 
Separate contributions to $B$ coming from diagrams (a), (b) and (c) of Fig.~~\ref{meddiag}
for the process $ \pi^{0}\rightarrow \gamma \gamma $ are as follows: 
\begin{eqnarray}
B_{(a)} &=&\frac{g_{A}g_{\rho }g_{\omega }(\kappa _{\rho }+1)}{m_{N}(4m_{N}^{2}-m_{\pi }^{2})}, \\
B_{(b)} &=&\frac{g_{N\Delta \pi }g_{N\Delta \rho }g_{\omega }F_{\pi }m_{\pi }%
}{9M_{\Delta }^{2}m_{N}[(m_{N}^{2}-M_{\Delta }^{2})^{2}-m_{\pi
}^{2}m_{N}^{2}]}(m_{N}^{4}+M_{\Delta }m_{N}^{3}-2M_{\Delta }^{2}m_{N}^{2}- 
\notag \\
&&-m_{\pi }^{2}m_{N}^{2}-M_{\Delta }^{3}m_{N}+M_{\Delta }^{4}),  \notag \\
B_{(c)} &=&\frac{g_{N\Delta \pi }g_{N\Delta \rho }g_{\omega }F_{\pi }m_{\pi }%
}{27M_{\Delta }^{4}[(m_{N}^{2}-M_{\Delta }^{2})^{2}-m_{\pi }^{2}m_{N}^{2}]}%
(-4m_{N}^{5}-6M_{\Delta }m_{N}^{4}+8M_{\Delta }^{2}m_{N}^{3}+  \notag \\
&&+5M_{\Delta }^{3}m_{N}^{2}+(4m_{N}+3M_{\Delta })m_{\pi
}^{2}m_{N}^{2}-16M_{\Delta }^{4}m_{N}-11M_{\Delta }^{5}),  \notag
\end{eqnarray}
where $M_{\Delta }$ is understood to carry the width $i\Gamma _{\Delta }$.
With parameters (\ref{par}) we find numerically 
for the formal case $\Gamma _{\Delta }=0$, 
\begin{eqnarray}
B_{(a)}=97{\rm GeV}^{-3}, %
B_{(b)}=-3.1{\rm GeV}^{-3}, 
B_{(c)}=-102{\rm GeV}^{-3}.  
\end{eqnarray}
For the physical vacuum value of the $\Delta $
width, $\Gamma _{\Delta }=0.12~~{\rm GeV},$ we find
\begin{eqnarray}
B_{(a)}=97~~{\rm GeV}^{-3},%
B_{(b)}=-(2.6+1.5i)~~{\rm GeV}^{-3}, 
B_{(c)}=-(87+42i)~~{\rm GeV}^{-3},\nonumber\\
\end{eqnarray}
and 
\begin{eqnarray}
|g_{{\rm eff}}/g_{{\rm vac}}|^{2}=\left| 0.99+0.04i\right| ^{2}=0\,.99,
\label{gef1}
\end{eqnarray}
whereas for nucleons only (diagram (a)) we would have 
\begin{eqnarray}
|g_{{\rm eff}}^{(a)}/g_{{\rm vac}}|^{2}=0.91^{2}=0.82.
\label{gef2} 
\end{eqnarray}
The imaginary values come from the $i\Gamma_{\Delta}$ term.
From Eq.~\ref{gef1} and \ref{gef2} we see that the effects of the $\Delta $
act in the opposite direction than the nucleons, resulting in an almost
complete cancellation between diagrams (a) and (a+b+c) of Fig.~~\ref{meddiag}.

\subsubsection{The $\pi ^{0}\rightarrow\gamma \gamma ^{\ast }$ decay} 
\begin{figure}[h]
\begin{center}
\includegraphics[width=.35\textwidth]{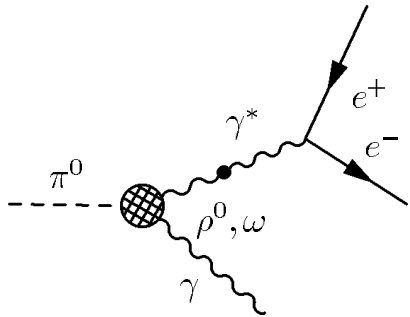}
\caption{\label{dalitzpi} The  $\pi ^{0}\rightarrow
\gamma \gamma ^{\ast }\rightarrow \gamma e^{+}e^{-}$ decay. Wavy lines indicates 
the $\rho$, $\omega$ or $\gamma$, dashed lines the pions, the hatched blob denotes the medium-modified vertex, 
the black dot is equal to $\frac{em_{\omega }^{2}}{g_{\omega }}$ 
or $\frac{em_{\rho }^{2}}{g_{\rho }}$ according to VDM.}
\end{center}
\end{figure}

The Dalitz decay of $\pi ^{0}$ into a photon and a lepton pair proceeds
through the decay into a real and a virtual photon, and subsequently the
decay of the virtual photon into the lepton pair, $\pi ^{0}\rightarrow
\gamma \gamma ^{\ast }\rightarrow \gamma e^{+}e^{-}$, see Fig.~~\ref{dalitzpi}.
The contributions to the constant $B$ from all diagrams of Fig.~~\ref{meddiag}
for the processes $\protect\pi ^{0}\rightarrow \protect\gamma_{(\omega)}\protect\gamma ^{\ast }$ and 
$\protect\pi ^{0}\rightarrow \protect\gamma_{(\rho)}\protect\gamma ^{\ast },$ are listed in Appendix A.

The virtual photon can
be either isovector ($\rho $-type) or isoscalar ($\omega $-type); thus $\gamma_{(\omega)}$ denotes a photon 
which couples as an isoscalar and $\gamma_{(\rho)}$ a photon which couples as an isovector.
We denote the virtuality of $\gamma ^{\ast }$ as $M_{\gamma*}$ and investigate the dependence
of $g_{\mathrm{eff}}$ on $M_{\gamma*}$. 
\begin{figure}[h]
\centerline{
\epsfysize = 11 cm \centerline{\epsfbox{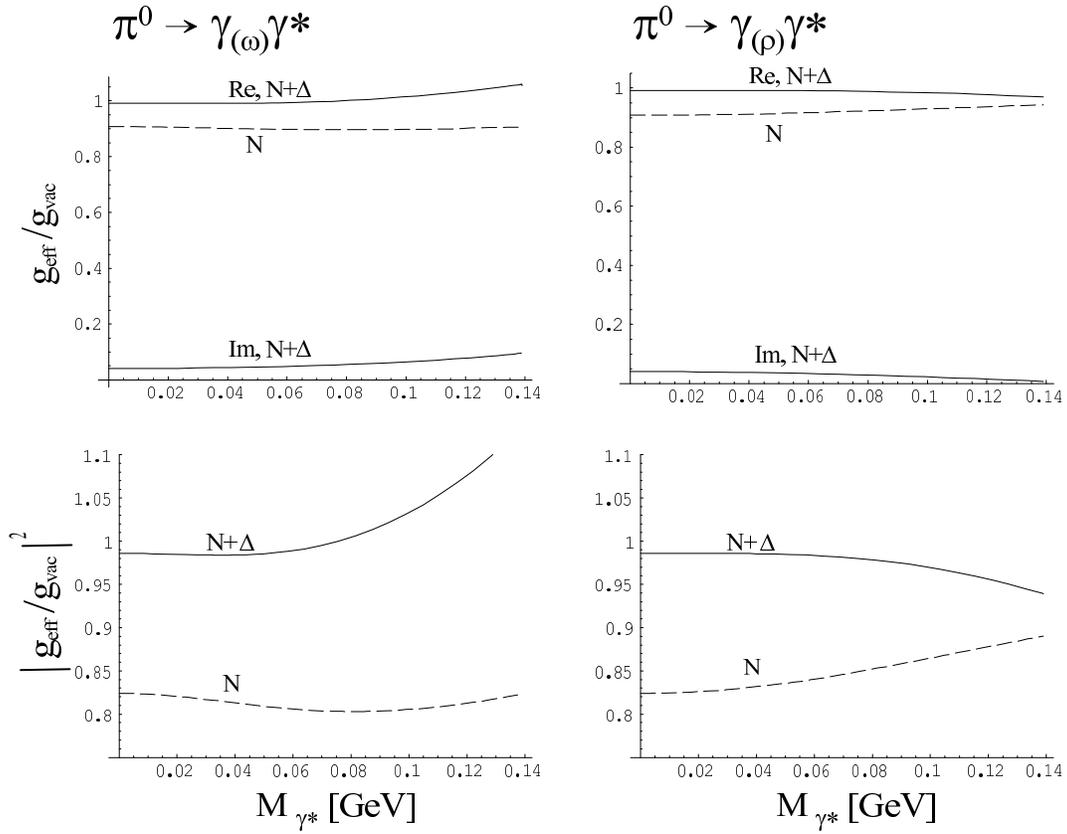}} \vspace{0mm}} 
\caption{\label{pggorimre}The quantity $g_{{\rm eff}}/g_{{\rm vac}}$ and $|g_{{\rm eff}}/g_{{\rm vac}}|^{2}$  
for the processes $\protect\pi ^{0}\rightarrow \protect\gamma_{(\omega)}\protect\gamma ^{\ast }$ and 
$\protect\pi ^{0}\rightarrow \protect\gamma_{(\rho)}\protect\gamma ^{\ast },$ plotted
as a function of the virtual mass of $\protect\gamma ^{\ast }$, $M_{\gamma^{\ast}}$.
The solid lines show the results of the full calculation with all diagrams (a+b+c) of Fig.~~\ref{meddiag} included,
while the dashed lines correspond to the calculations with nucleon diagram (a) only. 
The upper plots include results of real and imaginary parts. 
All plots are done for the case $\protect\rho _{B}=\protect\rho _{0}$ 
and $\Gamma _{\Delta }=120~~{\rm MeV}$.}
\end{figure}
The results both for the process $\protect\pi ^{0}\rightarrow \protect\gamma_{(\omega)}\protect\gamma ^{\ast }$ 
and $\protect\pi ^{0}\rightarrow \protect\gamma_{(\rho)}\protect\gamma ^{\ast },$  
are displayed in Fig.~~\ref{pggorimre}. All calculations are done at the saturation density 
$\protect\rho _{B}=\protect\rho _{0}$ and with the vacuum value of the $\Delta$ width, 
$\Gamma _{\Delta }=120~~{\rm MeV}$. 
In the two upper plots of Fig.~~\ref{pggorimre} the solid lines show the 
real and imaginary parts of the full calculations, 
which means that we take into account all diagrams of Fig.~~\ref{meddiag} (with nucleons and delta isobars), 
whereas the dashed lines show the calculations with nucleons only which, of course, are real. 
The $\Delta$ resonance has imaginary contribution for $Im\Gamma_{\Delta}$, whereas for 
nucleons the result is always real.  

In the two bottom plots of the same figure, 
the solid and dashed curves indicate full calculations with deltas and nucleons, respectively. 
The results displayed in Fig.~~\ref{pggorimre}, indicate that the change of $M_{\gamma*}$ in the 
allowable kinematic range from $0$ to $m_{\pi }$ has 
very little effect on the ratio of $g_{{\rm eff}}/g_{{\rm vac}}$ and $|g_{{\rm eff}}/g_{{\rm vac}}|^{2}$.
A somewhat different behavior for the isoscalar and isovector photons
at $M_{\gamma^{\ast}}$ approaching $m_{\pi }$ is noticed. 
We observe that curves corresponding to the calculations with the 
$\Delta$ isobar shift the curve slightly up for $\pi^0\to\gamma_{(\omega)}\gamma$ decay, whereas for isovector 
photons the curves 
are pulled down, when approaching $m_{\pi}$ value. On the other hand for both of those decays 
the shape of curves without the $\Delta$ isobar is similar.
By comparing the curves 
corresponding to the calculation with the $%
\Delta $ (diagrams (a+b+c) of Fig.~~\ref{meddiag}) and with nucleons only (diagram (a)),
it is visible that they act in opposite directions in diagrams (a) 
and (a+b+c), and it holds for the whole kinematically-available range.

\begin{figure}[tb]
\centerline{
\epsfysize = 12 cm \centerline{\epsfbox{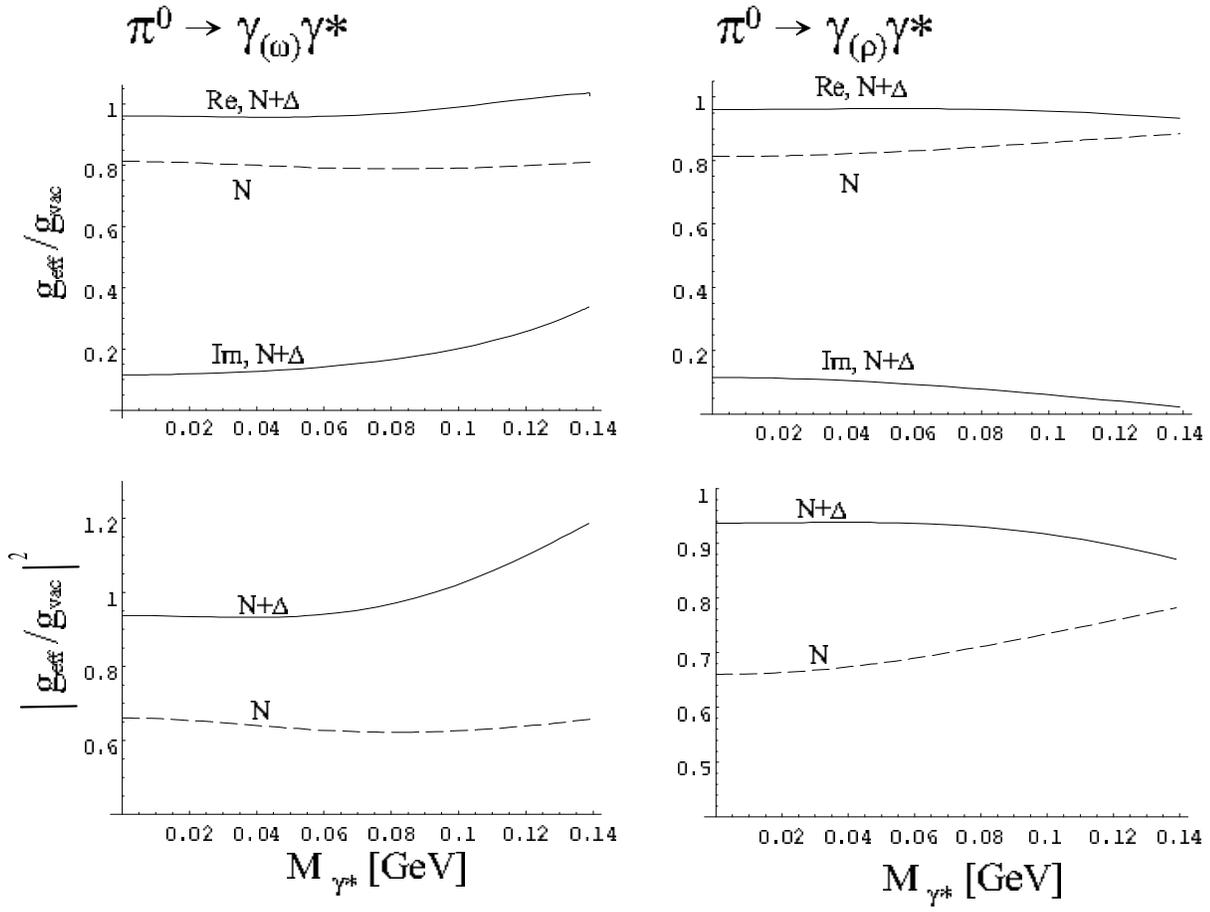}} \vspace{0mm}} 
\caption{\label{pgg70per}Same as Fig.~~\ref{pggorimre}, but for the values of $m_{N}$ 
and $M_{\Delta}$ reduced to 70\% of their vacuum values.} 
\end{figure}

In Fig.~~\ref{pgg70per}, we show the result of the calculation with fixed $\Gamma _{\Delta }$, 
but with $m_{N}$ and $M_{\Delta}$ scaled down to 70\% of their vacuum value. 
Such a reduced baryon mass is typical at the nuclear saturation density and is predicted by the 
Brown-Rho scaling hypothesis, see Chapter 1. 
The reduced mass of $\Delta$ is expected to behave similarly to the nucleon case.
We notice that $g_{\rm eff}$ is practically unchanged, 
and Fig.~~\ref{pggorimre} and Fig.~~\ref{pgg70per} show a similar behavior. 

\subsubsection{The $\omega \rightarrow \pi ^{0}\gamma ^{\ast }$
and $\rho ^{a}\rightarrow \pi ^{a}\gamma ^{\ast }$ decays} 

Next, we consider the processes $\omega \rightarrow \pi ^{0}\gamma ^{\ast }$
and $\rho ^{a}\rightarrow \pi ^{a}\gamma ^{\ast }$, where $\omega $ and $%
\rho $ are on the mass shell. In these cases the following vacuum values
of the coupling constants are used, $g_{\omega \rightarrow \rho \pi }=-1.13$ and $%
g_{\rho \rightarrow \omega \pi }=-0.76$. They are calculated from the experimental
partial decay widths 
\begin{eqnarray}
\Gamma _{\omega\pi\gamma }=717\mathrm{keV} 
~~~~\rm{and} ~~~~\Gamma _{\rho\pi\gamma }=79\mathrm{keV}.
\end{eqnarray}
Let us concentrate on the first decay. 
The decay width for the process $\omega \rightarrow \pi ^{0}\gamma$ has the form
\begin{equation}
\Gamma_{\omega\pi\gamma}=\sum_{\epsilon_{(1)} \epsilon_{(2)}}\frac{4\pi \mathcal{M}^2 p_{\rho}^2}
{32\pi^2 3m_{\omega}^2},
\end {equation}
where
\begin{equation}
\mathcal{M}=g_{\omega\pi\gamma}\epsilon ^{\mu \nu p q}\epsilon^{\mu}_{(1)}\epsilon^{\nu}_{(2)} 
=\frac{e g_{\omega\pi\rho}}{F_{\pi} g_{\rho}}
\epsilon ^{\mu \nu p q}\epsilon^{\mu}_{(1)}\epsilon^{\nu}_{(2)},
\end{equation}
and with the momentum of the $\rho$ 
\begin{equation}
p_{\rho}^2 = \frac{(m_{\omega}^2 + m_{\rho}^2 - m_{\pi}^2)^2}{4 m_{\omega}^2} - m_{\rho}^2.
\end{equation}
Putting together all the needed parameters from Eq.~~\ref{par} we 
obtain the physical value of the coupling constant, $g_{\omega\pi\rho}=-1.13$. In a similar way we obtain 
$g_{\rho\pi\omega}$. 
Note that these coupling constants differ from each other, as well as from $g_{\pi
\omega \rho }=-\frac{g_{\rho }g_{\omega }}{e^{2}}g_{\pi \gamma
\gamma }=-1.36$ used earlier for the $\pi ^{0}$ decay. These differences may
be attributed to form-factor effects: the virtuality of the vector mesons is
changed from 0 in the $\pi^{0}\to\gamma\gamma$ decay to the on-shell value in the vector meson decay.
For the case $\bf q=0$ the contributions to $B$ from Eq.~\ref {amini} corresponding to the diagram (a,b,c)
from Fig.~\ref{meddiag}, can be written as
\begin{eqnarray}
A^{\mu \nu }_{(i)}=\frac{i}{F_{\pi }}\frac{e^{2}}{%
g_{\rho }g_{\omega }}\left( g_{\pi \rho \omega }+\rho _{B}\frac{N_{(i)}}{D_{(i)}}\right) \epsilon
^{\nu \mu pQ}, \label {amini1}
\end{eqnarray}
The formulas for $N_{(i)}$ and $D_{(i)}$ are quite cumbersome, 
which reflects the presence of many terms in the Rarita-Schwinger propagator. We give them in Appendix B.
\begin{figure}[tb]
\centerline{
\epsfysize = 12 cm \centerline{\epsfbox{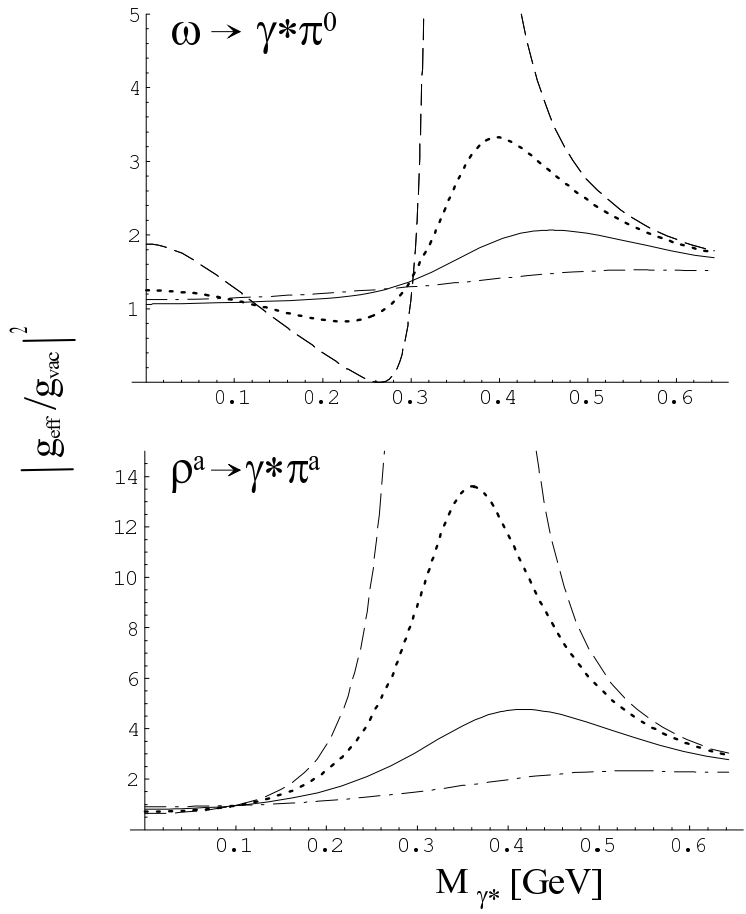}} \vspace{0mm}} 
\caption{\label{dens}Top: $|g_{{\rm eff}}/g_{{\rm vac}}|^{2}$ for the decay $\protect%
\omega \to \protect\gamma^\ast \protect\pi^0$ at the nuclear saturation
density, plotted as a function of the virtuality of the photon, $M_{\gamma^{\ast}}$. The
dashed, dotted, solid and dot-dash lines correspond to the cases, $\Gamma _{\Delta }=0$, 
$\Gamma _{\Delta }=60~{\rm MeV}$, 
$\Gamma _{\Delta }=120~{\rm MeV}$, $\Gamma _{\Delta }=240~{\rm MeV}$, respectively. Bottom: 
the same for the decay 
$\protect \rho^{a} \to \protect\gamma^\ast \protect\pi^a$.}   
\end{figure}

There is an interesting phenomenon in the decays of vector mesons, related
to the analyticity of the amplitudes of Fig.~~\ref{meddiag}(b) and (c) in the virtuality of $%
\gamma^{\ast}$. From the denominators of diagrams 
(b) and (c) of Fig.~~\ref{meddiag}, see Appendix B for the case $\Gamma_{\Delta }=0$, we can see that the
amplitudes from Eq.~~\ref{amini1} have a pole at the value of $M_{\gamma^{*}}$ equal to
\begin{equation}
M_{0}=\sqrt{\frac{m_{v}m_{N}^{2}+m_{\pi }^{2}m_{N}+m_{v}^{2}m_{N}-M_{\Delta
}^{2}m_{v}+m_{\pi }^{2}m_{v}}{m_{N}}}\approx 0.34{\rm GeV}, \label{pole}
\end{equation}
where $m_{v}=m_{\rho}$ or $m_{\omega}$. The above analytic structure is manifest in the numerical calculation 
presented below. For $\Gamma_{\Delta}\ne 0$ the pole at \ref{pole} moves away from the real axis. Thus, 
analyticity is important and immediately leads to large changes of the vertex function near the poles.
This formal feature of the presence of the pole can be clearly seen in the plots 
of Fig.~\ref{dens} for the case of $\Gamma_{\Delta }=0$ (dashed lines).  For the 
values of $\Gamma_{\Delta }=60~{\rm MeV}$ (dotted line), $\Gamma_{\Delta }=120~{\rm MeV}$, the physical value 
(solid line) and 
$\Gamma_{\Delta }=240~{\rm MeV}$ (dot-dashed line) the pole is washed-out, but its traces are still visible, 
with the curves reaching maxima around $M_{0}=0.4~~{\rm GeV}$. 
The upper value of the virtual mass is $M_{\gamma^{\ast}}= m_{\omega}-m_{\pi}=0.64~~{\rm GeV}$. 
The differences between all the curves for various values of $\Gamma_{\Delta}$
in the range of  $M_{\gamma^{\ast}}$ between $0.2$ and $0.6~~{\rm GeV}$ shows that the results are sensitive 
to the assumed value for $\Gamma_{\Delta }$. We can observe that at low values of M$_{\gamma^{\ast}}$  the
effective coupling constant remains practically unchanged. However, 
at higher values of M$_{\gamma^{\ast}}$, above $0.2~~{\rm GeV}$, the value of $g_{{\rm eff}}$ is significantly 
larger than in the vacuum. For the square of the coupling we find an
enhancement by a factor of $\sim 2$ for the $\omega$ decay, and by a factor
as large as $\sim 5$ for the $\rho$ decay. All these numbers are quoted at the 
saturation density $\rho_{B}=\rho_{0}$.

\begin{figure}[tb]
\centerline{
\epsfysize = 12 cm \centerline{\epsfbox{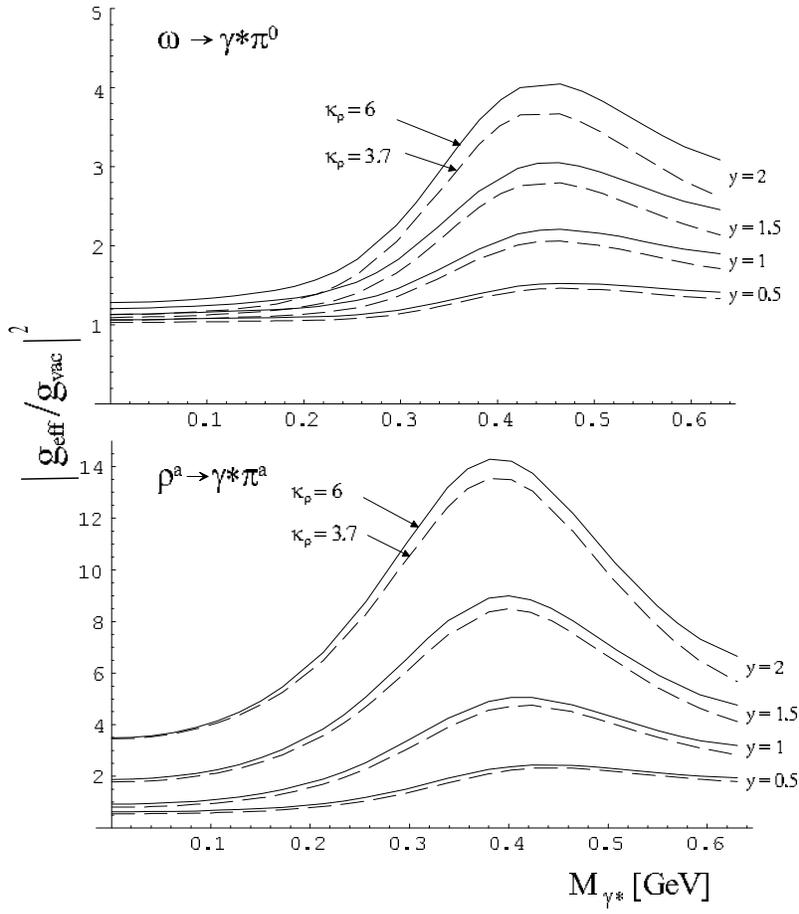}} \vspace{0mm}} 
\caption{\label{omrho3}Top: $|g_{{\rm eff}}/g_{{\rm vac}}|^{2}$ for the decay $\protect%
\omega \to \protect\gamma^\ast \protect\pi^0$ at the nuclear saturation
density, plotted as a function of the virtuality of the photon, $M_{\gamma^{\ast}}$. The
solid lines correspond to the case with $\kappa_{\rho}=6$, dashed lines for $\kappa_{\rho}=3.7$, $y=\frac{\rho_{B}}{\rho_{0}}$.
Bottom: the same for the $\rho^a \to \pi^a \gamma^\ast$ decay.}
\end{figure}

In Fig.~~\ref{omrho3} we display the results for different values 
of the baryon density, $\rho_{B}$. For simplicity we denote the saturation density as
\begin{equation}
y=\frac{\rho_{B}}{\rho_{0}}.
\end{equation}
We have done the calculations for both these decays for $\kappa_{\rho}=3.7$ (dashed lines) 
and $\kappa_{\rho}=6$ (solid lines). 
We can see that the effective coupling constant is more enhanced with growing density ratio $y$.

Our method can be trusted numerically only at sufficiently low values of the baryon density, such that the 
leading-density approximation holds. For the masses or condensates mentioned in Chapter 1 the nuclear 
saturation density its low enough to use the low density approximation.
On the other hand, we can see from Fig.~~\ref{omrho3} that the 
effects are large already at the saturation density.  
 
\begin{figure}[tb]
\centerline{
\epsfysize = 11 cm \centerline{\epsfbox{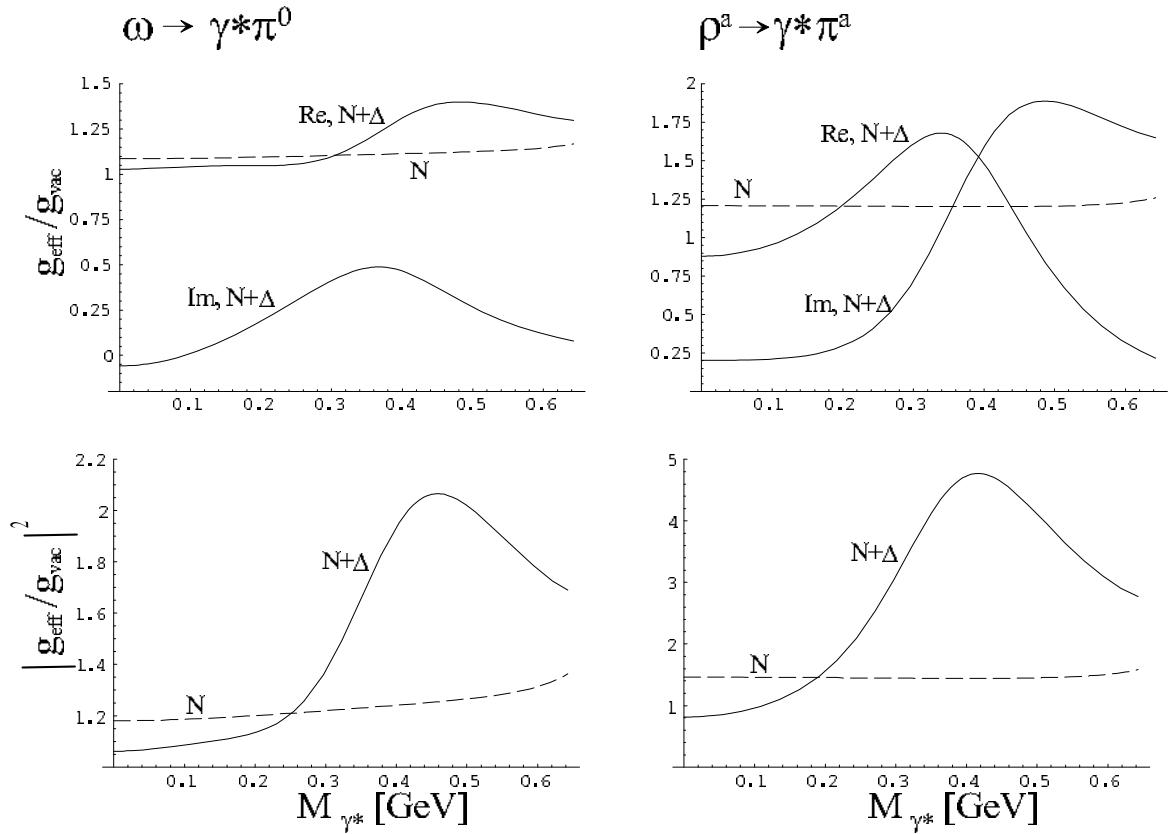}} \vspace{0mm}} 
\caption{\label{omrhoall}The ratio of the effective coupling constant to its vacuum value 
at the nuclear  saturation density for the decays $\protect%
\omega \to \protect\gamma^\ast \protect\pi^0$  and 
$\protect \rho^{a} \to \protect\gamma^\ast \protect\pi^a$, plotted
as a function of the virtuality of the photon, $M_{\gamma^{\ast}}$.
The solid and the dashed lines correspond to the  calculations 
with all diagrams of Fig.~~\ref{meddiag}, and calculations with diagrams including only nucleons, respectively. 
Top plots show the real and imaginary parts for all considered cases. 
We use $\Gamma _{\Delta }=120~~{\rm MeV}$.}   
\end{figure}

Therefore, numerical results at high densities obtained from our consideration 
are strongly indicative of possible large effects in the Dalitz decays of vector mesons in medium.
In the $\omega \rightarrow \pi ^{0}\gamma ^{\ast }$
and $\rho ^{a}\rightarrow \pi ^{a}\gamma ^{\ast }$ decays, we compare 
the calculation with and without the $\Delta$ isobar, see Fig.~~\ref{omrhoall}.

In the figure we plot the ratio $g_{{\rm eff}}/g_{{\rm vac}}$ and $|g_{{\rm eff}}/g_{{\rm vac}}|^{2}$
as a function of the virtual photon mass, $M_{\gamma^{\ast}}$ at the saturation density $y=1$ and for 
$\Gamma _{\Delta}=120~~{\rm MeV}$, same as we have done in the previous case of the  
$\pi\to\protect\gamma\gamma^{\ast}$ decay.
In two upper plots we show the imaginary and real (solid line) 
part of the calculations including nucleons and delta isobars, and the calculation with nucleons only 
(dashed lines).
In Fig.~~\ref{omrhoall} we can see that the real and imaginary functions of the considered processes, 
$\protect \omega \to \protect\gamma^\ast \protect\pi^0$  
and $\protect \rho^{a} \to \protect\gamma^\ast \protect\pi^a$, are changed 
in comparison to Fig.~~\ref {pggorimre}. 
The features of the dashed lines, corresponding to the calculations with nucleons only, 
remain practically unchanged. Both lower plots indicate that 
for the case with the $\Delta$, the change of $M_{\gamma^{\ast}}$ in the kinematic range above 0.2~~GeV 
has a visible effect on the effective coupling constant. 
We have already mentioned this when discussing Fig.~~\ref{omrho3}. 

\begin{figure}[tb]
\centerline{
\epsfysize = 11 cm \centerline{\epsfbox{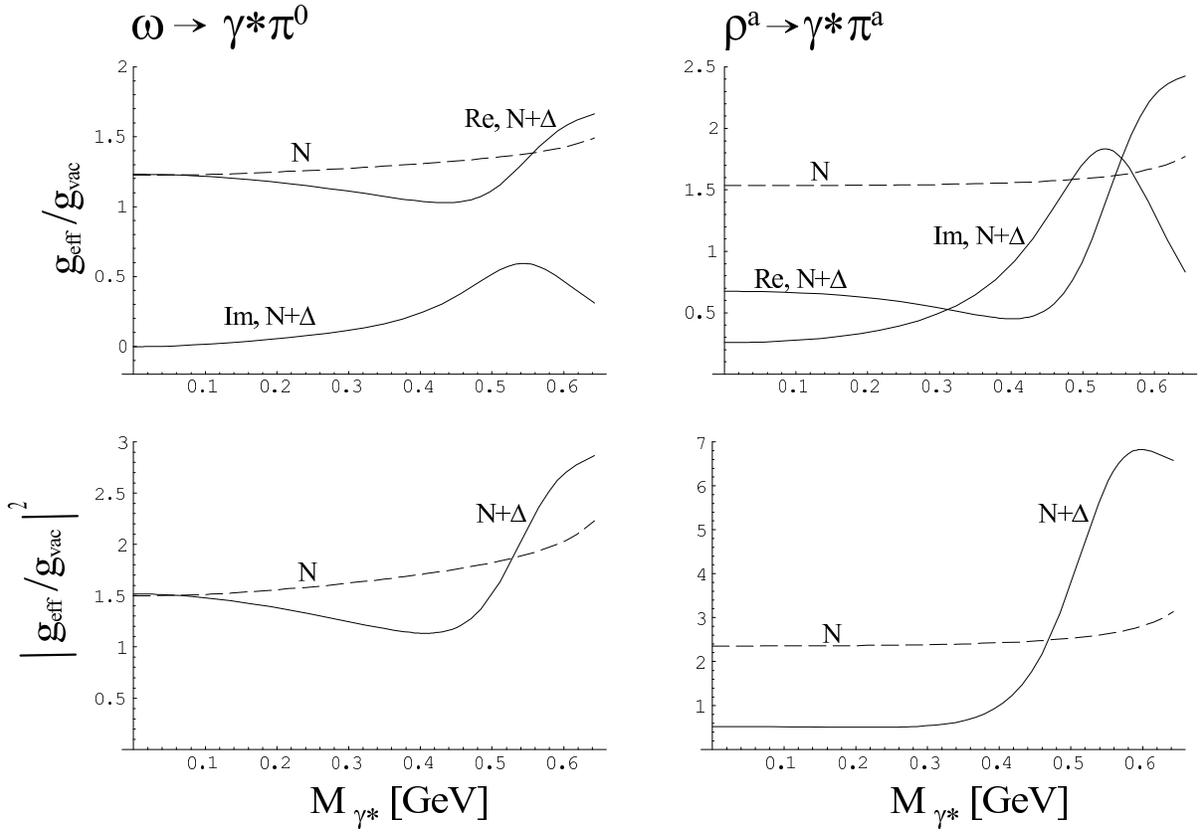}} \vspace{0mm}} 
\caption{\label{omrhoall70per} Same as Fig.~~\ref{omrhoall}, but for the values of $m_{N}$ 
and $M_{\Delta}$ reduced to 70\%.
} 
\end{figure}

Finally, we complete our discussion of the $\pi\omega\rho$ vertex at $\bf{q}=0$ with Fig.~~\ref{omrhoall70per}, 
which shows the result of the calculation for 
the values $m_{N}$ and $M_{\Delta}$ reduced to 70\%. In this case there are significant differences 
compared to the unreduced mass, Fig.~~\ref{omrhoall}. We can see that the 
effective coupling constant $\rm{g_{eff}}$ is enhanced by about $1.5$ for the 
$\omega \rightarrow \pi ^{0}\gamma ^{\ast }$ decay and by about $6$ 
for $\rho ^{a}\rightarrow \pi ^{a}\gamma ^{\ast }$ 
decay, and shifted to higher values of $M_{\gamma^{\ast}}$, above $0.4~~{\rm GeV}$. 

\section {Decay of particles moving with respect to the medium}

So far we have presented our results for the $\pi\omega\rho$ vertex 
for the decays of particles at rest with respect to the medium and 
we have pointed out large medium effects already  
at the nuclear saturation density. Now, we will introduce the general kinematic case 
of finite three-momentum of the decaying particle with respect to the medium, 
and calculate the widths and spectral functions in the transverse and longitudinal polarization channels.

In view of the substantial increase of the coupling constant for
the $\omega $ and $\rho $ decays when these particles are at rest with
respect to the medium, we expect that the effect survives for particles 
moving with respect to the medium. 
We analyze the $\omega \rightarrow \pi ^{0}\gamma ^{\ast }$ and $%
\rho ^{a}\rightarrow \pi ^{a}\gamma ^{\ast }$ processes in the case where the 
$\omega $ and $\rho $ meson move with a non-zero momentum with respect
to the medium, denoted by $\bf{q}$. The $\omega $ and $\rho $ particles can have transverse or
longitudinal polarization, defined by quantizing the spin along the
direction of $\bf{q}$. In medium the properties of these particles are different for these two
polarizations; in particular, their widths $\Gamma ^{L}$ and $\Gamma ^{T}$ are
different. 

\subsubsection{Kinematics}
 
\begin{figure}[h]
\begin{center}
\includegraphics[width=1 \textwidth]{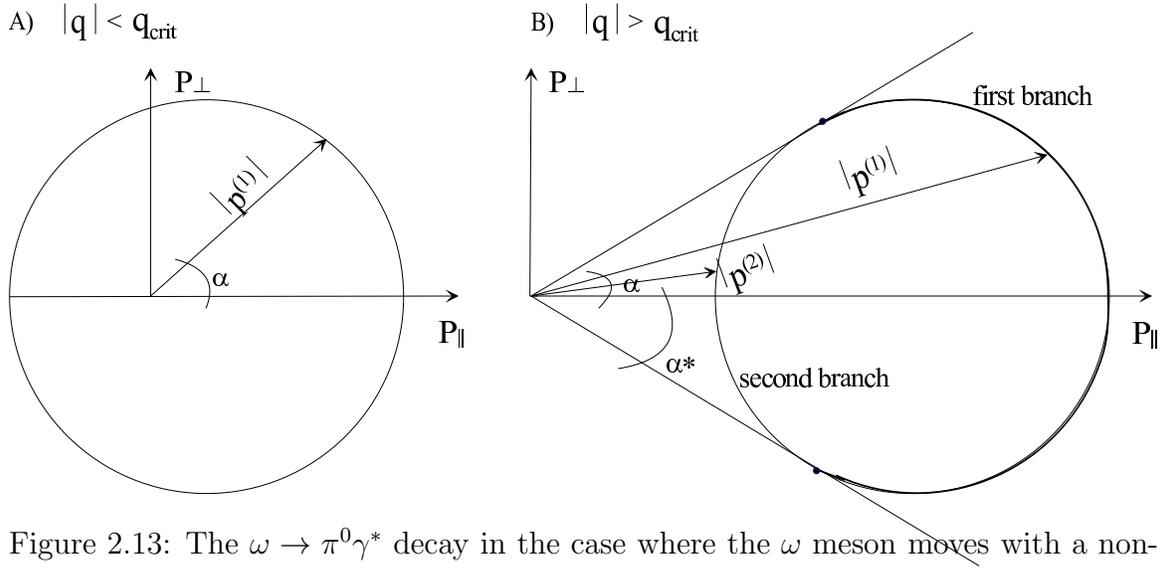}
\caption{\label{kin}The $\omega \rightarrow \pi ^{0}\gamma ^{\ast }$ decay in the case where the $\omega$ 
meson moves with a non-zero momentum $|\mathbf{q}|$ with respect to the medium.    
Figure A): the situation when $|\mathbf{q}|$ is smaller than $q_{\mathrm{crit}}$, and the values of $|\mathbf{p}|$ 
belong to the first branch of Eq.~\ref{twosol}. 
The angle $\alpha$ (between momentum $|\mathbf{p}|$ and $|\mathbf{q}|$) 
takes values from -$\pi$ to $\pi$. Figure B): for each value of $\alpha$ in the allowable range there are 
two branches. 
The second branch appears only for $|\mathbf{q}|$ above the critical value.}  
\end{center}
\end{figure}

We consider the kinematics for the decay $\omega \rightarrow \pi ^{0}\gamma ^{\ast }$ 
in the rest frame of the medium (not in the rest frame of the decaying particle as it is usually done 
in calculations in the vacuum). 
Thus the four-velocity of the medium is $u=(1,0,0,0)$, and 
\begin{eqnarray}
q^{2}=m_{\omega }^{2},~~~~ 
q\cdot u=q_{0}=\sqrt{m_{\omega }^{2}+|\mathbf{q}|^{2}},~~~~ 
p\cdot u=p_{0},\nonumber\\ 
p\cdot q=p_{0}q_{0}-\cos \alpha |\mathbf{p}||\mathbf{q}|,~~~~ 
p^{2}=p_{0}^{2}-|\mathbf{p}|^{2}=M_{\gamma^{\ast}}^{2}, 
\end{eqnarray}
where $q$ and $p$
are the momenta of the $\omega $ and $\gamma ^{\ast }$ , respectively. 
Using the energy conservation:
\begin{eqnarray}
E_{\omega }=E_{\gamma ^{\ast }}+E_{\pi }, 
\end{eqnarray}
where 
\begin{eqnarray}
E_{\omega }=\sqrt{m_{\omega }^{2}+|\mathbf{q}|^{2}},~~~~ 
E_{\gamma ^{\ast}}=\sqrt{M_{\gamma^{\ast}}^{2}+|\mathbf{p}|^{2}},\nonumber\\ 
E_{\pi }=\sqrt{m_{\pi }^{2}+|\mathbf{p}|^{2}+|\mathbf{q}|^{2}-2\cos\alpha |\mathbf{p}||\mathbf{q}|}, 
\end{eqnarray}
we find in general two solutions for the momentum $|\mathbf{p}|$:
\begin{eqnarray}
&&|\mathbf{p}^{(1,2)}| =\frac{|\mathbf{q}|\cos \alpha (m_{\pi
}^{2}+M_{\gamma^{\ast}}^{2}+m_{\omega }^{2})}{2(M_{\gamma^{\ast}}^{2}+\mathbf{q}^{2}\sin ^{2}\alpha )}\pm \nonumber\\
&&q_{0}\sqrt{m_{\pi }^{4}-2m_{\pi }^{2}(M_{\gamma^{\ast}}^{2}+m_{\omega
}^{2})+M_{\gamma^{\ast}}^{4}+m_{\omega }^{4}-M_{\gamma^{\ast}}^{2}(4\sin^2 \alpha|\mathbf{q}%
|^{2}+2m_{\omega }^{2})},\label{p1,p2} \nonumber\\
&&p_{0}^{(1,2)}=\sqrt{M_{\gamma^{\ast}}^{2}+\left( \mathbf{p}^{(1,2)}\right)^{2}} \label{twosol},
\end{eqnarray}
We consider two cases which are shown in Fig.~~\ref{kin}.

Depending on the value of $|\mathbf{q}|$ we may have a case where only
one branch for $|\mathbf{p}|$ is present, or where two branches appear. When $|%
\mathbf{q}|$ is smaller than some critical value $q_{\mathrm{crit}}$ we have
the situation of Fig.~~\ref{kin}(A). The values of $|\mathbf{p}|$ belong to one
branch, and $\alpha $ takes all values between $-\pi $ and $\pi $. When $|%
\mathbf{q}|$ is larger than $q_{\mathrm{crit}}$ then we have the situation
depicted in Fig.~~\ref{kin}(B). For each value of $\alpha$ in the allowable range we have two
branches. The angle $\alpha $ has a maximum value $\alpha ^{\ast }$ when $\ |%
\mathbf{p}^{1}|=|\mathbf{p}^{2}|$. Solving the equation corresponding to this condition we find that
\begin{equation}
\alpha ^{\ast }=\arcsin \left( \frac{1}{2|\mathbf{q}|M_{\gamma^{\ast}}}\sqrt{m_{\pi
}^{4}-2(M_{\gamma^{\ast}}^{2}+m_{\omega }^{2})m_{\pi }^{2}+(M_{\gamma^{\ast}}^{2}-m_{\omega }^{2})^{2}}%
\right) \hspace{0.3cm}.  \label{alfacrit}
\end{equation}
The transition between the behavior of Fig.~~\ref{kin} A) and B) occurs for $q=q_{%
\mathrm{crit}}$. The value of $q_{\mathrm{crit}}$ follows from the condition  
$|\mathbf{p}^{2}|=0$ for $\alpha =0$ as is clear from Fig.~~\ref{kin}. 
This condition yields, from Eq.~\ref{twosol}   
\begin{equation}
\frac{\sqrt{m_{\pi }^{4}-2(M_{\gamma^{\ast}}^{2}+m_{\omega }^{2})m_{\pi
}^{2}+(M_{\gamma^{\ast}}^{2}-m_{\omega }^{2})^{2}}}{2M_{\gamma^{\ast}}}= q_{\mathrm{crit}}.
\label{qcrit}
\end{equation}

\subsubsection{Transverse and longitudinal polarizations}
Now we are equipped with appropriate kinematic expressions to consider the width of 
the transversely and longitudinally polarized $%
\omega $ meson due to the decay $\omega \rightarrow \pi ^{0}\gamma ^{\ast }$. 
The expression for the width has the form
\begin{eqnarray}
\Gamma _{\omega \rightarrow \pi ^{0}\gamma ^{\ast }}=\frac{1}{n_{s}}%
\sum_{ss^{\prime }}\frac{1}{2q_{0}}\int \frac{d^{3}p}{(2\pi )^{3}2p_{0}}\int 
\frac{d^{3}p^{\prime }}{(2\pi )^{3}2p_{0}^{\prime }}|\mathcal{M}_{ss^{\prime
}}|^{2}(2\pi )^{4}\delta ^{(4)}(q-p-p^{\prime }),  
\label{szer1}
\end{eqnarray}
where $s$ is the polarization of the $\omega $, $s^{\prime }$ is the
polarization of the $\gamma ^{\ast }$, $n_{s}$ is the number of spin states
of the $\omega $ meson,  and$\ |\mathcal{M}_{ss^{\prime }}|^{2}$ is the
transition amplitude.
We perform the phase-space integral with the kinematics discussed above and obtain
\begin{eqnarray}
&&\Gamma _{\omega \rightarrow \pi ^{0}\gamma ^{\ast }} =\frac{1}{n_{s}}%
\sum_{s}\frac{1}{2q_{0}}\left[ \theta (q_{\mathrm{crit}}-q)\sum_{b=1}\int_{0}^{\pi
}d\alpha  \label{szer2}\right. \nonumber\\
&& \left. +\theta (q-q_{\mathrm{crit}})\sum_{b=1,2}\int_{0}^{\alpha ^{\ast }}d\alpha
\right] \sin \alpha \frac{(\mathbf{p}^{(b)})^{2}}{8\pi
p_{0}^{(b)}(q_{0}-p_{0}^{(b)})|a^{(b)}|}|\mathcal{M}_{ss^{\prime }}|^{2},
\end{eqnarray}
where $\sum_{b}$ is the sum over the one or two branches, and expressions 
\begin{equation}
a^{(1,2)}=\left. \frac{d(q_{0}-\sqrt{M^{2}+r^{2}}-\sqrt{m_{\pi
}^{2}+r^{2}-2r|\mathbf{q}|\cos \alpha +\mathbf{q}^{2}})}{dr}\right| _{r=|%
\mathbf{p}^{(1,2)}|},  \label{deldir}
\end{equation}
are the factors from the Dirac delta function $\delta (E_{\omega }-E_{\rho }-E_{\pi })$.

The $\omega$ meson has transverse polarization described by the
polarization vectors $\varepsilon _{(\pm )}^{\mu }$, and $\varepsilon _{(0)}^{\mu }$. 
Using explicit forms of these polarization vectors one finds the following formulas \cite{BFH1,disp}:
\begin{eqnarray}
&&-\varepsilon _{(+)}^{\mu \ast }\varepsilon _{(+)}^{\nu }-\varepsilon
_{(-)}^{\mu \ast }\varepsilon _{(-)}^{\nu } =\nonumber\\
&&g^{\mu \nu }-u^{\mu }u^{\nu }-%
\frac{(q^{\mu }-q\cdot u\ u^{\mu })(q^{\nu }-q\cdot u\ u^{\nu })}{q\cdot
q-(q\cdot u)^{2}}\equiv T^{\mu \nu }(q),  \label{T} \nonumber\\
\\
&&-\varepsilon _{(0)}^{\mu \ast }\varepsilon _{(0)}^{\nu } =-\frac{q^{\mu
}q^{\nu }}{q\cdot q}+u^{\mu }u^{\nu }+\frac{(q^{\mu }-q\cdot u\ u^{\mu
})(q^{\nu }-q\cdot u\ u^{\nu })}{q\cdot q-(q\cdot u)^{2}}\equiv L^{\mu \nu
}(q), \label{L} \nonumber
\end{eqnarray}
The tensors $T^{\mu \nu }$ and $L^{\mu \nu }$ are defined with such signs as
to form projection operators, {\em i.e. }, $T^{\mu \nu }T_{\nu }^{\cdot
\alpha }=T^{\mu \alpha }$, $L^{\mu \nu }L_{\nu }^{\cdot \alpha }=L^{\mu
\alpha }$, and $T^{\mu \nu }L_{\nu }^{\cdot \alpha }=0$. Furthermore, we
have $T^{\mu \nu }q_{\nu }=0$ and $L^{\mu \nu }q_{\nu }=0$, which reflects  
current conservation, as well as $T^{\mu \nu }u_{\nu }=0$. 

 By summing over all polarizations, we find the usual expression
\begin{equation}
-\sum_{s=\pm ,0}\varepsilon _{(s)\mu }^{\ast }\varepsilon _{(s)\nu }=T^{\mu
\nu }+L^{\mu \nu }=g^{\mu \nu }-\frac{q^{\mu }q^{\nu }}{q\cdot q}=Q^{\mu \nu
}(q).  \label{sumTL}
\end{equation}
\begin{figure}[tb]
\centerline{
\epsfysize = 11 cm \centerline{\epsfbox{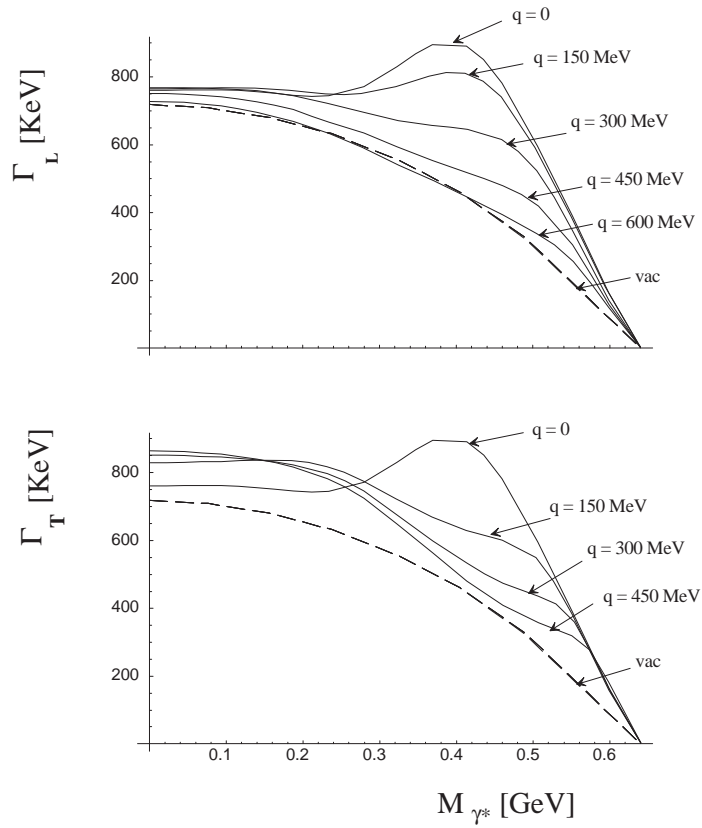}} \vspace{0mm}} 
\caption{\label{pedyom} The longitudinal (top) and transverse (bottom) widths for 
the $\omega \rightarrow \pi ^{0}\gamma ^{\ast }$
decay, plotted as a function of the virtuality 
of the photon, $M_{\gamma^\ast}$. The dashed and solid lines show the vacuum 
and medium cases for $\rho_{B}=\rho_{0}$, respectively.}
\end{figure} 

Now we return to an expression for the width, Eq.~~\ref{szer1}. We write the transition amplitude
\begin{eqnarray}
\frac{1}{n_{s}}\sum_{ss^{\prime }}|\mathcal{M}_{ss^{\prime }}|^{2} &=& \frac{1}{
n_{s}}\sum_{ss^{\prime }}[\varepsilon _{\left( s^{\prime }\right) \mu
}A^{\mu \nu }\varepsilon _{\left( s\right) \nu }\varepsilon _{\left(
s\right) \nu ^{\prime }}^{\ast }A^{\nu ^{\prime }\mu ^{\prime }\ast
}\varepsilon _{\left( s^{\prime }\right) \mu ^{\prime }}^{\ast }] \nonumber\\
&=&\frac{1}{%
n_{s}}\sum_{s}A^{\mu \nu }\varepsilon _{(s)\nu }^{(\omega )}\varepsilon
_{(s)v^{\prime }}^{(\omega )\ast }A^{\nu ^{\prime }\mu ^{\prime }\ast
}(-Q_{\mu \mu ^{\prime }}(p)),  \label{M}
\end{eqnarray}

\begin{figure}[tb]
\centerline{
\epsfysize = 11 cm \centerline{\epsfbox{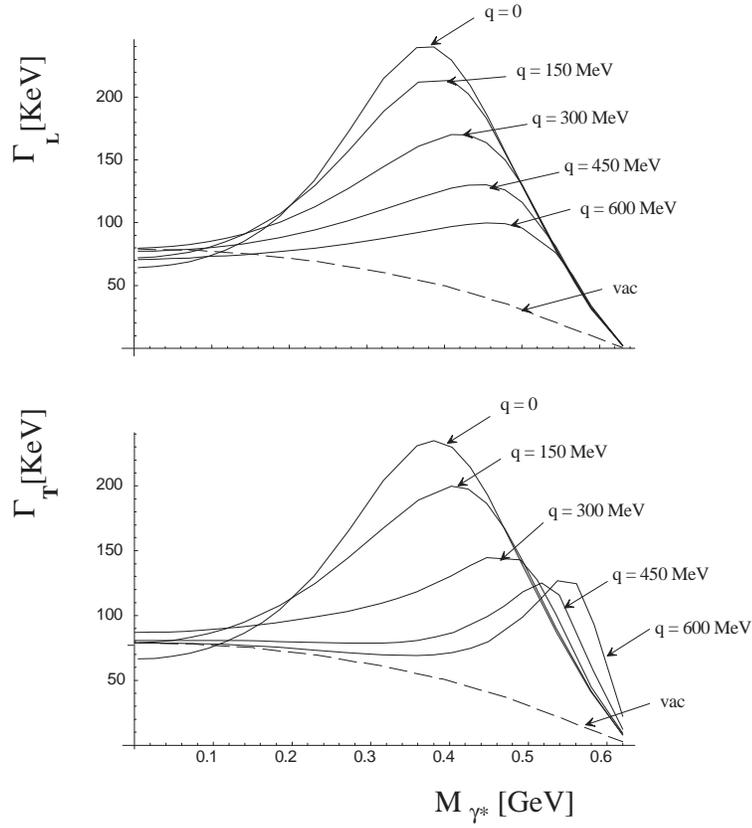}} \vspace{0mm}} 
\caption{\label{pedyrho} The longitudinal (top) and transverse (bottom) width for the 
$\rho ^{a}\rightarrow \pi ^{a}\gamma ^{\ast }$ decay, plotted as a function of the virtuality 
of the photon, $M_{\gamma^\ast}$. The dashed and solid lines show the vacuum and 
the medium cases, respectively.}
\end{figure}

where $\varepsilon _{\left( s^{\prime }\right) \mu }$ is a polarization
vector for $\gamma ^{\ast }$ and $\varepsilon _{\left( s\right) \nu }$ is
the polarization vector for the $\omega $ meson. Summing over $s=0$, and $s=\pm $  
we obtain, respectively,
\begin{eqnarray}
|\mathcal{M}_{L}|^{2} &=&L_{(\omega )}^{\nu \nu ^{\prime }}Q_{\left( \gamma
^{\ast }\right) }^{\mu \mu ^{\prime }}A_{\mu \nu }A_{\mu ^{\prime }\nu
^{\prime }}^{\ast }  \label{ML} \nonumber\\
|\mathcal{M}_{T}|^{2} &=&\frac{1}{2}T_{(\omega )}^{\nu \nu ^{\prime
}}Q_{\left( \gamma ^{\ast }\right) }^{\mu \mu ^{\prime }}A_{\mu \nu }A_{\mu
^{\prime }\nu ^{\prime }}^{\ast }.  \label{MT}
\end{eqnarray}
Combining Eq. \ref{szer2} and \ref{MT} we may  write the final expression
\begin{eqnarray}
&&\Gamma _{\omega \rightarrow \pi ^{0}\gamma ^{\ast }}^{P}(|\mathbf{q}|) =%
\frac{1}{n_{P}}\frac{1}{2q_{0}} \left[ \theta (q_{\mathrm{crit}}-|\mathbf{q}%
|)\sum_{b=1}\int_{0}^{\pi }d\alpha +  \right. \\
&& \left. +\theta (|\mathbf{q}|-q_{\mathrm{crit}})\sum_{b=1,2}\int_{0}^{\alpha
^{\ast }}d\alpha \right] \sin \alpha \frac{(\mathbf{p}^{(b)})^{2}}{8\pi
p_{0}^{(b)}(q_{0}-p_{0}^{(b)})|a^{(b)}|}|\mathcal{M}_{P}|^{2},  \label{szer}
\nonumber
\end{eqnarray}

\subsubsection{Results for transverse and longitudinal widths}

Our numerical results for the $\omega \rightarrow \pi ^{0}\gamma ^{\ast }$
decay are shown in Fig.~~\ref{pedyom}. We present the dependence of 
widths for the transverse and longitudinal 
polarization on the virtuality of the photon, $M_{\gamma^\ast}$. We have used here 
the constant $\Gamma_{\Delta}=120~~{\rm MeV}$. Calculations are done for 
different values of $|\bf{q}|$. 
From Fig.~~\ref{pedyom} we can conclude that the transverse and longitudinal widths 
decrease with momentum $|\bf q|$, and  
the medium effect is weakened with increasing momentum $|\bf q|$. However, we can observe that 
the effect remains substantial 
for $|\bf q|$ up to $\sim 200~~\rm{MeV}$, the relevant value for temperatures typical in heavy-ion collisions. 
As it is commonly known in a fireball formed in relativistic 
heavy-ion collisions, the momenta are lower than a typical temperature $\sim 150~~\rm{MeV}$.

We present analogous results for the $\rho ^{a}\rightarrow \pi ^{a}\gamma ^{\ast }$ decay  
in Fig.~~\ref{pedyrho}. We notice again that while $|\bf q|$ increases 
from $0$ to $600~~{\rm MeV}$, the medium effect is weakened, and 
both the transverse and longitudinal widths decrease. 

We see that the medium effect is large for the typical momenta 
up to $\sim 200~~\rm{MeV}$. For higher momenta this effect is lower; however it still remains visible compared 
to the vacuum result, see the dashed line in Fig.~\ref{pedyrho}. 

\chapter{Summary of Part I}
The main conclusion of our investigations in this part of the thesis is that the medium effects 
on the $\pi\omega\rho$ coupling in the discussed Dalitz processes $\omega\rightarrow \pi ^{0}\gamma ^{\ast }$,  
$\rho ^{a}\rightarrow \pi ^{a}\gamma ^{\ast }$ are large. Moreover, 
these medium effects come dominantly from the processes where the $\Delta$ 
isobar is excited in the intermediate state. 
On the other hand, for the $\pi ^{0}\rightarrow \gamma\gamma ^{\ast }$ the effects of the $\Delta$ 
isobar cancel almost exactly the effects of the nucleon particle-hole excitations, such that the  
medium effect is small. The Dalitz yields from the $\pi^0$ decays are not altered by the medium for 
two reasons: first, virtually all pions decay outside of the fireball due to their long lifetime; second, 
the width for this decay is practically unaltered by the medium.

At the nuclear saturation density, for the Dalitz decays from the $\rho$ and $\omega$ mesons, 
the value of the effective coupling constant 
is enhanced  compared to the vacuum value. The increased coupling constant greater by about a factor of 2 
for the $\omega$ meson and by about a factor of 5 for the $\rho$, leads directly to large 
widths of the $\omega$ and $\rho$ mesons in medium.  

Next, we have analyzed the case where the decaying particle ($\rho$ or $\omega$) 
moves with respect to the medium. In this case we have to  
look separately at the longitudinal and transverse polarizations. The properties of mesons 
are different for these polarizations. We investigated the dependence of the widths for transverse and 
longitudinal polarization on the invariant mass of dileptons. This 
leads to the conclusion that the medium effects decrease with growing magnitude of three-momentum, 
but remain significant for typical momenta in heavy-ion collisions, $q \lesssim 200~\rm{MeV}$.  

The enhancement of the coupling constant directly affects the calculations of the dilepton production 
in relativistic heavy-ion collisions. 
An increased value of width results in an 
increased dilepton yield. In the next part we will 
apply our model to evaluate the dilepton production from the $\rho$ and $\omega$ decays 
in heavy-ion collisions.

\part{Dilepton production rate}

\chapter{Electromagnetic signals from hot and dense matter}
In this part of the thesis we use directly the results of Part I.
We focus on the production of dileptons in hadronic decays, which  
is an area of great interest from a theoretical point of view. We will investigate 
the theoretical yields from the Dalitz decays of vector mesons in the region 
of $0.2-0.6~\rm {GeV}$, where many existing calculations have problems to explain the CERES 
and HELIOS experiments \cite{ceres, helios, li, LiKoBrown1, RappRev,Dom, bratko0,rapp, 
RappDil}. Our purpose is to see to what extent the medium modification of the $\pi\omega\rho$ 
vertex influences the dilepton production, and may be significant in understanding the experimental 
data in the region of low invariant masses.

As we mentioned in the Introduction, electron pairs are produced in all stages of ultrarelativistic
heavy-ion collisions. 
Because of the small electromagnetic coupling
constant, they carry undistorted information about the densest and hottest phases of the 
heavy-ion collisions where chiral symmetry restoration and/or quark gluon plasma (QGP) formation are expected.

The first pairs of leptons are created during the stage when two
nuclei pass through each other, while the last pairs are formed when the hadrons are
decoupled and move freely to the detectors. After decoupling of hadronic
matter the greater part of dileptons is produced by the electromagnetic decays of hadronic resonances.
For pair masses below 140 MeV/c$^{2}$ the $\pi ^{0}$ Dalitz decay dominates the spectrum,
whereas at higher masses the $\omega $, $\rho $, $\eta$, $\eta'$ and $\phi$ decays should be most significant. 

The dilepton mass spectra are very important for the study of the in-medium properties of vector mesons, 
which we have discussed in detail in Part I.

As we have pointed out in the Introduction, experimental measurements of dilepton mass spectra can be divided 
into three mass regions. The low 
mass region, below $1 ~{\rm GeV}$, is dominated by hadronic interactions and hadronic decays at
freeze-out. It is particularly sensitive to in-medium modifications of the light hadrons. 
The intermediate mass region lies between $1~{\rm GeV}$ and about $2.5~{\rm GeV}$, where the 
contribution from the thermalized QGP might be seen. 
We want to mention that the physical picture of QGP structure has changed in the last year. 
Now, many theoretical developments \cite{sQGPshur, sQGPgul} suggest  
that the matter produced at RHIC, in the high temperature region $T_{c}<T<4T_{c}$, 
is a strongly coupled quark gluon plasma (sQGP) contrary to the previous expectations which were based on  
weakly coupled quasiparticle gas. In order to understand what 
sQGP is, one may have a look at two other examples of the strongly coupled systems such as trapped ultra-cold atoms 
in a large scattering length regime or supersymmetric Yang-Mills gauge theories at strongly coupling limit 
(Conformal Field Theory with a non-running coupling at finite temperature). 
Finally, the high mass region at and above $m_{J/\Psi}$ 
is important in connection with the suppressed $J/\Psi$ production with respect to the background 
from the Drell-Yan process of the quark-antiquark annihilation. 

So far, the experimental measurements of dilepton yields in heavy-ion collisions have mainly been 
carried out at the CERN SPS facility by three collaborations: the CERES (Cherenkov Ring Electron Spectrometer) 
collaboration has measured dielectron spectra in the low-mass region \cite{ceres,ceres2,ceres3}, 
the HELIOS-3 collaboration has specialized in dimuon spectra up to the $J/\Psi$ region \cite{helios}, 
and the NA38/NA50 \cite{na38a,na38b,na50}
collaborations have been dedicated to study of dimuon spectra in the intermediate and high-mass region, emphasizing 
the $J/\Psi$ suppression. Dilepton data have also been taken by the DLS collaboration at 
BEVALAC \cite{ bevalac1, bevalac2} at much lower bombarding energies of about $1-2~{\rm A~~GeV}$.

The study of the low mass dileptons between $0.2-0.6 \rm{GeV}$ has recently become of considerable interest. 
In this region the dilepton continuum originates from the Dalitz decays of neutral mesons 
such as $\pi^{0}$, $\eta$, $\eta' \to e^{+}e^{-} \gamma$ and $\omega \to e^{+}e^{-}\pi^{0}$. 
The resonance peaks are due to direct decays, $\omega$, $\rho$, $\phi \to e^{+}e^{-}$. In central heavy-ion 
collisions the low mass dilepton enhancement is observed by the CERES and HELIOS-3 collaborations. This 
enhancement has been studied in various approaches, ranging from hydrodynamical models to complicated 
transport models. The calculations based on the cocktail model, which are designed to describe the 
proton-induced reactions, work very well for light-heavy systems, but fail to explain the 
observed phenomenon in nucleus-nucleus collisions. Various medium effects, such as the dropping vector 
meson masses, as first proposed by Brown and Rho \cite{brscale}, the modification of the rho meson spectral 
functions, or the enhanced production of $\eta$ have been proposed to explain the observed enhancement.

In the intermediate-mass region from about $1~{\rm GeV}$ to about $2.5~{\rm GeV}$ the excess of dileptons 
has also been observed by both the HELIOS-3 and NA38/NA50 collaborations. This enhancement is observed 
in central S+W, S+U, and Pb+Pb collisions in comparison to proton-induced reactions 
\cite{helios, na38a,na38b,na50}. 
The intermediate-mass dilepton spectra are particularly useful in the search for QGP.

The $J/\Psi$ suppression in the high-mass region above $2.5~{\rm GeV}$ is a very interesting problem. It was 
interpreted as a signal of QGP formation by Matsui and Satz already in 1986 \cite{satz}. 
Nowadays, there is an indication from experimental data that up to central S+Au collisions pre-resonance 
absorption in nuclear matter is sufficient to account for the observed $J/\Psi$ suppression \cite{gavin}.
However, data for central Pb+Pb collisions from NA50 collaboration show an additional strong $J/\Psi$ 
suppression, which is interpreted as a signal of color deconfinement \cite{kharzeev,jpblaizot}.

Below we discuss mainly the enhancement of the low-mass dileptons, because  
we have found, in Part I, that the considered $\pi\omega\rho$ coupling constant is sizeable enhanced 
in the medium just in the region between $0.2-0.6 \rm{GeV}$ of the invariant mass of dileptons. 
Hence in this region 
the modifications of the vertex directly influence the dilepton production rate.

\section{Overview of experimental measurements}
The CERN SPS is the first machine used for systematic investigations of the dilepton production in
ultrarelativistic hadron-nucleus and nucleus-nucleus collisions. In this section we review the 
CERN-SPS dilepton experiments CERES, HELIOS, and NA38/NA50. 
Since the dilepton spectra are usually 
compared to the expected contributions from hadron decays, the co-called {\em hadronic cocktail}, we begin this 
section by a short description of the hadronic cocktail model.

\subsubsection{Hadronic cocktail}
The hadronic cocktail is used  
to simulate the relative abundance of dielectrons produced by hadron decays in  proton-nucleus and 
nucleus-nucleus collisions at the invariant mass range covered by the CERES acceptance ($m_{ee}<2~\rm{GeV}$). 
This invariant mass range is dominated by the decay of light scalar and vector mesons, 
such as $\pi^{0}$, $\eta$, $\eta'$, $\rho^{0}$, $\omega$, and $\phi$. The hadronic cocktail treats 
proton-nucleus and nucleus-nucleus collisions as a superposition of individual nucleon-nucleon collisions and 
gives a reference to compare with the experimental observed dilepton yield. As input for this simulation 
one needs to know differential cross sections, the widths of all decays including dileptons in the final state, 
and the decay kinematics for all involved particles. 

The proton-nucleus collisions are approximated by a superposition of nucleon-nucleon collisions, 
i.e. the proton-nucleus yield is obtained by rescaling the nucleon-nucleon yield
with the mean charge-particle multiplicity of the colliding system. 
For Pb-Au collisions the relative cross sections ($\sigma/\sigma_{\pi^0}$, where $\sigma$ is the cross 
section for a particular hadron decay into the electron pairs) are taken from a thermal 
model, \cite{braunmunz}. 
In order to compare with experiments the hadronic cocktail contribution is divided by the total 
number of charged particles 
within the acceptance of the detector. The sum of all contributions from hadron decays is calculated with 
Monte-Carlo event generator. The invariant mass spectrum is calculated including all known hadronic sources. 
The particle ratios of these sources are assumed to be independent of the collision system and to scale with the 
number of produced particles. Their $p_{\perp}$-distribution are generated assuming $m_{\perp}$-scaling, 
\cite{bourg} based on the pion $p_{\perp}$ spectrum from different experiments and fitted \cite{alber} 
to the formula:
\begin{eqnarray}
\frac{dN}{dp_{\perp}} \sim p_{\perp}~~exp[-\frac{m_{\perp}}{T_{\rm{eff}}}],
\end{eqnarray}
where for SPS at energy $158~\rm{A~GeV}$ the inverse slope parameter is parameterized as 
\begin{equation}
T_{\rm{eff}}=175~\rm{GeV}+0.115~m [\rm{GeV}],
\end{equation}
with meson mass indicated as m. The rapidity distribution is a fit to measured data. All Dalitz decays 
are treated according to the  Krall-Wada \cite{KW} expressions 
with the experimental transition form factors  
taken from Ref.~\cite{Land}. The vector meson decays are generated using the expressions 
derived by Gounaris and Sakurai \cite{sakurai}.

Charm production is not taken into account since it is negligible in the low-mass range. Finally the laboratory 
momenta of the electrons are constructed with the experimental resolution and acceptance.

The differences between hadron cocktail and experiment indicate the existence of in-medium effects or 
the violation of the scaling behavior.

\subsection{CERES/NA45 experiments}
The only experiment which  measures the low-mass dilepton pairs up to about $1.2~{\rm GeV}$ 
is the CERES/NA45 experiment.
This collaboration carries out systematic measurements of dilepton spectra in proton-induced 
reactions. 
\begin{figure}[htb]
\centerline{
\vspace{0mm} ~\hspace{0cm} 
\epsfxsize = 8.6 cm \centerline{\epsfbox{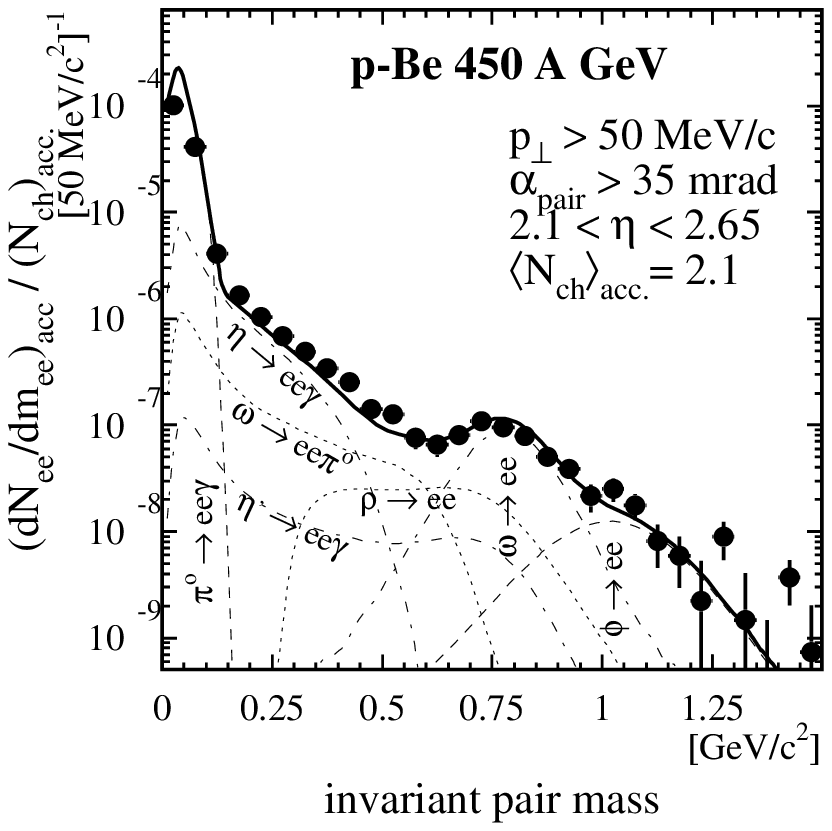}} \vspace{0mm}
\vspace{0mm} ~\hspace{-8.cm}
\epsfysize = 7.5 cm \centerline{\epsfbox{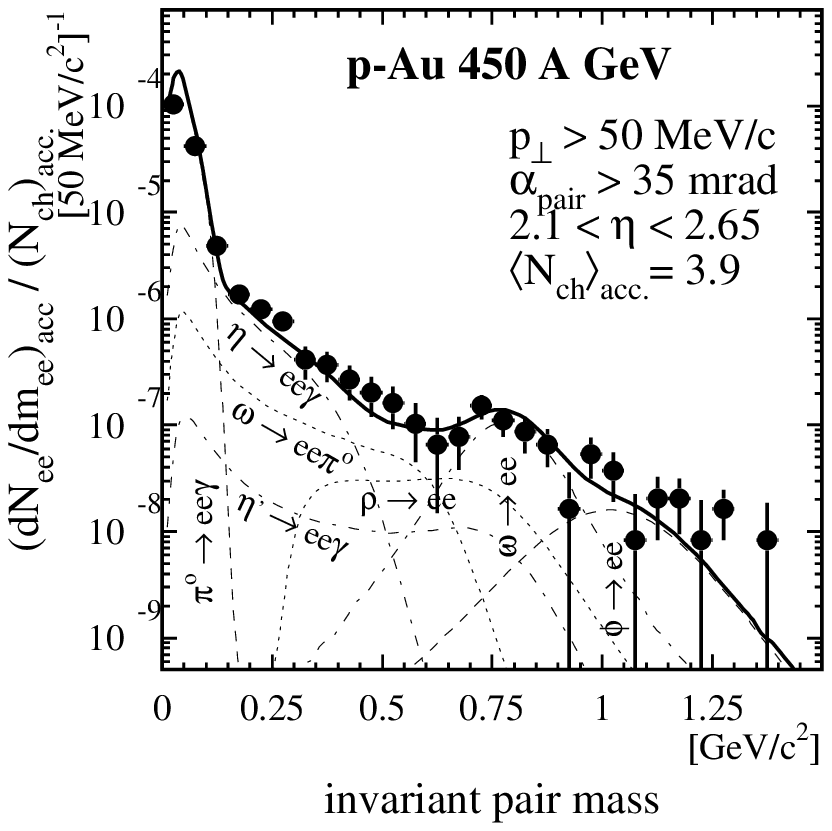}} \vspace{0mm}} 
\caption{\label{p450} Dilepton invariant mass spectrum of p+Be and p+Au collisions at the incident 
proton energy of $450~{\rm GeV}$, 
taken from \cite {PhD}. 
The expected sources from hadron decay (solid lines) and the experimental points are in 
excellent agreement.}
\end{figure}
Figure \ref{p450}, taken from \cite{ PhD}, shows the  
data on dielectrons of proton-induced collisions at $450~{\rm GeV}$ in comparison to the expected 
contributions of hadron decays, the {\em hadronic cocktail}. 
For instance, the dashed curves in Fig~\ref{p450} give the expected dielectron spectra from hadron decays  
for p-Be and p-Au reactions. The total 
expected cocktail of hadronic sources is denoted by the solid line. We observe that the dilepton 
yield is fitted by the cocktail model very well. 
\begin{figure}[htb]
\centerline{
\vspace{0mm} ~\hspace{0cm} 
\epsfxsize = 7.5 cm \centerline{\epsfbox{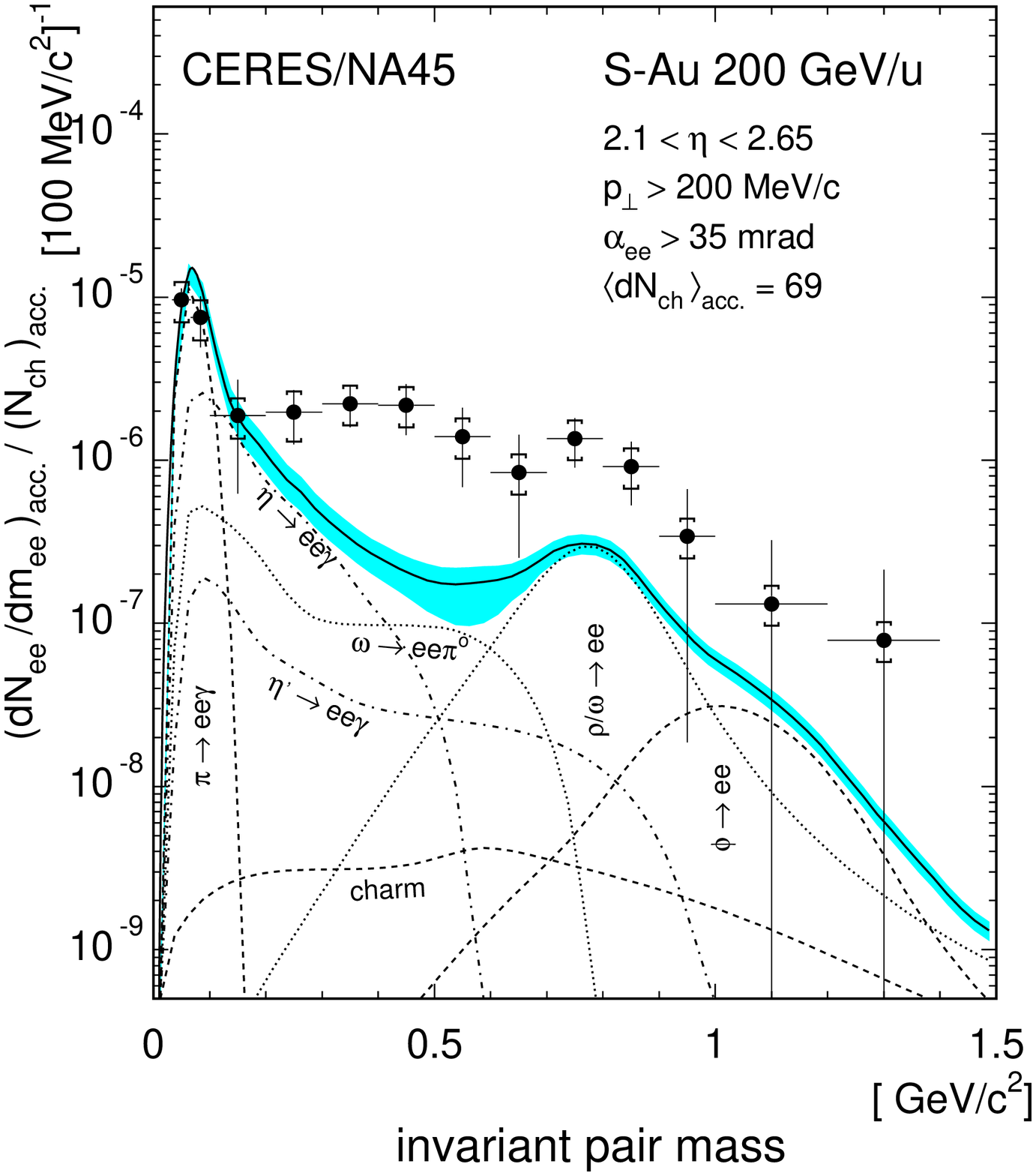}} \vspace{0mm}
\vspace{0mm} ~\hspace{-8.cm}
\epsfysize = 8.5 cm \centerline{\epsfbox{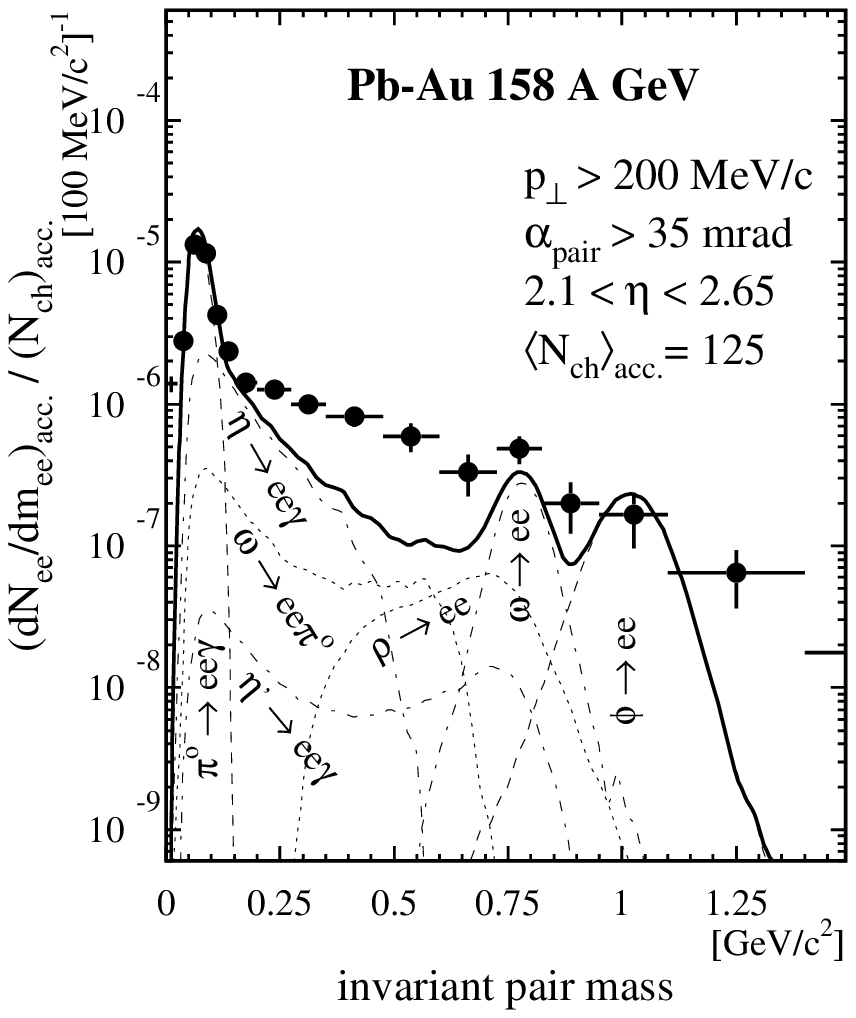}} \vspace{0mm}} 
\caption{\label{SAuPbAu} Dilepton invariant mass spectrum of S+Au and Pb+Au collisions at 
$200~{\rm GeV}$ and $158~{\rm GeV}$ from CERES collaboration, \cite{PhD}. The data is compared 
to expected cocktail of hadronic sources. 
A visible increase of dilepton production is observed in the region $0.2-0.6~{\rm GeV}$.} 
\end{figure}
The situation is different for 
dilepton spectra in central S+Au and Pb+Au collisions at $200~{\rm A~GeV}$ and $158~{\rm A~GeV}$, 
respectively, see Fig.~\ref{SAuPbAu} from \cite{PhD}. The dielectron yield significantly exceeds the expectations extrapolated from 
proton-induced reactions. The enhancement is most pronounced in the invariant mass region 
$0.2-0.6~{\rm GeV}$. In both cases the difference between the experimental and the cocktail model is 
about a factor of 6, whereas in p-induced reactions the spectrum from $0.2$ to $0.6~{\rm GeV}$ reproduces 
the data.
\begin{figure}[h]
\begin{center}
\includegraphics[width=0.6 \textwidth]{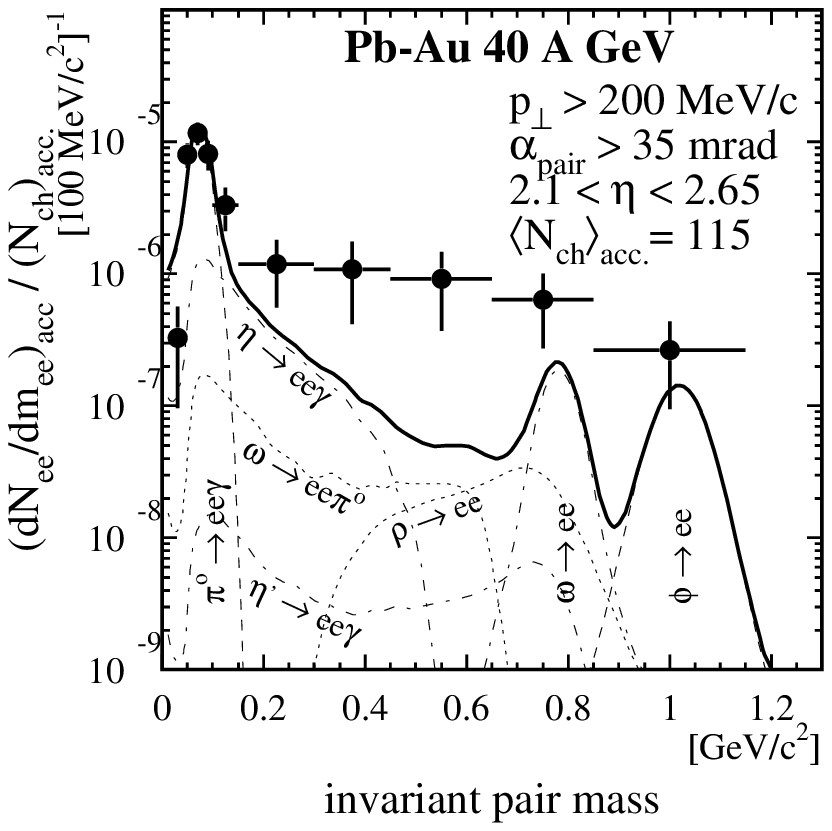}
\caption{\label{40GeV} Dielectron invariant mass spectrum of Pb+Au collisions at $40~{\rm A~GeV}$ 
taken from \cite{PhD}. The data is compared to the expected cocktail of hadronic sources (solid line).}
\end{center}
\end{figure}
The CERES collaboration has also studied the low-mass dielectrons in Pb+Au collisions 
at the lower energy of $40~{\rm GeV}$, see Fig.~\ref{40GeV}. 
Enhancement of the dilepton yield, related to the expected yield of hadronic sources, 
is even more pronounced than at the energy of $158~{\rm A~GeV}$ \cite{40gevlow}. 

The difficulty in explaining this experimental fact has become known as the \emph{dilepton puzzle} and is 
an important theoretical or experimental problem.

\subsection{HELIOS experiments}
In the HELIOS experiments there are three collaborations. 
The HELIOS-1 collaboration \cite{ helios-1} 
was the pioneer in the measurement of the dilepton spectrum at CERN SPS. It was the only 
experiment that measured both the $e^{+} e^{-}$ and $\mu^{+} \mu^{-}$ pair production in p+Be reactions at 
$450~{\rm GeV}$ as well as the first in which the hadron cocktail contribution to the dilepton spectra 
was analyzed. The conclusion is that the measured dielectron and dimuon spectra in p+Be collisions are 
described by the cocktail hadron decays, which was later confirmed by the CERES collaboration 
with much greater precision.
\begin{figure}[htb]
\centerline{
\vspace{0mm} ~\hspace{0cm} 
\epsfxsize = 7.5 cm \centerline{\epsfbox{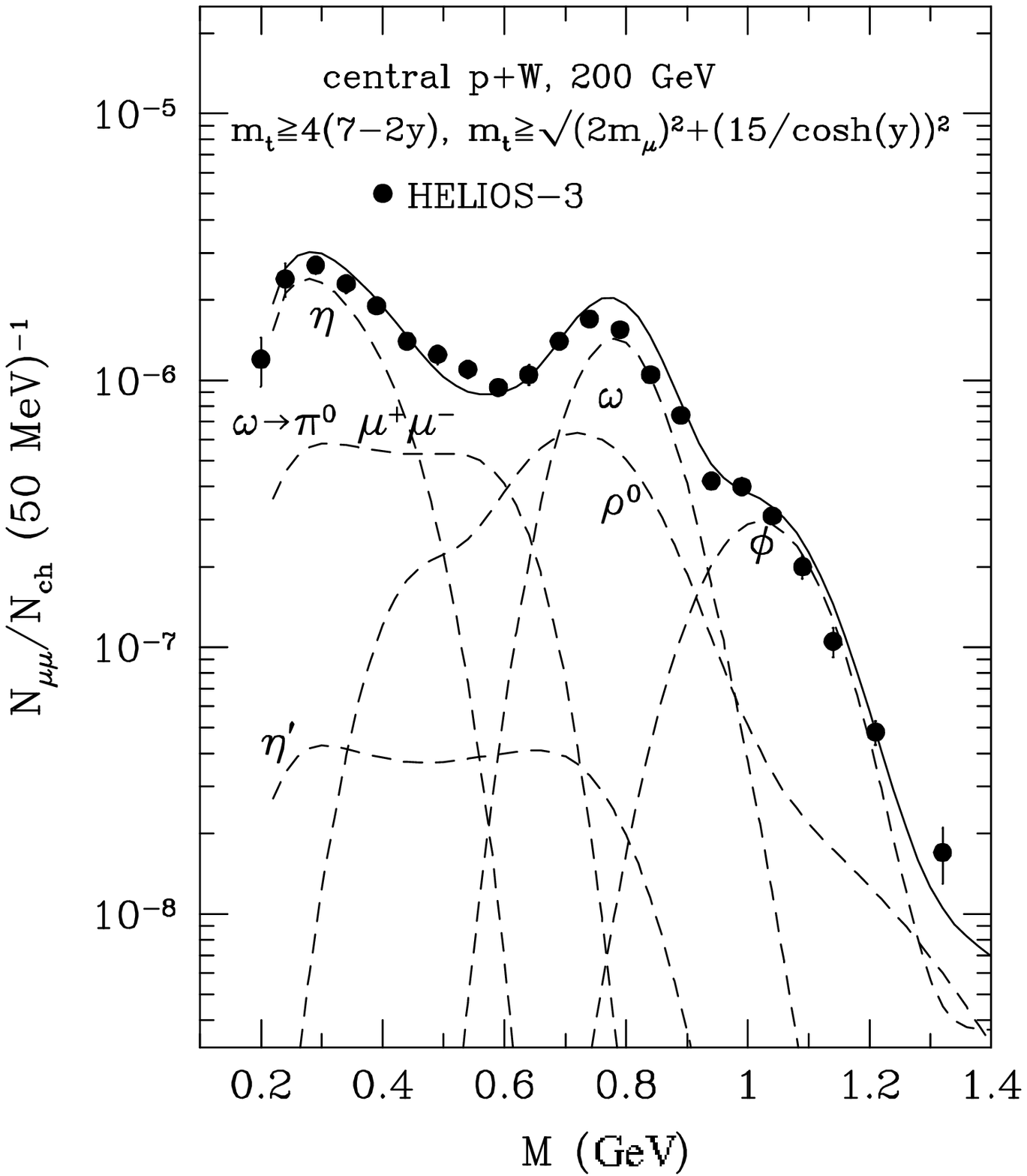}} \vspace{0mm}
\vspace{0mm} ~\hspace{-8.cm}
\epsfysize = 8.5 cm \centerline{\epsfbox{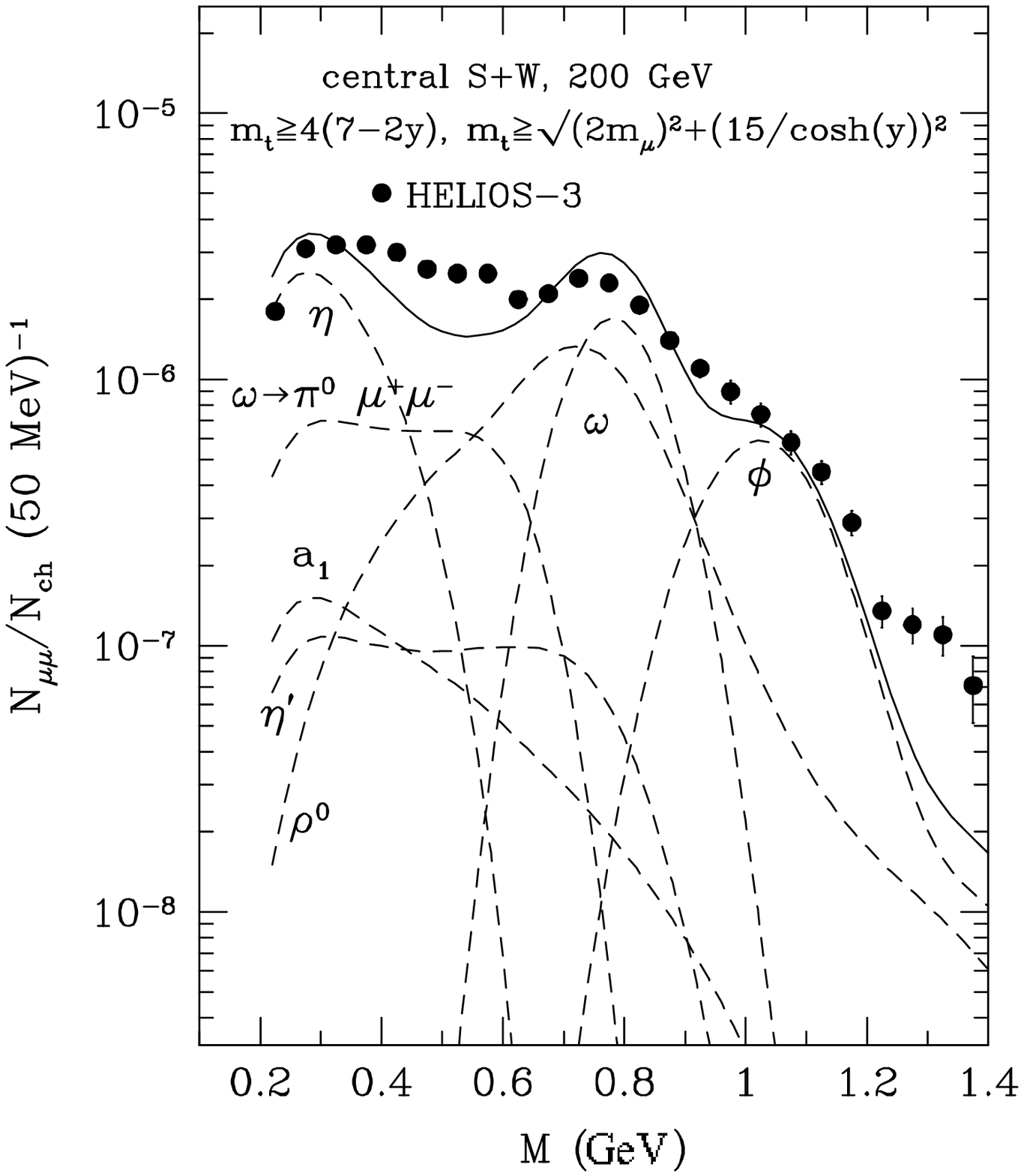}} \vspace{0mm}} 
\caption{\label{helios} Dilepton invariant mass spectrum of p+W (left) and S+W (right) collisions at 
$200~{\rm GeV}$ from HELIOS-3 collaboration. The data is compared to expected cocktail 
of hadronic sources, \cite{ sorge}.} 
\end{figure}
The HELIOS-2 collaboration \cite{ helios-2} was the first to measure the photon signals 
using the S and O beams at the CERN SPS energy. Very small direct photon 
enhancement (of about a few percent) was observed in these studies. This observation was later 
confirmed by the CERES and WA80 collaborations with better statistics.

The HELIOS-3 experiment \cite{ helios} is designed to study the dimuon production in p+W and S+W  
collisions at $200~{\rm GeV}$, and covers the low- and intermediate-mass region. In  
Fig.~\ref{helios}, taken from  \cite{ sorge}, we see the results for p+W collisions (left) and 
S+W collisions (right). For p+W collisions the low mass dimuons are mainly from the $\eta$ Dalitz 
decay, while the omega meson Dalitz decay is important in the mass region $0.2-0.6~{\rm GeV}$. The 
cocktail contribution is in agreement with the experimental points. 
For the S+W collisions the dimuon production is enhanced in both the 
low-mass and intermediate-mass regions, as compared to the proton-nucleus collisions. 

In the low dimuon invariant mass region the enhancement factor is smaller than that found by the 
CERES collaboration in central collisions. One of the possible reasons for this difference is that 
the HELIOS-3 experiment covers more forward  pseudorapidities, $3.7 \leq \eta \leq 5.2$, than the CERES 
experiment which is close to the midrapidity region, $2.1 \leq \eta \leq 2.65$. A significant 
excess in the dimuon yield in the intermediate-mass region is also reported, with an enhancement factor 
of about 2 \cite{ helios}. 

\subsection{NA38/NA50 experiments}
The NA38 and NA50 experiments specialize in the study of the $J/\Psi$ suppression in heavy-ion 
collisions. These collaborations measured the dimuon spectra in a mass region from the threshold to  
about $8~{\rm GeV}$. The measurements are carried out at energies of $200~{\rm A~GeV}$ for the p+W and 
S+U reactions (NA38), and $160~{\rm A~GeV}$ for the Pb+Pb reactions (NA50). 
A situation similar to that described for other experiments occurs. 

In proton-induced 
reactions the data are described with a rather good agreement as compared to the expected sources 
from hadron decay. On the other hand, for S+U and Pb+Pb reactions significant enhancement for low- 
and intermediate-mass region is observed. This shift is more visible in Pb+Pb collisions in the 
mass region around $2~{\rm GeV}$. 
The data from NA38/NA50 experiments have no centrality selection, whereas the CERES 
collaboration measures central collisions.

\subsection{Dilepton spectra at BEVALAC energies}
It is worth mentioning that the measurements of the dilepton production at much lower 
bombarding energies, in the 
range $1-5~{\rm A~GeV}$, have been performed by the DLS collaboration at LBL BEVALAC 
\cite{ bevalac1, bevalac2}. In this experiment a different temperature and density regime is probed. 
The dielectron spectra 
were measured for p+Be collisions at 1, 2.1 and $4.9~{\rm A~GeV}$, for Ca+Ca reactions at 1 and 
$2~{\rm A~GeV}$ and for Nb+Nb at $1.05~{\rm A~GeV}$. In the low invariant mass region from 
$0.2-0.6~{\rm A~GeV}$ production of dielectrons is found to be enhanced compared to estimates 
based on the theoretically known dilepton sources. 

For ultra-relativistic reactions (SPS) the low-mass dilepton excess is explained by the reduction of 
the $\rho$-meson mass in a dense medium. The in-medium 
broadening of the $\rho$-meson  
is also expected to be sufficient to explain the dilepton yield at SPS energies. 
These explanations fail for the DLS data. This fact is known as the 'DLS puzzle'. 
The reason probably lies in the fact that the possible pQCD contributions and sufficient 
$\pi^{+}\pi^{-}$ annihilations are absent at this energies.

\subsection{Other experiments}
The new NA60 experiment studies the production of open charm and the production of dimuons instead of dielectrons 
in proton-nucleus and nucleus-nucleus collisions at CERN SPS. This experiment is a continuation of the 
NA38 and NA50 experiments but with significantly enhanced detector technology. 
The aim of the NA60 experiment is to examine various possible signatures of the transition 
from hadronic to deconfined partonic matter \emph{e.g.}, charmonium suppression, dimuons from 
thermal radiation, and modifications of vector meson properties. The NA60 experiment is continuing with 
$^{115}In$ beams at least up to the year 2005, Ref.~\cite{na60}.

Dilepton spectra were also measured at KEK in p+A reactions at a beam energy of $12~{\rm A~GeV}$ \cite{ kek}. 
Also here considerable enhancement over the expected yield from hadronic decays was observed below 
the $\rho$-meson peak. Explanations of the experimental spectrum within a dropping mass scenario or the 
broadening of vector meson are not proper in this case as they were not at the Bevalac energies.
Similar experiments with much better statistics are planned at GSI SIS by the HADES collaboration \cite{gsi}.

Summarizing this section, in the low-mass region between 0.2-0.6~\rm{GeV} of the invariant mass of the dilepton, 
experiments reported an unexpected enhancement of dileptons in nucleus-nucleus collisions with comparison to 
the extrapolation from more elementary proton-induced reactions. This has been a topic of great interest from 
a theoretical point of view, to which the next section is devoted.

\section{Overview of theoretical models}

In this section we would like to review various theoretical explanations that have been put forward 
to explain the experimental data discussed in the previous section.
A lot of theoretical efforts has been devoted to understanding 
the experimental results in the low-mass region from the CERES and HELIOS experiments. 
Theorists have proposed possible explanations of the 
dilepton puzzle based on transport model calculations, many-body approaches, and the 
effects of dropping rho meson mass and/or increasing width. 

The main contributions to $e^{+}e^{-}$ pairs with mass below 300~\rm{MeV} are: the direct 
leptonic decay of vector mesons such as $\rho^{0}$, $\omega$ and $\phi$, the pion-pion annihilation 
which proceeds through the $\rho^{0}$ meson, the kaon-antikaon annihilation that proceeds through 
the $\phi$ meson, and the Dalitz decay of $\pi^{0}$, $\eta$, $\omega$ and $\rho$.

The CERES and HELIOS-3 data are in reasonable agreement with each other in proton-induced reactions.
They are well explained by the Dalitz and direct vector meson decays. 
In nucleus-nucleus collisions the substantial enhancement of low-mass dileptons in the mass region 
$0.2-0.6~{\rm GeV}$ remains unexplained. 
The difference between proton-induced and nucleus-nucleus 
reactions may lie in additional contributions from pion-pion and kaon-antikaon annihilation to dilepton 
production in heavy-ion collisions. Authors of \cite{li, LiKoBrown1, sorge, CassingDil} 
have included the contributions from the pion-pion annihilation but neglected any medium effects. 

The theoretical results are still in disagreement 
with data in the mass region of interest, which leads to the suggestion that medium modifications of 
vector mesons are needed \cite{Urban0,rapp,RappDil}. It has been found that the models with the use of in-medium vector meson masses 
describe the CERES data much better than models with vacuum meson masses. The same situation occurs 
for the dimuon spectra from the central nucleus-nucleus collisions measured by the HELIOS-3 collaboration 
\cite{ sorge, CassingKralik}.

The rho meson decays dominantly into two pions. The medium modifications of the pion dispersion 
relation and the rho-nucleon scattering contribute to the spreading of the rho meson spectral function,  
which results in enhancement of the production of low-mass dileptons from in-medium rho meson decay. 
This broadening of the rho meson spectral function provides a better description for the CERES 
\cite{rapp,RappDil} and HELIOS-3 data.

Many  calculations and mechanisms have been proposed for the explanation of the low-mass dilepton enhancement. 
Kapusta et al. \cite{ KapustaMc} and Hung et al. \cite{ Huang} 
proposed that the in-medium reduction of the $\eta$ 
and/or $\eta'$ masses leads to an enhanced production of these mesons, which in turn  
increases the low-mass dilepton yield. The authors of \cite{ GaleKap, Song} studied the effects 
of an in-medium pion dispersion relation on dilepton production and they indeed found that the dilepton 
yield is enhanced near $0.4~{\rm GeV}$ invariant mass. The contribution from the pion-rho 
annihilation has been studied in \cite{ haglin, murray}. 
This contribution was too small to explain the observed enhancement. 
Hung and Shuryak \cite{ HungDil} also emphasized the effect of dropping $a_{1}$ meson mass, 
which is the chiral partner of the rho meson, on the very low-mass dileptons. On the other hand, 
Kluger {\emph et al.} \cite{ kluger} studied dilepton production rates from the disoriented chiral 
condensate (DCC) state, and found that the dilepton yield with mass below 300~\rm{MeV} from the DCC state 
can be significantly larger than that from the equilibrium state.

In other words, the low-mass dilepton enhancement reported by the collaborations from the previous section 
required the introduction of pion-pion annihilation and medium modifications of the meson properties. 
The dropping rho meson mass scenario, as implied by the Brown-Rho scaling conjectures and QCD sum rule 
calculations, provides a possible explanation of the observed enhancement. On the other hand, the medium 
modification of the rho meson spectral functions due to its interaction with baryons and medium 
modification of coupling constants also provides a reasonable description of experimental data.

Here we have discussed various theoretical approaches. Next in this thesis we follow the lines of 
Brown-Rho \cite{brscale} and Rapp-Wambach \cite{rapp,RappDil} and analyze medium modification of vector mesons.

\chapter{Formalism of the dilepton production from vector mesons}
We are interested in the $\pi\omega\rho$ coupling constant which enters into the 
Dalitz decay of the $\omega$ and $\rho$ meson. 
The Dalitz $\rho$ decay is usually neglected in other analyses because of its
absence in the Particle Data Book \cite{pdbook}. Nevertheless, it may occur and while it is less significant 
in the vacuum than the $\omega$ Dalitz decay, we will show that in the medium it becomes equally important.

We point out that we work at zero temperature. The authors of Ref.~\cite{urban} have analyzed the finite 
temperature effects for the $\rho\pi\pi$ coupling. Such effects turn out to be very small, therefore in 
our analysis they are also neglected. 

In our calculations we use the Vector Meson Dominance Model 
to estimate the dilepton yields from vector-meson decays. Apart from the Dalitz decays, an important source of 
additional dileptons in the region near twice the pion mass is the $\pi\pi$ annihilation. In this Chapter 
we first briefly review the VMD model. 
Next, we describe the structure of three-body
Dalitz decays. 

In order to calculate the dilepton spectrum one has to assume the evolution model of the 
collision system. Therefore, we review the Rapp-Wambach \cite{Rappevol, RappShur} model of the hydrodynamic 
expansion of the fire cylinder, which we are going to apply in our calculations. In section 4.4 we discuss the 
kinematics for both direct and Dalitz decays with inclusion of effects of medium expansion in heavy-ion 
collisions. Thus we get the yield of dileptons produced during the expansion. 
In this analysis the Bose enhancement is neglected for brevity. Finally, we present our results 
and compare them to the CERES experimental data.

\section{Vector Dominance Model (VDM)}
\begin{figure}[h]
\begin{center}
\includegraphics[width=.42\textwidth]{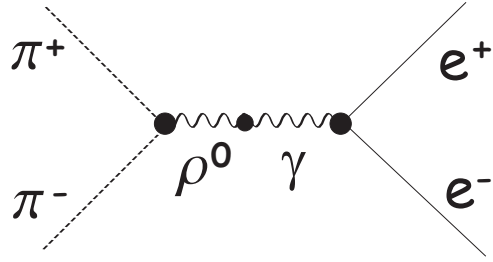}
\caption{\label{pipi} Vector Dominance Model of dilepton production from pion annihilation.}
\end{center}
\end{figure}

The dilepton production rate per unit time and volume from the annihilation of pions, Fig.~\ref{pipi},  
is given by the formula
\begin{eqnarray}
\frac{dN_{l+l-}}{d^{4}x} = \sum_{spins}\int dP_{1}dP_{2}dQ_{1}dQ_{2}f(p_{1})f(p_{2})(2\pi )^{4}
\delta ^{(4)}(p_{1}+p_{2}-q_{1}-q_{2})|{\mathcal{M}_{\pi^{+}\pi^{-} \to {l^{+}l^{-}}}}|^{2}.\nonumber \\
\label{pipidil}
\end{eqnarray}
The sum runs over spins of leptons, where $p_{1}$ and $p_{2}$ are the four-momenta 
of the annihilating pions and $q_{1}$ and $q_{2}$ are the four-momenta of the outgoing leptons. 
The $dP$ and $dQ$ indicate the integration measures 
\begin{eqnarray}
 dP_{i}=\frac{d^{3}p}{(2\pi)^3 2 E_{p_{i}}}, ~~dQ_{i}=\frac{d^{3}q}{(2\pi)^3 2 E_{q_{i}}},  
\end{eqnarray}
where $E_{p_{i}}=\sqrt{p_{i}^{2}+m_{\pi}^2}$ and $E_{q_{i}}=\sqrt{q_{i}^{2}+m_{l}^2}$, $i=1,2$. We use the following 
normalization for the lepton spinors, $\bar{u}u =\bar{v}v=2m_{l}$.
In Eq.~\ref{pipidil} $f(p)$ is the pion distribution function of the form
\begin{eqnarray}
f(p)=\left[\exp {\frac{E_{p}- \mu_{\pi}}{T}}-1 \right]^{-1},
\end{eqnarray}
with the pion chemical potential $\mu_{\pi}$, incorporated in several works. 
The positive pion potential increases the number of pions over the equilibrium value, as discussed later on 
in section 5.3.

The matrix element $\mathcal{M}_{\pi^{+}\pi^{-} \to l^{+}l^{-}}$ is calculated in the 
framework of the VDM. According to this model the electromagnetic interactions of
hadrons are described by the intermediate coupling of hadrons to vector mesons, Fig~\ref{pipi}. The
annihilation of the pions ($\pi ^{+}\pi ^{-}\rightarrow l^{+}l^{-}$) can be understood as the 
formation of a vector meson by the annihilation of pions and, the sequential decay of the vector meson 
into a lepton pair. Therefore, pions regain an electromagnetic form 
factor. The matrix element for this process has the structure
\begin{eqnarray}
\mathcal{M}_{\pi^{+}\pi^{-} \to l^{+}l^{-}}=  \frac{1}{q^{2}}eL^{\mu} 
\sum_{V}\frac{e}{f_{V}}\frac{m_{V}^{2}}{m_{V}^{2}-q^{2}-im_{V}\Gamma_{V}} g_{\pi\pi V} (p_{1}-p_{2})_{\mu},
\label{mel}
\end{eqnarray}
where the term $m_{V}^{2}/(m_{V}^{2}-q^{2}-im_{V}\Gamma_{V})$ comes from the vector meson propagator 
\begin{eqnarray}
G_{V}(q)=\frac{g^{\mu\nu}-\frac{q^{\mu} q^{\nu}}{m_{V}^2}}{q^2-m_{V}^2 -im_{V}\Gamma_{V}}.
\end{eqnarray} 
The quantities $m_{V}$ and $\Gamma_{V}$ are the mass and width of the vector meson $V$, the coupling of the 
vector meson to the virtual photon are described by the $f_{V}$ parameter, 
$g_{\pi\pi V}$ is the effective coupling constant for the $\pi\pi V$ interaction, and $1/q^{2}$ comes from the 
photon propagator such as 
\begin{eqnarray}
G_{\gamma}^{\mu\nu}=\frac{g^{\mu\nu}-\frac{q^{\mu} q^{\nu}}{q^2}}{q^2}
\end{eqnarray}  
with $q$ denoting the four-momentum of the virtual photon ($q^{2}=M^{2}$, the invariant mass squared of the lepton pair).
 
The quantity $L^{\mu}$ is the  leptonic current which may be expressed in terms of the Dirac bispinors,
\begin{eqnarray}
L^{\mu}(q_{1},s_{1};q_{2},s_{2})=2m_{l} \bar{u} (q_{1},s_{1})\gamma^{\mu}v(q_{2},s_{2}),
\end{eqnarray}
where $s_{1}$ and $s_{2}$ denote the spin variables of the outgoing leptons.
The term \begin{eqnarray}
\sum_{V}\frac{e}{f_{V}}\frac{m_{V}^{2}}{m_{V}^{2}-q^{2}-im_{V}\Gamma_{V}} 
g_{\pi\pi V}(p_{1}-p_{2})_{\mu}
\end{eqnarray}
from Eq. \ref{mel}, describes the coupling of the photon to two incoming pions via the electromagnetic 
form factor in the VDM including also the propagator of the intermediate vector meson. 

\section{Dalitz decays}
The structure of the three-body Dalitz decays of the form $A \to B l^{+}l^{-}$ is as follow: 
a particle decays $A$ into $B$ and a vector meson $V$ which is subsequently converted into a virtual 
photon according to the vector dominance principle. Finally, the photon decays into the 
$e^{+}e^{-}$ pair, see Fig~\ref{dalitz}. This process is usually called the 'internal conversion' of a virtual photon 
into a lepton pair.
For these processes, where either a vector meson ($V$) or pseudoscalar meson ($P$) decays via 
$V \to P l^{+}l^{-}$ or $P \to V l^{+}l^{-}$, the transition amplitude has the invariant form
\begin{eqnarray}
\mathcal{M}_{A \to B l^{+}l^{-}}=L^{\mu} \epsilon^{\alpha \beta \gamma \mu} p_{\alpha}q_{\beta} \epsilon_{\gamma} 
\frac{e^{2}}{q^{2}} f_{AB}(M^{2}) \label{M},
\end{eqnarray}
where
$L^{\mu}$ is a leptonic current, $p_{\alpha}$ the four-momentum of meson $B$, $q$ the four-momentum 
of the virtual photon $\gamma^{*}$, $q^{2}=M^2$ the invariant mass squared of the lepton pair, 
$f_{AB}(M^2)$ is the electromagnetic form factor describing the $A \to B$ transition, and 
$\epsilon^{\alpha \beta \gamma \mu}$ is a totally antisymmetric tensor. We easily obtain the useful decomposition
\begin{eqnarray}
|{\mathcal{M}_{A \to B {l^{+}l^{-}}}}|^{2}=|{\mathcal{M}_{A\to B \gamma^{*}}}|^{2}\times
\frac{1}{M^4}|{\mathcal{M}_{\gamma^{*} \to {l^{+}l^{-}}}}|^{2},
\end{eqnarray}
whereby we can see why this process is called the internal conversion. 
Here 
\begin{eqnarray}
|{\mathcal{M}_{\gamma^{*} \to {l^{+}l^{-}}}}|^{2}=e^2 L^{\mu}(L^{\mu'})^*
\end{eqnarray}
and 
\begin{eqnarray}
|{\mathcal{M}_{A\to B \gamma^{*}}}|^{2}=A_{\mu}(A_{\mu'})^*,
\end{eqnarray}
where 
\begin{eqnarray}
A_{\mu}=\epsilon^{\alpha \beta \gamma \mu} p_{\alpha}q_{\beta} \epsilon_{\gamma} e f_{AB}(M^{2}).
\end{eqnarray}
Using the decomposition \ref{M} for the differential decay distribution we can write 
\begin{eqnarray}
\frac{d \Gamma_{A \to B l^{+}l^{-}}}{dM^{2}}=\frac{\Gamma_{\gamma^* \to l^{+} l^{-}}}
{\pi M^{4}} \Gamma_{A \to B \gamma^{*}},
\label{abll}
\end{eqnarray}
where for the width of the internal conversion of the photon we find
\begin{eqnarray}
\Gamma_{\gamma^{*} \to l^{+}l^{-}}=\frac{2 \alpha_{QED}}{3} \sqrt{1-\frac{4m_{l}^2}{M^2}}
(2m_{l}+M^2),
\label{abll1} 
\end{eqnarray}
and
\begin{eqnarray}
\Gamma_{A \to B \gamma^{*}}=\Gamma_{A \to B \gamma}|f_{AB}(M^2)|^2 \frac{[m_{A}^{2}-(m_{B}+M)^2]^{3/2}
[m_{A}^{2}-(m_{B}-M)^2]^{3/2}}{(m_{A}^{2}-m_{B}^{2})^{3}}
\label{abll2} 
\end{eqnarray}
is related to the decays where the photon is real. The normalized form factor for the $A \to B\gamma^*$ 
transition has a form
\begin{eqnarray}
|f_{AB}(M^2)|^2=\frac{m_{\rho}^4+m_{\rho}^2\Gamma_{\rho}^2}{(m_{\rho}^2-M^2)^2+m_{\rho}^2\Gamma_{\rho}^2}. 
\label{fabM}
\end{eqnarray}
It has the property $|f_{AB}(M^2)|^2 \to 1$ for $M \to 0$, for increasing $M$ the form factor also 
increases. After putting all Eqs~\ref{abll}, \ref{abll1}, \ref{abll2} and \ref{fabM} together we arrive at the 
differential decay distribution which we will use to calculate the mass spectrum of dileptons 
from Dalitz decays. For more details see \cite{PKoch}.

\section{Model of the hydrodynamic expansion of the fire-cylinder}
The description of medium effects on the dilepton production in heavy-ion collisions
requires modeling of the time evolution of the collision system. The
theoretical models can be divided into: transport models
\cite{ sorge, weise}, hydrodynamical approaches \cite{ HungDil, BaierDil, murray}, and
thermal fireball models \cite {rapp, Rappevol}. 

In this work our calculations are based on the fire cylinder hydrodynamic expansion 
model of Ref.~\cite{ Rappevol, RappShur}. Lepton pairs are formed in a fire cylinder which 
undergoes a hydrodynamic expansion.
The hadron-chemical analysis of a large body of hadronic heavy-ion data has shown that the final 
observed particle abundances at the SPS energies are consistent with a common chemical freezeout at 
temperatures of around $175~{\rm MeV}$ and baryon chemical potentials of around $270~{\rm MeV}$. 
In the two-freezeout scenario, during the  
subsequent expansion and cooling the system continues to strongly interact via elastic collisions 
maintaining the thermal equilibrium until the thermal freezeout. The absence of pion-number changing 
reactions then induces a finite pion chemical potential. To incorporate these features, Rapp 
{\emph{et al.}}, have proposed a simple expanding fireball model \cite{ Rappevol, RappShur}. 
This model assumes that the system is in thermal equilibrium up to time $t_{\max }$ , when the thermal 
freeze-out occurs. The velocity field characterizing the expansion depends on space-time coordinates 
in the following way:
\begin{eqnarray}
v_{\shortparallel }(t,z) &=&(v_{z}+a_{z}t)\frac{z}{z_{\max }(t)}, \label{vsp} \\
v_{\perp }(t,r) &=&(v_{r}+a_{r}t)\frac{r}{r_{\max }(t)}, \nonumber
\end{eqnarray}
where
\begin{eqnarray}
z_{\max }(t) &=&z_{0}+v_{z}t+\frac{1}{2}a_{z}t^{2},\label{zrmax} \\ 
r_{\max }(t) &=&r_{0}+v_{r}t+\frac{1}{2}a_{r}t^{2}, \nonumber
\end{eqnarray}
are the boundaries of the system at time $t$. We use the following
parameters of expansion: 
$t_{\max }=11~{\rm fm}$  is fireball lifetime, $z_{0}=4.55~{\rm fm}$ is equivalent to formation time 
or initial temperature, 
$r_{0}=4.6~{\rm fm}$  corresponds to the initial transverse nuclear overlap radius, 
$v_{r}=0$, and the parameters $v_{z}=0.5$, $a_{z}=0.023$ fm$^{-1}$ and $a_{r}=0.05$ fm$^{-1}$ 
are adjusted to the final observed flow velocities.
The time dependencies of the temperature and the baryon density are as follows
\begin{eqnarray}
T(t) &=&(210 \rm{MeV})\exp \left( -\frac{t}{18.26\rm{fm}}\right) , \\
\rho _{B}(t) &=&\frac{260}{V(t)}  \label{rhoB}, 
\end{eqnarray}
where 260 denotes the number of participating baryons, and
\begin{eqnarray}
V(t) &=&2\pi z_{\max }(t)r_{\max }^{2}(t),
\end{eqnarray}
is the volume of the fire-cylinder at time $t$.

For the time dependence of the pion chemical potential, $\mu(t)$, we assume, following Ref.~\cite{ rapp},  
the linear rise from $20~{\rm MeV}$ at $t=0$ to $80~{\rm MeV}$ at $t=t_{max}$. 

\section{Kinematics for dilepton yield}

\subsubsection{Direct decays}
\begin{figure}[t]
\begin{center}
\includegraphics[width=0.4\textwidth]{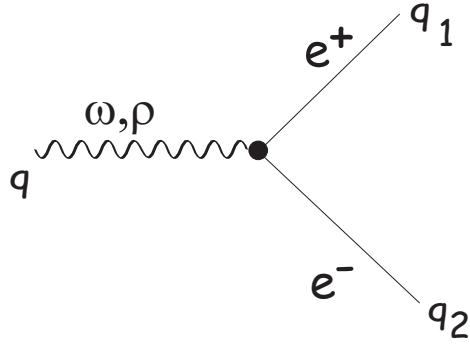}
\caption{\label{dirmom} The assignment of momenta in the $\omega\to e^{+}e^{-}$ decay, where $q$ is the 
incoming momentum of the $\omega$ or $\rho$ meson, $q_{1}$ and $q_{2}$ are the outgoing momenta of the leptons}
\end{center}
\end{figure}

A dominant source of dileptons in the low-mass region is the direct leptonic decays of vector mesons. 
Although we focus on Dalitz decays for dileptons of mass 
in the region of interest, one also considers the $\rho^{0}$ and $\omega$ mesons, see Fig.~\ref{dirmom}. 
The dilepton rate formula for the $\rho^{0} \to e^{+}e^{-}$ decay is
\begin{equation}
\frac{dN ^{e+e-}}{d^4 x dM^2}=\int \frac{d^{3}q}{(2\pi ^3)} \Gamma_{\rho \to e^{+}e^{-}}
f_{\rho}(q \cdot u(x)) 
\frac{1}{\pi} \frac{3 M \Gamma_{\rho}}{(m_{\rho}^{2}-M^2)^2+(M \Gamma_{\rho})^2}  ,  
\label{rhodir}
\end{equation}
where $M$ is the invariant mass of the lepton pair, $x$ is a space time point, and $\Gamma_{\rho}$ 
is the total rho-meson width. 
The quantity $\Gamma_{\rho \to l^{+}l^{-}}$ is the decay width for $\rho^{0} \to e^{+}e^{-}$,
\begin{equation}
\Gamma_{\rho \to l^{+}l^{-}}=\frac{4\pi \alpha_{QED}^{2} m_{\rho}^4}{3g_{\rho} M^5}
(1-\frac{4m_{l}^{2}}{M^2})^{1/2} (M^2 + 2m_{l}^{2}),
\end{equation}
where $\alpha_{QED}$ denotes the fine structure constant, and $m_{l}$ is the mass of the lepton. 
The function $f_{\rho}(q \cdot u)$ in Eq.~\ref{rhodir} is the thermal Bose-Einstein distribution 
of the $\rho$ mesons in the thermal model, 
\begin{equation}
f_{\rho}(q \cdot u(x))=\exp[-(\frac{q \cdot u(x) - 2\mu_{\pi}(x)}{T(x)})] \label{BE},
\end{equation}
with the pion chemical potential $\mu_{\pi}$.

\subsubsection{Expansion of the fireball}
\begin{figure}[t]
\begin{center}
\includegraphics[width=0.65 \textwidth]{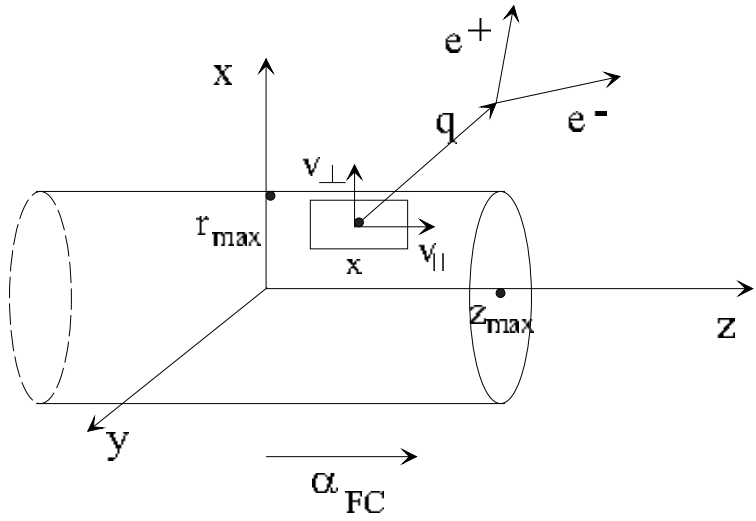}
\caption{\label{det} The hydrodynamic expansion of the cylinder moving in the lab system with rapidity 
$\alpha_{FC}$ in its own center-of-mass system. $e^{+}, e^{-}$ are the leptons formed in a fire cylinder, 
whereas $\rm{v_{\perp}}$, $\rm{v_{\parallel}}$ are the velocities, which depend on space-time, and 
$z_{max}$ and $r_{max}$ are the boundaries of the system at time $t$; see section 4.3.}
\end{center}
\end{figure}

To be as realistic as possible, we include the effects of the expansion of the medium formed in a
relativistic heavy-ion collision, Fig~\ref{det}. The lepton pairs are produced in a fire
cylinder which moves in the lab system with rapidity $\alpha_{FC}$.
The fire cylinder expands hydrodynamically by itself. In particular, Eqs.~\ref{vsp} and \ref{zrmax} 
describe this expansion in the fire cylinder's center-of-mass system (CM). In such a case
it is convenient to rewrite Eq. (\ref{rhodir}) in 
variables which are more convenient for the kinematics of the emission process and the
geometry of the experiment. We introduce $M_{\perp }=\sqrt{%
M^{2}+q_{\perp }^{2}}$, the transverse mass of the dilepton pair, $y^{lab}$,
the rapidity of the pair measured in the lab system, ${\bf u}_{\perp }$, the
transverse four-velocity of the fluid element producing dileptons, and $%
\alpha ^{lab}$, the rapidity of this fluid element in the lab. With these
variables we have in the cylinder's rest frame
\begin{equation}
q\cdot u=M_{\perp }\sqrt{1+u_{\perp }^{2}}\hbox{cosh }(y^{lab}-\alpha
^{lab})-{\bf q}_{\perp }\cdot {\bf u}_{\perp }.  \label{qu}
\end{equation}
The velocity of the fluid element in the lab is a relativistic superposition
of the velocity of the fire cylinder in the lab and the hydrodynamic flow
considered in the CM system. Thus we have 
\begin{equation}
\alpha ^{lab}=\alpha +\alpha _{FC}=\hbox{arctanh }v_{||}+\alpha _{FC},%
\hspace{0.5cm}u_{\perp }={\frac{v_{\perp }\hbox{cosh}(\alpha )}{\sqrt{%
1-v_{\perp }^{2}\hbox{cosh}^{2}(\alpha )}}}.  \label{alabuperp}
\end{equation}
The velocities $v_{||}$ and $v_{\perp }$ are now defined in the
CM system and they depend on time and space coordinates.

Next, we analyze the kinematic constraint of the CERES experiment, see Appendix C. The
experimental acceptance cuts can be included using the function 
\begin{eqnarray}
&&Acc (M,y^{lab},q_{\perp }) ={\frac{\int d^{2}q_{1\perp }d^{2}q_{2\perp
}dy_{1}dy_{2}\,\,\,{\bf \phi }\,\,\,\delta (E_{q}-E_{q_{1}}-E_{q_{2}})\delta
^{(3)}({\bf q}-{\bf q}_{1}-{\bf q}_{2})}{\int d^{2}q_{1\perp }d^{2}q_{2\perp
}dy_{1}dy_{2}\delta (E_{q}-E_{q_{1}}-E_{q_{2}})\delta ^{(3)}({\bf q}-{\bf q}%
_{1}-{\bf q}_{2})}},  \nonumber \\
&&  \label{Phi}
\end{eqnarray}
where ${\bf q}_{1,\,2}$ are the momenta of the emitted electrons, $y_{1,\,2}$
are the electron rapidities, and ${\bf \phi }$ is a product of step
functions which enforces the experimental setup conditions: $2.1=<y_{1,\,2}<2.65$, 
$q_{1,\,2}^{\,\perp }>$ 200 MeV, and $\theta _{ee}>$ 35mrad, where $\theta_{ee}$ is the angle between 
the directions of the leptons.
We can assume that the rapidities and pseudorapidities 
of the electrons are equal because of the smallness of the electron mass.
The construction of the function $Acc (M,y^{lab},q_{\perp })$ 
requires numerical calculation of a two-dimensional integral of 
a function involving a product of step functions, which can be done by a Monte Carlo method.  
With the inclusion of the experimental acceptance cuts, the dilepton
production rate is 
\begin{eqnarray}
\frac{dN^{e+e-}}{d^{4}x dM} &=& 2M
\int {\frac{d^{2}q_{\perp }}{(2\pi )^{3}}}\int dy^{lab} Acc(M,y^{lab},q_{\perp }) \times \nonumber \\ 
&\times& \frac{1}{\pi} \frac{3 M \Gamma_{\rho}}{(m_{\rho}^{2}-M^2)^2+(M \Gamma_{\rho})^2} \Gamma
_{\rho \rightarrow l^{+}l^{-}} f_{\rho }  \label{Acc} 
\end{eqnarray}

To calculate the dilepton spectrum, we have to assume a model of the hydrodynamic 
expansion of the fire cylinder, as described in the previous subsection. 
Finally, the yield of dileptons produced during the expansion is 
\begin{equation}
{\frac{dN^{e+e-}}{dM}}=\int\limits_{0}^{t_{max}}dt\int%
\limits_{0}^{r_{max}(t)}2\pi
rdr\int\limits_{-z_{max}(t)}^{z_{max}(t)}dz\left( {\frac{dN}{%
d^{4}x\,dM}}\right) ,  \label{expyield}
\end{equation}
where $\frac{dN}{(d^{4}x dM)}$ is given by Eq. (\ref{Acc}) with elements defined in Eqs. \ref{rhodir}, \ref{BE}, 
\ref{Phi}. 
Similarly, we calculate the dilepton rate formula for the $\omega\to l^{+} l^{-}$ decay.

\subsubsection{Dalitz decays}
\begin{figure}[h]
\begin{center}
\includegraphics[width=0.4\textwidth]{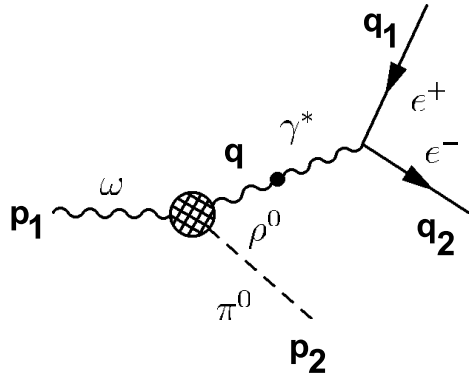}
\caption{\label{mom} The assignment of momenta in the $\omega\to\pi^{0}e^{+}e^{-}$ decay, where $p_{1}$ is the 
incoming momentum of the $\omega$, $p_{2}$ is outgoing momentum of the pion, and $q$ is the outgoing momentum 
of the virtual photon. The outgoing momenta of the leptons are indicated by $q_{1}$ and $q_{2}$.}
\end{center}
\end{figure}

In order to calculate the Dalitz decay $\omega\to\pi^{0}e^{+}e^{-}$, 
we choose the following kinematics, Fig~\ref{mom}: 
$p_{1}$ is the incoming momentum of the $%
\omega $, $q$ is the outgoing momentum of the virtual photon, and $%
p_{2}=p_{1}-q$ is the outgoing momentum of the pion, where $q^{2}=M^{2}$,
and $M$ denotes the invariant mass of the lepton pair. 

The particles are on the mass shell, hence $%
p_{1}^{2}=m_{\omega }^{2}$, $p_{2}^{2}=m_{\pi }^{2}$, and $p_{1}\cdot q=%
\frac{1}{2}(m_{\omega }^{2}+M^{2}-m_{\pi }^{2})$. 
In Fig.~\ref{mom}, $q_{1}$ and $%
q_{2}$ are the outgoing momenta of the leptons. The dilepton production rate
from the Dalitz decays of the $\omega $ is given by the general formula
\begin{eqnarray}
\frac{dN^{e+e-}}{d^{4}x} &=&\sum_{ss^{\prime }}\sum_{p}\int dP_{1}f_{\omega
}(p_{1}\cdot u)dP_{2}\left[ 1+f_{\pi }(p_{2}\cdot u)\right] \int
dQ_{1}dQ_{2}(2\pi )^{4}\times \nonumber\\
\label{dilrate}
&&\delta ^{(4)}(p_{1}-p_{2}-q_{1}-q_{2})|\mathcal{M}^{(p)}|^{2}
\end{eqnarray}
where $x$ is a space-time point, $f_{\omega }(p_{1}\cdot u)$ and $f_{\pi
}(p_{2}\cdot u)$ are the distribution functions for the $\omega $ meson and the
pion, respectively, $u$ is the four velocity of the medium, $dP_{i}=\frac{%
d^{3}p_{i}}{(2\pi )^{3}2E_{p_{i}}}$, $dQ_{i}=\frac{d^{3}q_{i}m_{i}}{(2\pi
)^{3}E_{q_{i}}}$, $|\mathcal{M}|^{2}$ is the transition amplitude, $\sum_{ss^{\prime }}$ is a  
sum over spins of the leptons, and $\sum_{p}$ denotes the sum over the polarizations of the
decaying meson. The amplitude for these processes has the invariant form
\begin{equation}
\mathcal{M}^{(p)}=A_{\gamma \sigma }\varepsilon_{(p)}^{\gamma }(p_{1})G(q)\frac{%
em_{\rho }^{2}}{g_{\rho }}\frac{1}{q^{2}}eL^{\sigma }
\end{equation}
where $A_{\gamma \sigma }$ is given by Eq.(\ref{struct}), $%
\varepsilon^{\gamma }$ is the polarization vector of the $\omega $, $%
G(q)$ is the denominator of the rho-meson propagator, $\frac{em_{\rho }^{2}}{g_{\rho }}$ is the
conversion factor from the vector-dominance model, $\frac{1}{q^{2}}$ comes from the
virtual photon propagator, and $L^{\sigma }=\overline{u}(q_{1},s)\gamma
^{\sigma }$v$(q_{2},s^{\prime })$, with $u$ and v denoting the lepton
spinors. We can write the amplitude in a more convenient notation
\begin{equation}
\mathcal{M}^{(p)}=\widetilde{V}_{\sigma }^{(p)} \frac{e}{q^{2}}L^{\sigma }
\end{equation}
where $\widetilde{V}_{\sigma }^{(p)}$ is the $\omega \pi \gamma ^{\ast }$ vertex,
denoted as $\widetilde{V}_{\sigma }^{(p)}=V_{\sigma }^{(p)}G(q)\frac{em_{\rho }^{2}}{%
g_{\rho }}$ , and $V_{\sigma }^{(p)}=A_{\gamma \sigma }\varepsilon_{(p)}^{\gamma
}(p_{1})$ is the $\omega \rho \pi $ vertex. We obtain
\begin{equation}
|\mathcal{M}^{(p)}|^{2}=\widetilde{V}_{\sigma }^{(p)}\widetilde{V}_{\sigma ^{\prime
}}^{(p) \ast }\frac{e^{2}}{M^{4}}L^{\sigma }(L^{\sigma ^{\prime }})^{\ast }.
\end{equation}
Now, we rewrite the dilepton-rate formula Eq. \ref{dilrate} as
\begin{eqnarray}
&&\frac{dN^{e+e-}}{d^{4}x} =\sum_{s}\int dQ_{1}dQ_{2}(2\pi )^{4}\delta
^{(4)}(q-q_{1}-q_{2})L^{\sigma }(L^{\sigma ^{\prime }})^{\ast }4\pi \alpha
_{e.m.}\times \\
&&\sum_{p}\int d^{4}q\int dP_{1}f_{\omega }(p_{1}\cdot u)dP_{2}\left[
1+f_{\pi }(p_{2}\cdot u)\right] \delta ^{(4)}(p_{1}-p_{2}-q)\widetilde{V}%
_{\sigma }^{(p)}\widetilde{V}_{\sigma ^{\prime }}^{(p) \ast }\frac{1}{M^{4}},\nonumber
\end{eqnarray}
where $\alpha _{e.m.}=e^{2}/4\pi $ is the fine structure constant, $%
L^{\sigma }(L^{\sigma ^{\prime }})^{\ast }$ depends on $q_{1}$ and $q_{2}$, $%
\widetilde{V}_{\sigma }^{(p)}\widetilde{V}_{\sigma ^{\prime }}^{(p) \ast }$ depends on 
$p_{1}$ and $p_{2}$, and $M^{2}=q^{2}$. First we compute the integral over $%
dQ_{1}$ and $dQ_{2}$ in the above expression. It has the form
\begin{equation}
S^{\sigma \sigma ^{\prime }}=-4\pi \alpha _{e.m.}\int dQ_{1}dQ_{2}(2\pi
)^{4}\delta ^{(4)}(q-q_{1}-q_{2})L^{\sigma }(L^{\sigma ^{\prime }})^{\ast },
\end{equation}
where explicitly,
\begin{eqnarray}
&&L^{\sigma }(L^{\sigma ^{\prime }})^{\ast } =\overline{u}(q_{1},s)\gamma
^{\sigma }\text{v}(q_{2},s^{\prime })\overline{\text{v}}(q_{2},s^{\prime
})\gamma ^{\sigma ^{\prime }}u(q_{1},s)=\\
&&Tr\left[ \gamma ^{\sigma }\frac{q_{1}+m_{l}}{2m_{l}}\gamma ^{\sigma ^{\prime }}\frac{q_{2}-m_{l}}{2m_{l}}%
\right] =\frac{1}{m_{l}^{2}}(q_{1}^{\sigma } q_{2}^{\sigma ^{\prime
}}+q_{1}^{\sigma ^{\prime }} q_{2}^{\sigma }-g^{\sigma \sigma ^{\prime
}}(q_{1}\cdot q_{2}))-g^{\sigma \sigma ^{\prime }}.\nonumber
\end{eqnarray}
In general, we can decompose the symmetric tensor \ $S^{\sigma \sigma
^{\prime }}$, which depends on the vector\ $q$,\ as
\begin{equation}
S^{\sigma \sigma ^{\prime }}=AQ^{\sigma \sigma ^{\prime }}+B\frac{q^{\sigma
}q^{\sigma ^{\prime }}}{q^{2}},
\end{equation}
where $Q^{\sigma \sigma ^{\prime }}$ is a projection operator given by
Eq. \ref{sumTL}. However, due to gauge invariance, $B=0$. Indeed, contracting \ $%
S^{\sigma \sigma ^{\prime }}$\ with \ $q^{\sigma }q^{\sigma ^{\prime }}$we
find 
\begin{eqnarray}
S^{\sigma \sigma ^{\prime }}q_{\sigma }q_{\sigma ^{\prime }}4\pi \alpha
_{e.m.}\int dQ_{1}dQ_{2}(2\pi )^{4}\delta ^{(4)}(q-q_{1}-q_{2})&\times&\nonumber\\
\left[ \frac{1}{m_{l}^{2}}(2q_{1}\cdot qq_{2}\cdot q-q^{2}q_{1}\cdot q_{2})-q^{2}\right]
&=&Bq^{2}.
\end{eqnarray}
The term in square brackets is 0 due to the condition $q=q_{1}+q_{2}$\ and
the lepton on the mass shell condition $q_{1}^{2}=q_{2}^{2}=m_{l}^{2}$;
hence, as required by general principles, $B=0$\ . To extract $A$ we
contract $S^{\sigma \sigma ^{\prime }}$ with $g_{\sigma \sigma ^{\prime }}$,
obtaining 
\begin{equation}
A=\frac{1}{3}S^{\sigma \sigma ^{\prime }}g_{\sigma \sigma ^{\prime }}=4\pi
\alpha _{e.m.}\int dQ_{1}dQ_{2}(2\pi )^{4}\delta ^{(4)}(q-q_{1}-q_{2})\frac{1%
}{3m_{l}^{2}}(M^{2}+2m_{l}^{2})
\end{equation}

The phase-space integral is most easily obtained in
the rest frame of the decaying virtual photon of mass $M$. We get
\begin{eqnarray}
\int dQ_{1}dQ_{2}(2\pi )^{4}\delta ^{(4)}(q-q_{1}-q_{2}) =
\frac{m_{l}^{2}}{2\pi }(1-\frac{4m_{l}^{2}}{M^{2}})^{\frac{1}{2}},
\end{eqnarray} 
and the partial decay width for the virtual photon is
\begin{equation}
\Gamma _{\gamma ^{\ast }\rightarrow l^{+}l^{-}}=\frac{A}{2M}=\frac{\alpha
_{e.m.}}{3M}\left( 1-\frac{4m_{l}^{2}}{M^{2}}\right) ^{\frac{1}{2}%
}(M^{2}+2m_{l}^{2}).
\end{equation}
Thus
\begin{eqnarray}
\frac{dN^{e+e-}}{d^{4}x} &=&\sum_{p}\int d^{4}q\int dP_{1}f_{\omega }(p_{1}\cdot
u)dP_{2}\left[ 1+f_{\pi }(p_{2}\cdot u)\right]\nonumber\\ 
&\times&  \delta^{(4)}(p_{1}-p_{2}-q)
T^{P}(p_{1},q)\frac{2}{M^{3}}\Gamma _{\gamma ^{\ast }\rightarrow l^{+}l^{-}},
\end{eqnarray}
where we have defined 
\begin{equation}
T^{p}(p_{1},q)=\widetilde{V}_{\sigma }^{(p)}\widetilde{V}_{\sigma ^{\prime }}^{(p) \ast
}Q^{\sigma \sigma ^{\prime }}.
\end{equation}
A useful form of the dilepton-rate formula can be obtained through the use of
the identity 
\begin{equation}
\int d^{4}q=\int d^{4}q\int dM^{2}\delta (M^{2}-q^{2})=\int d^{3}q\int dM^{2}%
\frac{1}{2E_{q}}.
\end{equation}
We have
\begin{eqnarray}
&&\frac{dN^{e+e-}}{d^{4}xdM^{2}} =\sum_{p}\int \frac{d^{3}q}{2E_{q}}\int
dP_{1}f_{\omega }(p_{1}\cdot u)dP_{2}\left[ 1+f_{\pi }(p_{2}\cdot u)\right]
\times \nonumber \\
&&\delta ^{(4)}(p_{1}-p_{2}-q)\frac{2}{M^{3}}T^{p}(p_{1},q)\Gamma _{\gamma ^{\ast
}\rightarrow l^{+}l^{-}} =\sum_{p}\int dQ\int dP_{1}f_{\omega }(p_{1}\cdot u)\times \nonumber\\ 
&&dP_{2}\left[ 1+f_{\pi }(p_{2}\cdot u)\right] (2\pi )^{4}
\delta ^{(4)}(p_{1}-p_{2}-q)\frac{1}{\pi M^{3}}T^{p}(p_{1},q)\Gamma _{\gamma
^{\ast }\rightarrow l^{+}l^{-}}.
\end{eqnarray}
Explicitly, the quantity $T^{p}(p_{1},q)$ is equal to 
\begin{eqnarray}
T^{p}(p_{1},q)=A_{\gamma
\sigma }\varepsilon _{(p)}^{\gamma }(p_{1})A_{\gamma ^{\prime }\sigma
^{\prime }}^{\ast }\varepsilon _{(p)}^{\ast \gamma ^{\prime
}}(p_{1})Q^{\sigma \sigma ^{\prime }}(q)|F|^{2}.
\end{eqnarray}
The transversely polarized $\omega $
has two helicity states, with projection equal to $\pm 1$ on the direction of
q, while the longitudinally polarized $\omega $ has one helicity state,  
with the corresponding projection equal to $0$. Taking this into account we
find: 
\begin{eqnarray}
T^{L}(p_{1},q)=L^{\gamma \gamma ^{\prime }}Q^{\sigma \sigma ^{\prime
}}A_{\gamma \sigma }A_{\gamma ^{\prime }\sigma ^{\prime }}^{\ast }, 
\end{eqnarray}
and 
\begin{eqnarray}
T^{T}(p_{1},q)=T^{\gamma \gamma ^{\prime }}Q^{\sigma \sigma ^{\prime
}}A_{\gamma \sigma }A_{\gamma ^{\prime }\sigma ^{\prime }}^{\ast }, 
\end{eqnarray}
which we may rewrite as
\begin{equation}
\sum_{p}T^{p}(p_{1},q)=(T_{L}+2T_{T})|F|^{2},
\end{equation}
where
\begin{equation}
|F|^{2}=|G(q)|^{2}\frac{4\pi \alpha _{e.m.}m_{\rho }^{2}}{g_{\rho }^{2}}=%
\frac{4\pi \alpha _{e.m.}}{g_{\rho }^{2}}\left( \frac{m_{\rho }^{4}}{%
(m_{\rho }^{2}+M^2)^{2}+M^{2}\Gamma _{\rho }^{2}}\right) ,
\end{equation}
and $\Gamma_{\rho}$ is the width of the $\rho $ meson of virtual mass $M$. The width is dominated by the 
$\rho\to\pi\pi$ channel, hence $\Gamma _{\rho }=\frac{g_{\rho \pi \pi }^{2}}{48\pi M^{2}}(M^{2}-4m_{\pi
}^{2})^{\frac{3}{2}}$.
 Then we can write
\begin{eqnarray}
&&\frac{dN^{e+e-}}{d^{4}xdM^{2}} =\int \frac{d^{3}q}{(2\pi )^{3}2E_{q}}\int \frac{
d^{3}p_{1}}{(2\pi )^{3}2E_{p_{1}}}f_{\omega }(p_{1}\cdot u)\int \frac{
d^{3}p_{2}}{(2\pi )^{3}2E_{p_{2}}}(2\pi )^{4}\times \nonumber\\
&&\left[ 1+f_{\pi }(p_{2}\cdot u)\right] 
\delta (E_{p_{1}}-E_{p_{2}}-E_{q})\delta ^{(3)}(\overrightarrow{p_{1}}-
\overrightarrow{p_{2}}-\overrightarrow{q})\times \nonumber\\
&&(T_{L}+2T_{T})|F|^{2}\frac{1}{\pi M^{3}}\Gamma _{\gamma ^{\ast }\rightarrow l^{+}l^{-}}.
\end{eqnarray}
The distribution functions have the thermal form 
$f_{\omega }=f_{\omega }(p_{1}\cdot u)=\exp [-(p_{1}\cdot u)/T(t)]$ and 
$f_{\pi }=[1+f_{\pi }(p_{2}\cdot u)]=1+\exp [-(p_{2}\cdot u)/T(t)]$,
where $(p_{1}\cdot u)=E_{p_{1}}u_{0}-p_{1\parallel }u_{\parallel
}-p_{1\perp }u_{\perp }(\cos \alpha \cos \theta -\sin \alpha \sin \theta )$
, and $u_{0}=\sqrt{1+u_{\perp }^{2}+u_{\parallel }^{2}}$ . Integrating
our expression over the $p_{2}$ momentum and substituting $d^{3}q$ as $%
q_{\perp }dq_{\perp }dq_{\parallel }d\alpha $ and $d^{3}p_{1}$ as $%
p_{1\perp }dp_{1\perp }dp_{1\parallel }d\theta $, where the angle $%
\alpha $ is between $\overrightarrow{q_{\perp }}$ and $\overrightarrow{u_{\perp}}$,
and $\theta $ between $\overrightarrow{p_{1\perp}}$
and $\overrightarrow{q_{\perp }}$, we obtain
\begin{eqnarray}
\frac{dN^{e+e-}}{d^{4}xdM^{2}} &=&\int \frac{q_{\perp }dq_{\perp
}dq_{\parallel }d\alpha }{(2\pi )^{3}2E_{q}}\int \frac{p_{1\perp
}dp_{1\perp }dp_{1\parallel }d\theta }{(2\pi )^{3}2E_{p_{1}}}f_{\omega
}f_{\pi }\frac{\pi }{E_{p_{2}}}\times \nonumber\\
&&\delta (E_{p_{1}}-E_{p_{2}}-E_{q})(T_{L}+2T_{T})|F|^{2}\frac{1}{\pi M^{3}}%
\Gamma _{\gamma ^{\ast }\rightarrow l^{+}l^{-}}.
\label{integr}
\end{eqnarray}
The on-mass-shell conditions are 
\begin{eqnarray}
E_{p_{1\text{ }}}&=&\sqrt{m_{\omega }^{2}+p_{1\perp}^{2}+p_{1\parallel }^{2}},\\ 
E_{p_{2}}&=&\sqrt{m_{\pi }^{2}+p_{2}^{2}}=\sqrt{m_{\pi }^{2}+(\overrightarrow{p_{1}}+\overrightarrow{q})^2}
=\sqrt{m_{\pi}^{2}+(p_{1\parallel }-q_{\parallel })^{2}+p_{1\perp
}^{2}+q_{\perp }^{2}-2p_{1\perp }q_{\perp }\cos \theta },\nonumber\\ 
E_{q}&=&\sqrt{M^{2}+q_{\perp }^{2}+q_{\parallel }^{2}}\nonumber.
\end{eqnarray}
Thanks to energy conservation we can eliminate integration over the angle $\theta $, 
from the Dirac delta in Eq.~\ref{integr}. 
We get $\delta(\theta-\theta_{0})$, where
\begin{equation}
\cos \theta_{0} =\frac{-(M^{2}+m_{\omega }^{2}-m_{\pi }^{2}+2p_{1\parallel
}q_{\parallel }-2\sqrt{m_{\omega }^{2}+p_{1\parallel
}^{2}+p_{1\perp }^{2}}\sqrt{M^{2}+q_{\parallel }^{2}+q_{\perp }^{2}})}{%
2p_{1\perp }q_{\perp }}.
\end{equation}

We have to remember that $\cos \theta_{0} $ must be in the range $[-1,1]$. For
dilepton production from $\omega$-meson decays we can write:
\begin{equation}
\frac{dN^{e+e-}}{d^{4}xdM^2} =
\int \frac{\pi dq_{\perp }dq_{\parallel }d\alpha }{%
(2\pi )^{3}2E_{q}}\frac{dp_{1\perp }dp_{1\parallel }}{(2\pi
)^{3}2E_{p_{1}}}f_{\omega }f_{\pi }\frac{2~\Theta(1- \mid \cos \theta_{0} \mid)} {\mid \sin \theta_{0} \mid}
\frac{1}{\pi M^{3}}(T_{L}+2T_{T})|F|^{2}\Gamma _{\gamma ^{\ast
}\rightarrow l^{+}l^{-}},
\end{equation}
where the factor 2 in the numerator comes from the two solutions of the equation $\cos\theta=\cos\theta_{0}$. 

\begin{figure}[t]
\centerline{
\vspace{0mm} ~\hspace{0cm} 
\epsfxsize = 9 cm \centerline{\epsfbox{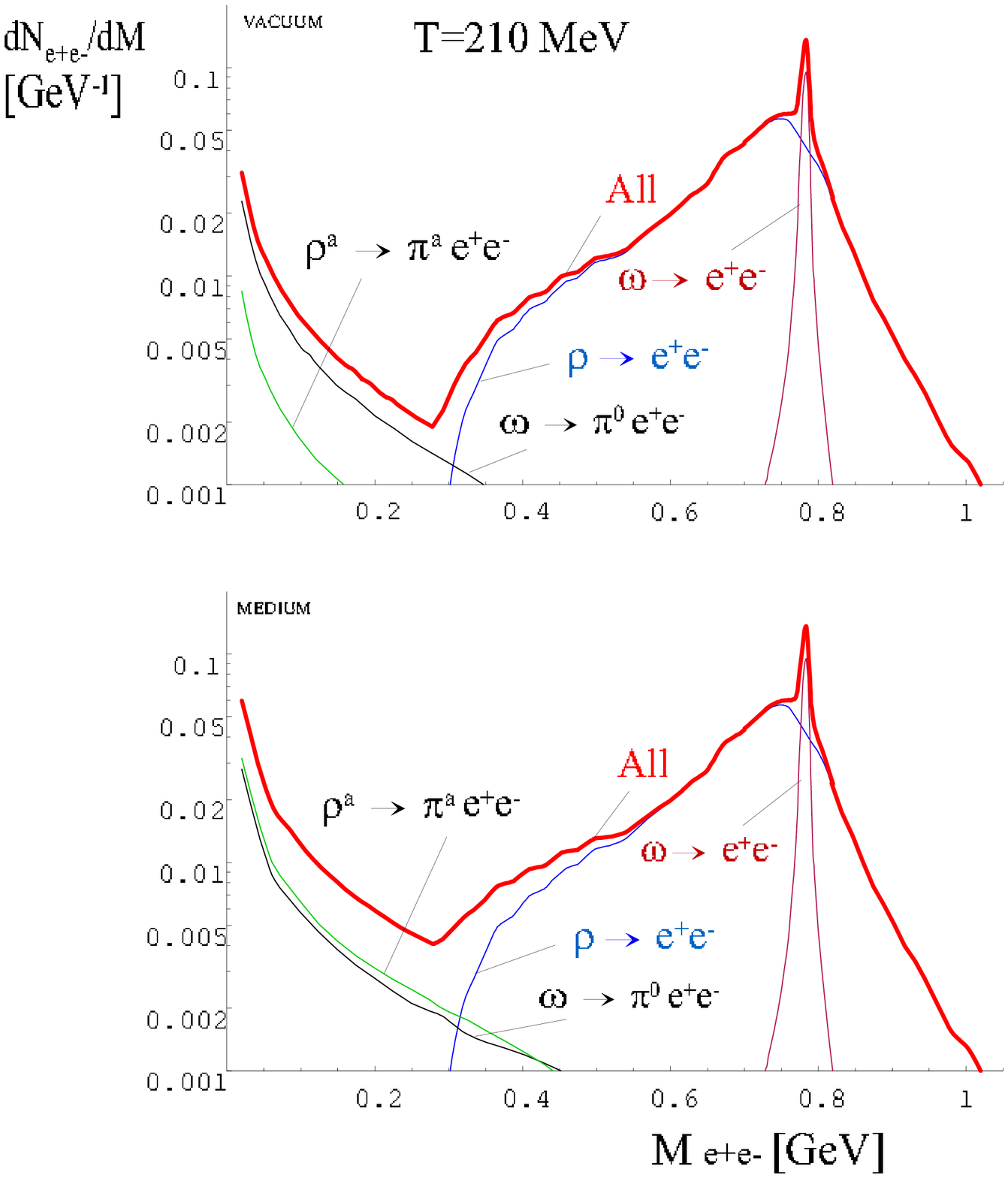}} \vspace{0mm}
\vspace{0mm} ~\hspace{-7.cm}
\epsfysize = 11 cm \centerline{\epsfbox{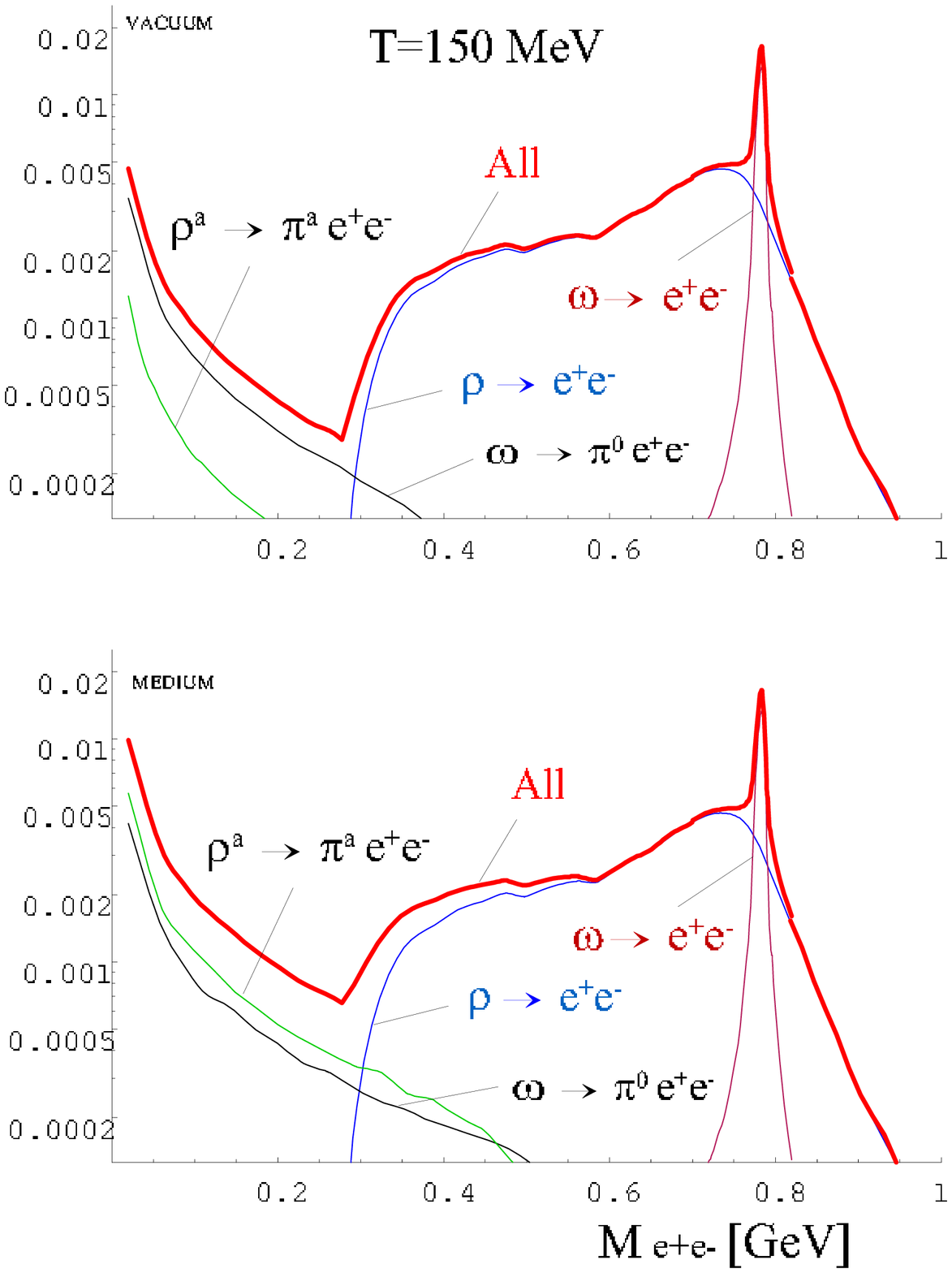}} \vspace{0mm}} 
\caption{\label{temp1}The dilepton yield  for the direct decays $\omega \to e^{+}e^{-}$ and 
$\rho \to e^{+}e^{-}$, and the Dalitz decays
$\omega \to  \pi^0 e^{+}e^{-}$ and $\rho^a \to \pi^a e^{+}e^{-}$, 
plotted as a function of the invariant mass, $M_{e+e-}$. The two upper plots 
are for the vacuum value of the coupling constant; lower plots include the medium-modified 
$\pi\omega\rho$ vertex. We show our results for two different temperatures,   
$T=210~{\rm MeV}$ (left), from Ref.~\cite{Rappevol} and $T=150~{\rm MeV}$ (right), \cite{ab2}. 
The thick line indicates the sum of all contributions.}
\end{figure}

Next, we apply the experimental acceptance cuts, which are included in the function
\begin{eqnarray}
&&Acc (M,q_{\parallel },q_{\perp })={\frac{\int d^{2}q_{1\perp }d^{2}q_{2\perp
}dq_{1 \parallel}dq_{2 \parallel}\frac{1}{E_{q1}E_{q2}}~~{\bf \phi }~~
\delta (E_{q}-E_{q_{1}}-E_{q_{2}})\delta
^{(3)}({\bf q}-{\bf q}_{1}-{\bf q}_{2})}{\int d^{2}q_{1\perp }d^{2}q_{2\perp
}dq_{1 \parallel}dq_{2 \parallel}\frac{1}{E_{q1}E_{q2}}~~\delta (E_{q}-E_{q_{1}}-E_{q_{2}})
\delta ^{(3)}({\bf q}-{\bf q}%
_{1}-{\bf q}_{2})}},  \nonumber \\
&&  \label{Acc2}
\end{eqnarray}
where the electron rapidities, see eq. \ref{Phi}, were rewritten as $dy=d q_{\parallel}/E_{q}$,
with $E_{q}=\sqrt{M^2 + q_{\perp}^2 + q_{\parallel}^2}$.
Finally, after using the fire-cylinder hydrodynamic expansion model of \cite{ Rappevol, RappShur} 
described in sec. 4.3,  
the yield of dileptons produced during the expansion is (the Bose enhancement is neglected for brevity) 
\begin{equation}
\frac{dN^{e+e-}}{dM^2}=\int_{0}^{t_{\max }}dt\int_{0}^{r_{\max }(t)}2\pi
rdr\int_{-z_{\max }(t)}^{z_{\max }(t)}dz\left( \frac{dN^{e+e-}}{d^{4}xdM^2}
Acc (M,q_{\parallel },q_{\perp }) \right).
\end{equation}
In the case of the Dalitz decay of the $\rho$ meson we apply the same formulas.

\chapter{Results of our model}

\begin{figure}[t]
\begin{center}
\includegraphics[width=1.0 \textwidth]{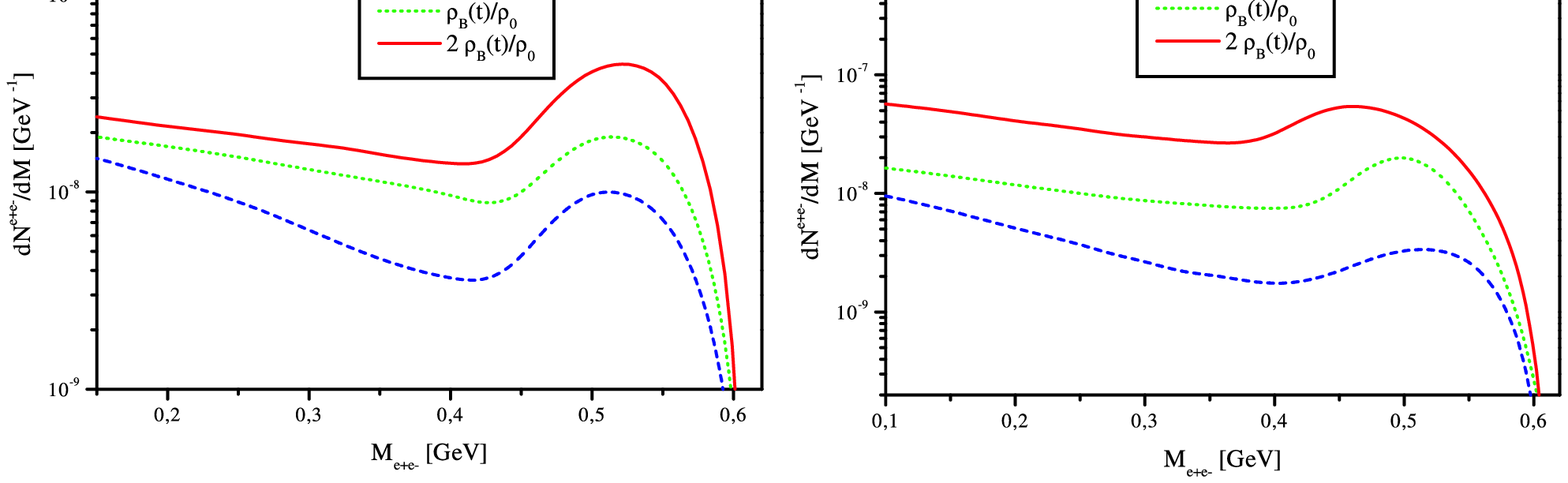}
\caption{\label{lin210}The dilepton yield  for the Dalitz decays. The dashed lines   
are for the vacuum value of coupling constant, whereas the solid lines include the medium modified 
$\pi\omega\rho$ vertex. Green and red lines show the results for different value of baryon density $\rho_{B}(t)$ 
and $2~\rho_{B}(t)$, respectively. We show our results for $T=210~{\rm MeV}$.}
\end{center}
\end{figure}
\begin{figure}[t]
\begin{center}
\includegraphics[width=1.0 \textwidth]{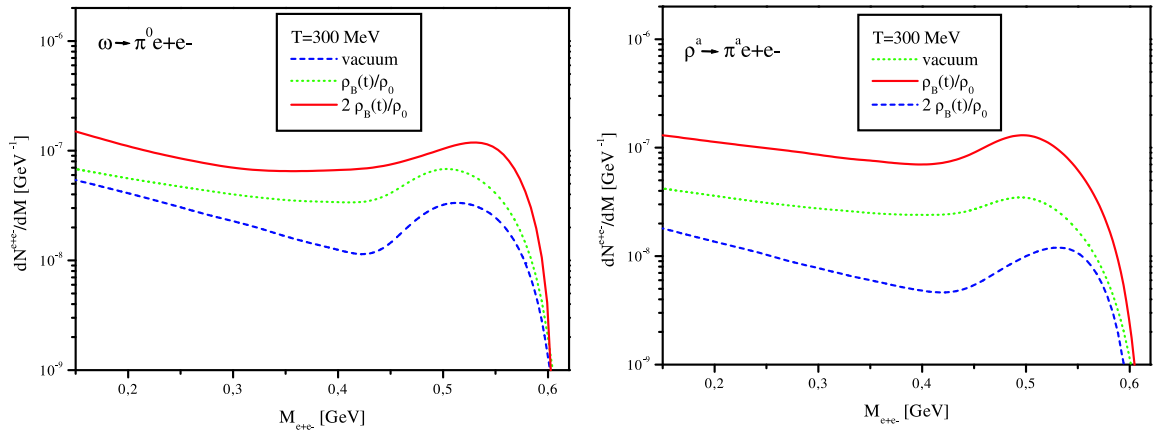}
\caption{\label{lin300} Same as Fig.~\ref{lin210}, but for the $T=300~{\rm MeV}$. 
Left side, the dilepton yield  for the Dalitz decay
$\omega \to  \pi^0 e^{+}e^{-}$; right side, for $\rho^a \to \pi^a e^{+}e^{-}$, 
plotted as a function of the, $M_{e+e-}$. 
Dashed lines correspond to the vacuum value of the coupling constant, solid lines include medium modified 
$\pi\omega\rho$ vertex. Green and red lines show the results for different values of the baryon 
density $\rho_{B}(t)$ and $2~\rho_{B}(t)$, respectively.}
\end{center}
\end{figure}

\begin{figure}[h]
\begin{center}
\includegraphics[width=0.82 \textwidth]{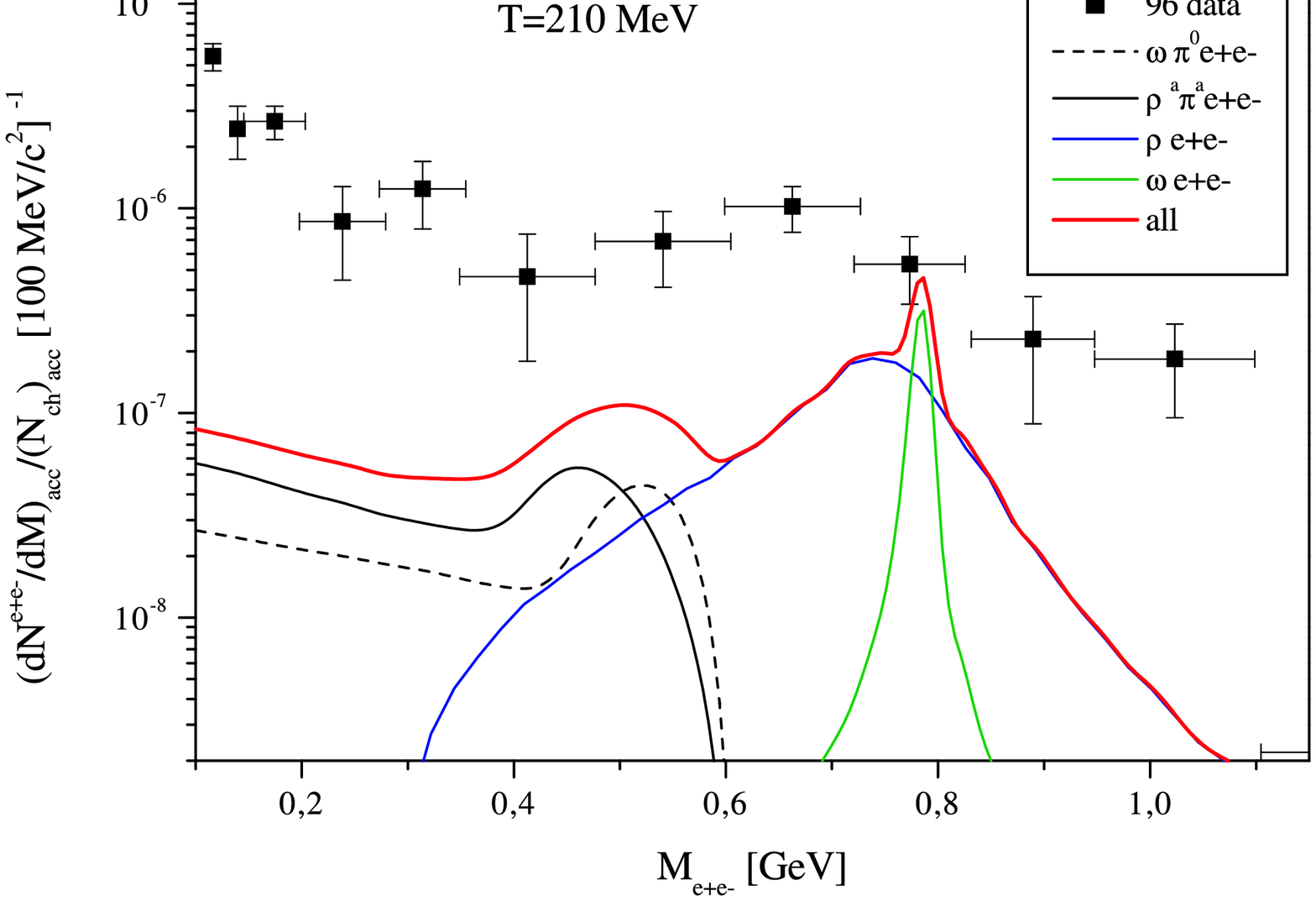}
\caption{\label{exp1} Dilepton emission rate for $158~\rm{A GeV}$ Pb+Au CERES experiment from the 
Dalitz and direct decays (for $p_{\perp}>0.2~\rm{GeV}$). The dashed and solid line correspond to the $\omega\to\pi^{0}e^{+}e^{-}$ and 
$\rho^{a}\to\pi^{a}e^{+}e^{-}$ decays, respectively. The solid blue and green 
line correspond to direct contributions $\rho\to e^{+}e^{-}$ and $\omega\to e^{+}e^{-}$. 
We show our results for temperature $\rm{T=210~MeV}$ and double normal value of the baryon density.}
\end{center}
\end{figure}
In the following we present our numerical results. We begin with calculations 
without expansion of the fire-cylinder and without the acceptance cuts, Eqs.~\ref{Phi} and \ref{Acc2},  
shown in Fig. \ref{temp1}. 
We do this in case to compare with other works. 
New and important result of this study is to applying in our 
calculations the CERES experimental kinematic cuts, what in other theoretical works is usually skipped. 

In Fig.~\ref{temp1} we plot the dilepton yield as a function of the invariant mass, $M_{e+e-}$. 
On the left-hand side we show the results for the temperature of $210~{\rm MeV}$ of Ref. \cite{Rappevol}; see 
section 4.3. 
\begin{figure}[h]
\begin{center}
\includegraphics[width=0.82 \textwidth]{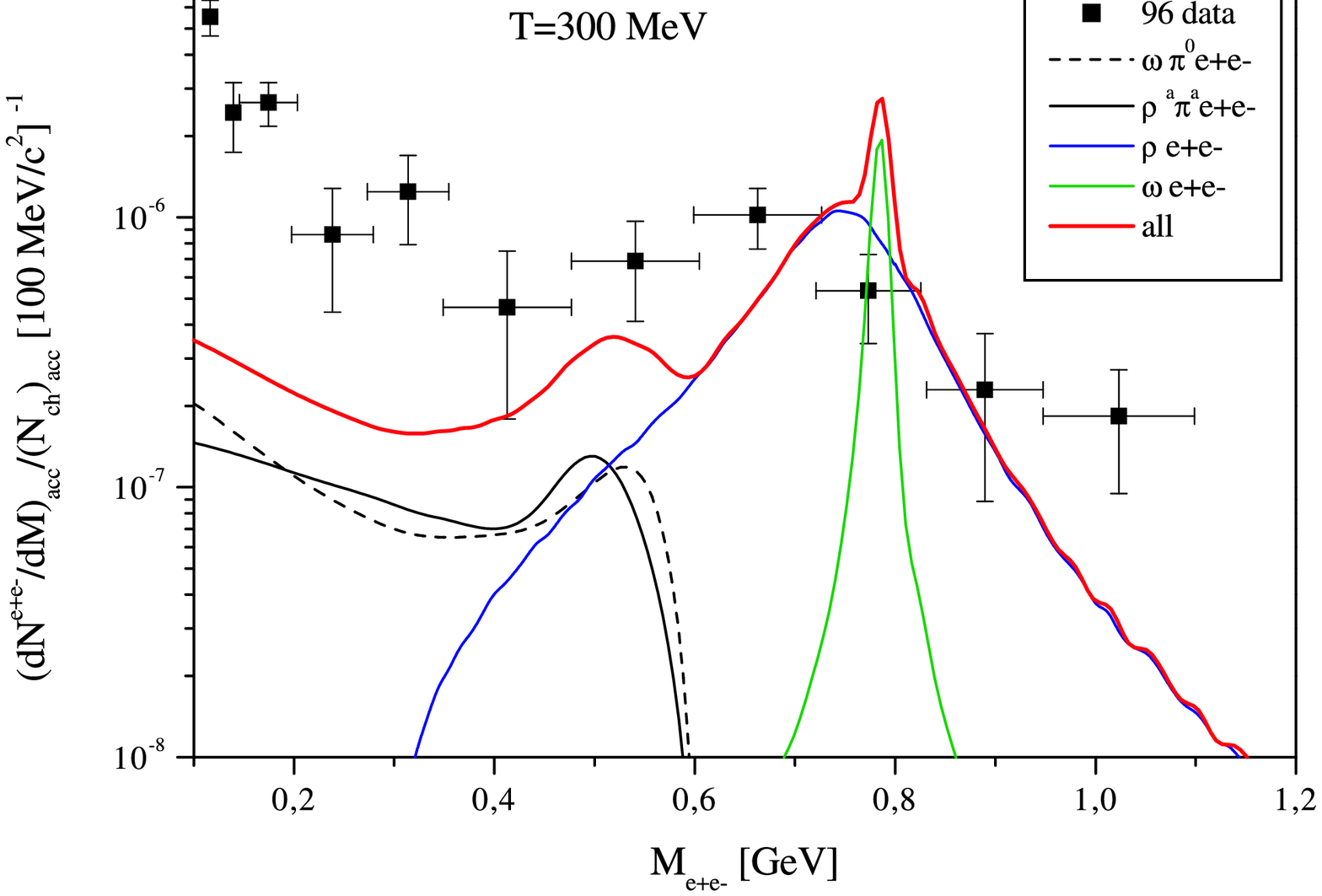}
\caption{\label{exp2} Same as Fig.~\ref{exp1} but for $\rm{T=300~MeV}$.}
\end{center}
\end{figure}
The upper plot is for the vacuum value of the $\pi\omega\rho$ coupling constant. 
We display the direct contributions, such as 
$\omega \to e^{+} e^{-}$ and $\rho \to e^{+} e^{-}$, which are indicated by pink and blue curves, 
and the Dalitz contributions of both $\omega$ (black curve) and $\rho$ (green curve) mesons. 
On the lower plot we present the same results but with medium modified $\pi\omega\rho$ 
vertex. The thick red line indicates the sum of all direct and Dalitz contributions.
We see that the Dalitz decays are enhanced by about a factor of $2$ compared to the vacuum values. 
On the right-hand side we show the results for a lower temperature of $150~{\rm MeV}$, with similar conclusions. 
It is visible that the direct decays are dominant in the region above $0.35~{\rm GeV}$ and that the number 
of produced particles increases with growing temperature. Here we also observe the peak from the direct decay of 
the omega meson, which is not observed in the data because of the too low experimental resolution.

Next, we include in our calculations the expansion effects and the CERES acceptance function. 
In Fig. \ref{lin210} we illustrate  
the dilepton yield for the $\omega \to  \pi^0 e^{+}e^{-}$ (left-hand side) and $\rho^a \to \pi^a e^{+}e^{-}$ 
(right-hand side) decays, plotted as a function of the invariant mass of dileptons, $M_{e+e-}$.
The plots are done for the maximum fireball temperature of $210~\rm{MeV}$ and for different values of 
the baryon density, $\rho_{B}(t)$ 
(green dotted curve) and $2~\rho_{B}(t)$ (red curve). In our calculations we use the Rapp-Wambach model 
of the fireball, but we change the temperature from T=210 to T=300 and increase two times the baryon density. 
This significantly enhances the medium effects.
By the blue dashed line we indicate the case with the vacuum value of the $\pi\omega\rho$ vertex.
We can observe that the medium effects from the Dalitz decays of vector mesons are clearly visible, 
over the entire region of $M_{e+e-}=0.2-0.6~\rm{GeV}$. 
For both Dalitz decays $\omega\to\pi^0 e^{+}e^{-}$ and $\rho^{a}\to\pi^{a}e^{+}e^{-}$ there is a difference 
of about a factor of $2$ for the case with normal baryon density compared to the vacuum value. 
Simultaneously, we notice that this enhancement factor rises to a higher value of about a factor of $5$ with the  
growing fireball density. 

We have a similar situation in Fig. \ref{lin300}, where the same results as in Fig.~\ref{lin210}, but for 
temperatures of $300~\rm{MeV}$, are presented.
The dilepton production from Dalitz decays is found to be strongly enhanced for the medium-modified vertex 
compared to the vacuum for this value of the temperature. 
This difference grows even larger, to about one order of magnitude, with the growing baryon density. 
It is naturally explained since we expect to observe more particles produced at higher temperatures 
than at lower ones.

We continue our discussion with Fig. \ref{exp1} and \ref{exp2}, where we present the dilepton spectra 
from the direct and Dalitz decays compared to the CERES experimental data, taken from Ref.~\cite{PhD}, for two 
different temperatures, T=210~\rm{MeV} and T=300~\rm{MeV}. Both plots are for the same density equal to 
twice the normal baryon density.
As we can see the medium modifications help, but not enough to reproduce the experimental data. However, 
for the higher temperature of the fireball, T=300~\rm{MeV}, results lie much closer to the data.
We notice that the medium effects from Dalitz decays of 
the $\omega$ and $\rho$ mesons are large and are specifically about two times larger than in vacuum. 
We see that the contributions from the 
$\rho^{a} \to \pi^{a} e^{+}e^{-}$ decay are equally important as the ones from the Dalitz decay 
of the $\omega$ meson and should not be neglected during the considerations connected with the 
low-mass dilepton spectrum, as is frequently done. 
It is visible from our analysis that in the region of interest, $0.2-0.6~\rm{GeV}$, 
where the existing calculations 
have problems explaining the experimental data, our effect helps but still fails to reproduce the experimental 
data.

We have to admit that there are other processes which contribute in the explored region of $M_{e+e-}$, like 
Dalitz decays of $\eta$ and $\eta'$ mesons which we do not include in our calculation. 
We have not taken into account other effects such as shift of the $\rho$ meson peak position or broadening 
of the widths.
We also do not show 
the 'cocktail' contribution of decays of hadrons on top of our results, which would not be 
consistent. 

After all, we stress that the $\pi\omega\rho$ coupling constant 
undergo substantial modification in the nuclear medium and directly 
influence the dilepton production rate, to which this thesis was devoted.

\chapter{Summary and discussion}
In this thesis we have considered the modifications of the $\pi\omega\rho$ coupling in 
dense nuclear matter and its influence on dilepton spectra in relativistic heavy-ion collisions 
coming from the Dalitz decays. 

In the first part we have analyzed the $\pi\omega\rho$ vertex 
in the context of the Dalitz decays of $\pi^{0}$, $\omega$ and $\rho$ particles. 
Our calculations were done in the framework of a fully relativistic hadronic theory with mesons, 
nucleons and $\Delta$ isobars. We have presented diagrams, propagators, typical vertices for the 
interactions with and without $\Delta$ resonance, and the in-medium tensor structure of the $\pi\omega\rho$ 
coupling constant. Presence of the additional four-vector, the four-velocity of the medium, made this structure 
more complicated. We have worked at zero temperature and 
in the leading-density approximation. We have considered decays of particles in the rest and the moving frame with 
respect to the medium. 

In the case where the decaying particles are moving with respect to the medium we have analyzed the kinematics in 
a convenient way, in the rest frame of the nuclear matter, not in the rest frame of decaying particle, as it 
frequently done in the vacuum calculations. Because each polarization of the decaying particles 
behaves differently the properties of these particles are different, in particular their widths. Thus, we have 
analyzed the dependence of the widths for transverse and longitudinal polarizations on the invariant mass 
of dileptons.

The main conclusion of our investigation is that  
the medium effects on the $\pi\omega\rho$ coupling are large and primarily come from the process where the
$\Delta$ is excited in the intermediate state. 
On the other hand, for the $\pi^{0}\to\gamma\gamma^{*}$ decay the effects of the $\Delta$ isobar cancel almost 
exactly the nucleon particle-hole excitation, such that the medium contribution is small. 
The dilepton yield from the $\pi^{0}$ decays are not changed by the medium, because virtually all pions, 
due to their long lifetime, decay outside the fireball and because the width for this decay is practically 
unaltered by the nuclear matter.
However, we have found a sizeable increase of the $\pi\omega\rho$ vertex compared to its vacuum value for the 
Dalitz decays from $\omega$ and $\rho$ mesons. This coupling constant is enhanced by about a factor of 2  
for the $\omega\to\pi^{0}\gamma^{*}$ decay and by a factor of 5 for the $\rho^{a}\to\pi^{a}\gamma^{*}$ decay. 
For the decaying particle moving with respect to the nuclear matter, the medium effects decrease with 
growing magnitude
of three momentum, but remain significant for a momenta up to $\sim 200~\rm{MeV}$, the relevant value for 
temperatures typical in heavy-ion collisions. This results confirm that the medium effects are 
substantial. 

Through-out our calculations we were following a relativistic covariant approach. We have also shown in detail 
how to obtain the leading-density approximation for dileptons involving loops with density-dependent 
nucleon propagators. 
In order to simplify our approach we have not included any form factors in the vertices, because they lead to 
fundamental problems with gauge symmetry conservation, in particular the Ward-Takahashi identities and 
current conservation are violated.
We have considered the Dalitz decay of the $\rho$ meson which was never analyzed before. 

In the second part we have applied our model to evaluate the dilepton production rate from the Dalitz decays of 
vector mesons in relativistic heavy ion collisions. In order to estimate the dilepton yields we have used the 
VDM and including expansion of the medium. In our calculations we have adopted the 
Rapp-Wambach model of the hydrodynamic expansion of the fire-cylinder, which includes longitudinal and 
transverse expansion.  
The kinematic constraints of the CERES experiment also is used. The construction 
of the acceptance function required a numerical calculation of a two-dimensional 
integral involving the product of step functions, which we have obtained by a Monte Carlo method. 

In order to compare our results with other works, we have shown the dilepton yield as a function 
of the invariant mass for calculations without expansion of the fire-cylinder and without the acceptance cuts. 
We have presented plots for two temperatures, $210~\rm{MeV}$ and $150~\rm{MeV}$, with conclusions that the 
Dalitz decays are enhanced by about a factor of 2 compared to the vacuum and that the number of particles 
increases with temperature. On the other hand, the direct decays are dominant in the region around $0.4~\rm{GeV}$.

Next, we have included the expansion and the CERES acceptance function. 
Our final plots are done for two fireball temperatures ($T=210~\rm{MeV}$ and $T=300~\rm{MeV}$) and for  
density equal to twice of the normal baryon density.
Comparing our results to the dilepton invariant mass spectrum measured by the CERES 
collaboration at CERN SPS we have found that the Dalitz decays are enhanced in the 
region of $0.2-0.6~\rm{GeV}$, and medium modifications of coupling constant 
increase the effect. 
It is worth pointing out that the contribution from Dalitz decay of the $\rho$ meson is equally important 
as the $\omega$. 
However our results still underestimate the experimental data. 

Our model together with other effects, which are not considered here, such as creation of  
dileptons from other decays ($\eta$ and $\eta'$), dropping masses or broadening widths, 
may be able to explain the present experimental data. The other effects should be taken into 
account in future work.

For our numerical analysis we have developed a program written in 'Mathematica' with the very useful 
FeynCalc package 
for algebraic calculation in elementary particle physics. We have included in the Mathematica code 
the acceptance function, which was calculated in Fortran.

The low-mass dilepton research is a quickly evaluating branch of theoretical and experimental physics. 
Inclusion of more sources of dileptons or detailed verifications of existed theoretical models, together with 
new experimental data will provide new insight into that relevant problem in relativistic heavy-ion collisions  
physics.

\chapter{Appendices}
\section{Appendix A}
The contributions to $B$ from diagrams (a), (b) and (c) of Fig.~~\ref{meddiag} for the process 
$\pi ^{0}\rightarrow\gamma(\omega) \gamma ^{\ast }$ are as follows:
\begin{eqnarray}
B_{(a)}^{\pi^{0}(\omega)} &=&\frac{m_{N} (\kappa_{\rho}+1)g_{A}g_{\rho }g_{\omega }m_{\pi }^2
(3 x^2 + m_{\pi}^2 )}{( 4 m_{N}^{2} - m_{\pi }^{2}) 
((m_{N} x^2+ m_{N} m_{\pi}^2)^{2}- x^4 m_{\pi }^2 )}, \notag \\
\bigskip
B_{(b)}^{\pi^{0}(\omega)} &=&-\frac{g_{N\Delta\pi}g_{N\Delta \rho}g_{\omega }F_{\pi }m_{\pi}}
{9 M_{\Delta}^2 m_{N}[(m_{N}x^2+ m_{N} m_{\pi }^2 )^2 - (m_{N}^2 - M_{\Delta}^2 + x^2)^2 m_{\pi }^2]}
((m_{N}^2 - M_{\Delta}^2) \times \notag \\
&&(m_{N}^2 + m_{N}M_{\Delta} - M_{\Delta}^2 + x^2) m_{\pi }^2 - (m_{N} x^2 (m_{N}^{3} - m_{N} M_{\Delta}^2 M_{\Delta} x^2)) 
- m_{N}^2 m_{\pi }^4), \nonumber
\end{eqnarray}
\begin{eqnarray}
B_{(c)}^{\pi^{0}(\omega)}&=& - \frac{g_{N\Delta\pi}g_{N\Delta \rho}g_{\omega }F_{\pi }}
{27 M_{\Delta}^4 m_{\pi }[( m_{N}^2 - M_{\Delta}^2 + x^2)^2 m_{\pi }^2 -( m_{N} m_{\pi }^ 2 + m_{N} x^2)^2]}
( -(m_{N}^2 (2 m_{N} + \notag \\
&&+3 M_{\Delta}) x^6)+ + x^2 ( 4 m_{N}^5 + 6 m_{N}^4 M_{\Delta} - 8 m_{N}^3 x^2 + 3 m_{N}^2(M_{\Delta}^3 - 
3 M_{\Delta} x^2) + \notag \\
&&+ 2 m_{N}(4 M_{\Delta}^4 - 2 M_{\Delta}^2 x^2 + x^4)+ 3 M_{\Delta} (M_{\Delta}^4 - 3 M_{\Delta}^2 x^2 
+ 2 x^4))m_{\pi }^2 + (4 m_{N}^5 + \notag \\ 
&&+6 m_{N}^4 M_{\Delta}+ 11 M_{\Delta}^5 - 17 M_{\Delta}^3 x^2 + 6 M_{\Delta} x^4 - 2 m_{N}^3 (4 M_{\Delta}^2 
+ x^2)- m_{N}^2 (5 M_{\Delta}^3 + \notag \\
&&+ 9 M_{\Delta} x^2) + 4 m_{N}(4 M_{\Delta}^4 
- 5 M_{\Delta}^2 x^2 + 2 x^4))m_{\pi }^4 - m_{N}(4 m_{N}^2 + 3 m_{N}M_{\Delta} + \notag \\
&&+ 2 x^2) m_{\pi }^6)), 
\notag \\
\end{eqnarray}
where $x=m_{\omega}$

The contributions to $B$ from  $\pi ^{0}\rightarrow\gamma(\rho) \gamma ^{\ast }$  decay are as follows:
\begin{eqnarray}
B_{(a)}^{\pi^{0}(\rho)} &=&\frac{g_{A} g_{\rho } g_{\omega } m_{\pi}^2 (2 m_{N}^2 x^2 
(3 + \kappa_{\rho}) + (x^2 \kappa_{\rho} + 2 m_{N}^2 (1 + \kappa_{\rho})) m_{\pi }^2)}
{2 m_{N}(4 m_{N}^2 - m_{\pi}^2)((m_{N}x^2 + m_{N} m_{\pi}^2)^2 - x^4 m_{\pi }^2)}, \notag \\
\bigskip
B_{(b)}^{\pi^{0}(\rho)} &=&\frac{g_{N\Delta\pi}g_{N\Delta \rho}g_{\omega}F_{\pi} m_{\pi}
(x^2 - m_{\pi }^2)(m_{N}^2 x^4 (m_{N}^3 - 2 M x^2 - m_{N}(M_{\Delta}^2 + 2 x^2))+ x^2}
{9 M_{\Delta}^2((M_{\Delta}^2 - m_{N}^2)^2 m_{\pi}^2 - (m_{N}x^2 - m_{N} m_{\pi }^2)^2)((m_{N}x^2 
+ m_{N}m_{\pi}^2)^2 - x^4 m_{\pi}^2)} \times \notag \\
&&(4 m_{N}^5 + m_{N}^4 M_{\Delta} + 3 m_{N} M_{\Delta}^4 + 2 M_{\Delta}^5 + m_{N}^3(-7 M_{\Delta}^2 
+ 6 x^2) - 3 m_{N}^2 (M_{\Delta}^3 - \notag \\
&&- M_{\Delta} x^2)) m_{\pi }^2 - m_{N}(m_{N}^4 + m_{N}^3 M_{\Delta} + M_{\Delta}^4 
+ m_{N}^2 (-2 M_{\Delta}^2 + 5 x^2) + \notag \\
&&+ m_{N}(-M_{\Delta}^3 + M_{\Delta} x^2)) m_{\pi }^4 + m_{N}^3 m_{\pi }^6), \notag \\
\bigskip
B_{(c)}^{\pi^{0}(\rho)} &=&\frac{g_{N\Delta\pi} g_{N\Delta \rho} g_{\omega} F_{\pi}
(-(m_{N}^2(2 m_{N} + 3 M_{\Delta}) x^6) + x^2 (4 m_{N}^5 + 6 m_{N}^4 M + 8 m_{N} M_{\Delta}^4} 
{27 M_{\Delta}^4 m_{\pi} ((M_{\Delta}^2 - m_{N}^2)^2 m_{\pi}^2 - (m_{N}x^2 - m_{N} m_{\pi }^2)^2)} + \notag \\ 
&&+ 3 M_{\Delta}^5 + 8 m_{N}^3 x^2 +  + 3 m_{N}^2 (M_{\Delta}^3 + 3 M_{\Delta} x^2)) m_{\pi }^2 
-(4 m_{N}^5 + 6 m_{N}^4 M_{\Delta} + \notag \\
&&+16 m_{N} M_{\Delta}^4 + 11 M_{\Delta}^5 - 2 m_{N}^3 (4 M_{\Delta}^2 - 5 x^2) + m_{N}^2(-5 M_{\Delta}^3 
+ 9 M_{\Delta} x^2)) m_{\pi }^4 + \notag \\
&&+m_{N}^2(4 m_{N} + 3 M_{\Delta}) m_{\pi }^6) 
\notag \\
\end{eqnarray}
with $x=m_{\rho}$.

\section{Appendix B}
\subsubsection{Numerators and denominators of Eq.~\ref{amini}}

For the case $\bf q=0$ the contributions to $B$ from Eq.\ref {amini} corresponding to the diagrams (a,b,c)
from Fig.\ref{meddiag}, can be written as
\begin{eqnarray}
A^{\mu \nu }_{(i)}=\frac{i}{F_{\pi }}\frac{e^{2}}{%
g_{\rho }g_{\omega }}\left( g_{\pi \rho \omega }+\rho _{B}\frac{N_{(i)}}{D_{(i)}}\right) \epsilon
^{\nu \mu pQ}, \label {amini1}
\end{eqnarray}
The formulas for $N_{(i)}$ and $D_{(i)}$ are very long.
Numerators $N_{(i)}$ become manageable in the formal case $\Gamma_{\Delta}=0$. 
Below we show the numerators for the diagrams with nucleons only, because those with $\Delta$ isobar 
are too long. We find 

\begin{eqnarray}
N_{(a)}^{\omega}&=&{g_A} {g_{\rho }} {g_{\omega }} 
( m_{N}^2 \kappa_{\rho }  {{m_{\pi }}}^8 ( m_{N}^2 - {{m_{\omega }}}^2) 
+2 m_{N}^2 {{m_{\pi }}}^6 ( -2 m_{N}^2 x^2 \kappa_{\rho }  + 
( x^2 \kappa_{\rho }  - 3 m_{N}^2 \times \notag \\
&&\times( 1 + \kappa_{\rho })){{m_{\omega }}}^2+ \kappa_{\rho } {{m_{\omega }}}^4)  - 
m_{N}^2 {( x^2 - {{m_{\omega }}}^2) }^2 ( -( m_{N}^2 x^4 \kappa_{\rho })  + 
x^2 ( x^2 \kappa_{\rho }  + \notag \\ 
&&+m_{N}^2 ( 6 + 4 \kappa_{\rho })){{m_{\omega }}}^2 + m_{N}^2 ( 2 + \kappa_{\rho } ) {{m_{\omega }}}^4) + 
{{m_{\pi }}}^4 ( 6 m_{N}^4 x^4 \kappa_{\rho }  + 2 m_{N}^2 x^2 ( -( x^2 \kappa_{\rho })+ \notag \\  
&&+ 3 m_{N}^2 ( 1 + \kappa_{\rho })) {{m_{\omega }}}^2 + 
( 2 m_{N}^2 x^2 + x^4 \kappa_{\rho }  + 2 m_{N}^4 ( 5 + 4 \kappa_{\rho }))
{{m_{\omega }}}^4 - m_{N}^2 \kappa_{\rho } {{m_{\omega }}}^6)  - \notag \\
&&-2 m_{N}^2 {{m_{\pi }}}^2 ( 2 m_{N}^2 x^6 \kappa_{\rho }  - 
x^4 ( x^2 \kappa_{\rho }  + 3 m_{N}^2 ( 1 + \kappa_{\rho }))
{{m_{\omega }}}^2+ x^2 ( -x^2 + 2 m_{N}^2 ( 5 + \notag \\
&&+ 4 \kappa_{\rho }))
{{m_{\omega }}}^4 + ( m_{N}^2 - x^2) ( 1 + \kappa_{\rho }){{m_{\omega }}}^6)), \nonumber 
\end{eqnarray}
\begin{eqnarray}
N_{(a)}^{\rho}&=&-({g_A} {g_{\rho }} {g_{\omega }} {{m_{\rho }}}^2
( 6 m_{N}^4 ( 1 + \kappa_{\rho }){{m_{\pi }}}^6 + m_{N}^2 {( x^2 - {{m_{\rho }}}^2 ) }^2 
( 2 m_{N}^2 x^2 ( 3 + \kappa_{\rho }) + ( x^2 \kappa_{\rho }  + \notag \\
&&+2 m_{N}^2 ( 1 + \kappa_{\rho })){{m_{\rho }}}^2)  
- {{m_{\pi }}}^4 ( 2 m_{N}^4 x^2 ( 3 + 5 \kappa_{\rho })  + m_{N}^2 ( x^2 ( 2 - 3 \kappa_{\rho })  + 
10 m_{N}^2 \times \notag \\
&&\times( 1 + \kappa_{\rho })){{m_{\rho }}}^2 + x^2 \kappa_{\rho }  {{m_{\rho }}}^4) 
+2 m_{N}^2 {{m_{\pi }}}^2 ( m_{N}^2 x^4 ( -3 + \kappa_{\rho })  - x^2 ( x^2 ( 1 + \notag \\
&&+2 \kappa_{\rho })  - 
2 m_{N}^2 ( 5 + 3 \kappa_{\rho })){{m_{\rho }}}^2+( -x^2 + m_{N}^2 ( 1 + \kappa_{\rho }))  {{m_{\rho }}}^4))).
\end{eqnarray}
The expressions for $D_{i}$ for the case $\Gamma_{\Delta}=0$  
for $\omega \rightarrow \pi ^{0}\gamma ^{\ast }$ decay, where $x=m_{\rho}$, are 
\begin{eqnarray}
D_{(a)}^{\omega}&=&
2(2 m_{N} - {m_{\omega }}) (2 m_{N} + {m_{\omega }})
(m_{N} x^2 - m_{N} {{m_{\pi }}}^2 - x^2{m_{\omega }} + m_{N}{{m_{\omega }}}^2) (m_{N} x^2 - \notag \\ 
&&-m_{N} {{m_{\pi }}}^2 + x^2 {m_{\omega }} + m_{N}{{m_{\omega }}}^2) ({{m_{\pi }}}^2
(m_{N} - {m_{\omega }})  + m_{N} (-x^2 + {{m_{\omega }}}^2)) \times \notag \\ 
&&\times({{m_{\pi }}}^2 (m_{N} + {m_{\omega }})  + m_{N} (-x^2 + {{m_{\omega }}}^2)), \nonumber
\end{eqnarray}
\begin{eqnarray}
D_{(b)}^{\omega}&=&9 M^2 {m_{\omega }}(-( m_{N} x^2)  + m_{N} {{m_{\pi }}}^2 + (m_{N}^2 - M_{\Delta}^2 + x^2)
{m_{\omega }} - m_{N} {{m_{\omega }}}^2)( m_{N} x^2 - \notag \\
&&-m_{N} {{m_{\pi }}}^2 +(m_{N}^2 - M^2 + x^2)
{m_{\omega }} + m_{N} {{m_{\omega }}}^2)( m_{N} x^2 + ( m_{N}^2 - M_{\Delta}^2){m_{\omega }} - m_{N}
\times \notag \\ 
&&\times{{m_{\omega }}}^2 + {{m_{\pi }}}^2( -m_{N}
+ {m_{\omega }})) ( -( m_{N} x^2) + ( m_{N}^2 - M_{\Delta}^2) {m_{\omega }} + m_{N} {{m_{\omega }}}^2 +\notag\\ 
&&+{{m_{\pi }}}^2( m_{N} + {m_{\omega }}), \nonumber
\end{eqnarray}
\begin{eqnarray}
D_{(c)}^{\omega}&=& 27 M_{\Delta}^4 {m_{\omega }}( -( m_{N} x^2) + m_{N} {{m_{\pi }}}^2 
+ ( m_{N}^2 - M_{\Delta}^2 + x^2)
{m_{\omega }} - m_{N} {{m_{\omega }}}^2) ( m_{N} x^2 - \notag \\ 
&&-m_{N} {{m_{\pi }}}^2 +( m_{N}^2 - M_{\Delta}^2 + x^2)
{m_{\omega }} + m_{N}{{m_{\omega }}}^2)( m_{N} x^2 + ( m_{N}^2 - M_{\Delta}^2) {m_{\omega }} - \notag \\
&&-m_{N} {{m_{\omega }}}^2 +{{m_{\pi }}}^2( -m_{N} + {m_{\omega }}))( -( m_{N} x^2) + ( m_{N}^2 
- M_{\Delta}^2)  {m_{\omega }} + m_{N} 
{{m_{\omega }}}^2 +\notag \\ 
&&+{{m_{\pi }}}^2 ( m_{N} + {m_{\omega }})) \notag \\
\end{eqnarray}
The expressions for $D_{i}$ for the case $\Gamma_{\Delta}=0$ are 
for the $\rho \rightarrow \pi ^{0}\gamma ^{\ast }$ decay, where  $x=m_{\omega}$

\begin{eqnarray}
D_{(a)}^{\rho}&=&2(2 m_{N} - {m_{\rho }})( 2 m_{N} + {m_{\rho }}) ( m_{N} x^2 - m_{N} {{m_{\pi }}}^2 - 
x^2 {m_{\rho }} + m_{N} {{m_{\rho }}}^2)( m_{N} x^2 - \notag \\ 
&&-m_{N} {{m_{\pi }}}^2 + x^2 {m_{\rho }} + 
m_{N} {{m_{\rho }}}^2)( {{m_{\pi }}}^2 ( m_{N} - {m_{\rho }}) + m_{N} ( -x^2 + {{m_{\rho }}}^2))\times \notag \\
&&\times({{m_{\pi }}}^2( m_{N} + {m_{\rho }}) + m_{N} ( -x^2 + {{m_{\rho }}}^2)), \nonumber
\end{eqnarray}
\begin{eqnarray}
D_{(b)}^{\rho}&=&9 M_{\Delta}^2( m_{N} + M_{\Delta} - {m_{\rho }})( m_{N} + M_{\Delta} + {m_{\rho }})
( m_{N} x^2 - m_{N} {{m_{\pi }}}^2 - x^2 {m_{\rho }} +\notag \\ 
&&+m_{N} {{m_{\rho }}}^2)( m_{N} x^2 - m_{N} {{m_{\pi }}}^2 + x^2 {m_{\rho }} + m_{N} {{m_{\rho }}}^2)
( m_{N} x^2 + ( m_{N}^2 - M_{\Delta}^2){m_{\rho }} - \notag \\
&&-m_{N} {{m_{\rho }}}^2 + {{m_{\pi }}}^2
( -m_{N} + {m_{\rho }}))( -( m_{N} x^2) + ( m_{N}^2 - M_{\Delta}^2) {m_{\rho }} + m_{N} {{m_{\rho }}}^2+\notag \\ 
&&+ {{m_{\pi }}}^2 ( m_{N} + {m_{\rho }})), \nonumber
\end{eqnarray}
\begin{eqnarray}
D_{(c)}^{\rho}&=&
27 M_{\Delta}^4 ( m_{N} + M_{\Delta} - {m_{\rho }})  {m_{\rho }} ( m_{N} + M_{\Delta} 
+ {m_{\rho }})( m_{N} x^2 + ( m_{N}^2 - M_{\Delta}^2)
{m_{\rho }} - \notag \\ 
&&-m_{N} {{m_{\rho }}}^2 +{{m_{\pi }}}^2 ( -m_{N} + {m_{\rho }})) ( -( m_{N} x^2)  + 
( m_{N}^2 - M_{\Delta}^2) {m_{\rho }} + m_{N} {{m_{\rho }}}^2 + \notag \\
&&+{{m_{\pi }}}^2 ( m_{N} + {m_{\rho }})). \notag \\
\end{eqnarray}

\section{Appendix C}

\subsubsection{Ceres Acceptance function}
\begin{figure}[h] 
\begin{center}
\includegraphics[width=0.4 \textwidth]{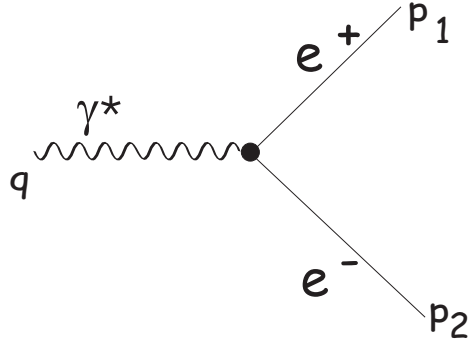}
\caption{\label{kinacc}The decay of $\gamma^*$ into a lepton pair.}  
\end{center}
\end{figure}

In this appendix we develop the expression for the acceptance function of the CERES experiment.
The kinematics cuts include the pseudo rapidity range, reflecting the angular acceptance of the detector cuts for 
transverse momenta of the measured $e^{+}$ and $e^{-}$, as well as the minimum angle between the directions 
of the $e^{+}$ and $e^{-}$. The kinematics of the decay is depicted in Fig.~\ref{kinacc}. 

We use the energy and momentum conservation
\begin{eqnarray}
E &=& E_{1}+E_{2},\nonumber \\
q_{\parallel}&=&p_{\parallel1} + p_{\parallel2}\label{enmomcons},
\end{eqnarray}
where
\begin{eqnarray}
E&=&m_{\perp} \rm{cosh(y)},~~~~~~E_{1,2}=p_{\perp1,2} \rm{cosh( y_{1,2})},\nonumber\\
q_{\parallel}&=&m_{\perp} \rm{sinh(y)},~~~~~~p_{\parallel1,2}=p_{\perp1,2} \rm{sinh(y_{1,2})}\label{e&q},
\end{eqnarray} 
with $\rm{y}$ denoting the rapidity of $\gamma^{*}$, and $m_{\perp}=\sqrt{m^2+q_{\perp}^2}$. 
Next, from the relation~$q_{\perp}=p_{\perp1}+p_{\perp2}$ we obtain
\begin{eqnarray}
p_{\perp2}=\sqrt{q_{\perp}^2+p{_{\perp1}}^2-2q_{\perp}p_{\perp1} \cos{\xi}} \label{pp2},
\end{eqnarray}
where $\xi$ is the angle between $q_{\perp}$ and $p_{\perp1}$. Solving Eqs.~\ref{enmomcons} with the help of 
Eq.~\ref{e&q} we obtain the solution for the rapidities of $e^{+}$ and $e^{-}$ of the form
\begin{eqnarray}
\hbox{sinh}(y_{1})^{(1),(2)}&=&\frac{1}{2 m_{\perp}p_{\perp1}} \hbox{sinh}(y)(m_{\perp}^2
+p_{\perp1}^2-p_{\perp2}^2)\pm \nonumber \\
&\pm& \sqrt{m_{\perp}^4 - 2(p_{\perp1}^2+p_{\perp2}^2)m_{\perp}^2 + (p_{\perp1}^2-p_{\perp2}^2)^2} 
\hbox{cosh}(y)\label{branch1},
\end{eqnarray}
and
\begin{eqnarray}
\hbox{sinh}(y_{2})^{(1),(2)}&=&\frac{1}{2 m_{\perp}p_{\perp2}} \hbox{sinh}(y)(m_{\perp}^2
-p_{\perp1}^2+p_{\perp2}^2)\pm \nonumber \\
&\pm& \sqrt{m_{\perp}^4 - 2(p_{\perp1}^2+p_{\perp2}^2)m_{\perp}^2 + (p_{\perp1}^2-p_{\perp2}^2)^2} 
\hbox{cosh}(y) \label{branch2}.
\end{eqnarray}
\begin{figure}[t] 
\begin{center}
\includegraphics[width=.9 \textwidth]{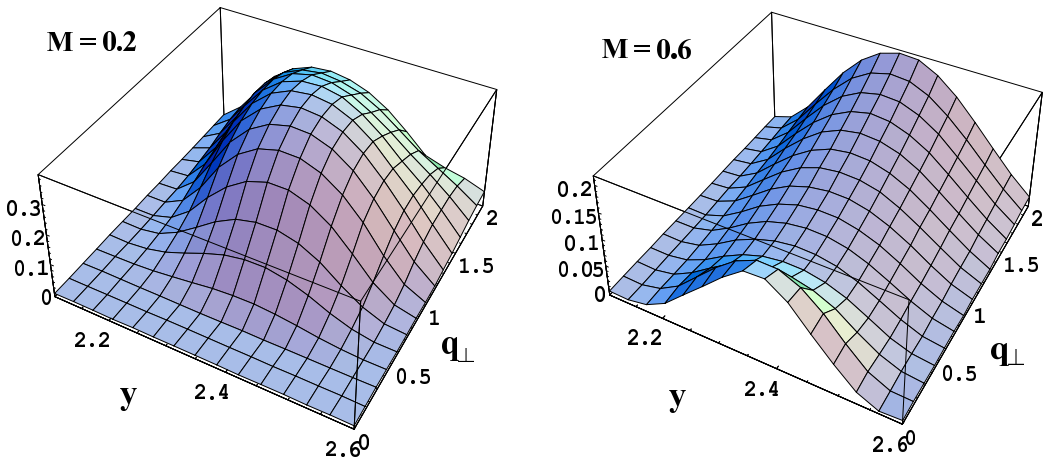}
\caption{\label{plotacc}The Ceres acceptance function for different value of $M$, left-hand side $M=0.2$, 
right-hand side $M=0.6$.}  
\end{center}
\end{figure}
The (1) and (2) label the two possible branches (the situation is similar to the case discussed in Chapter 2, 
section 2.3.)
These solutions are real if the condition 
\begin{eqnarray}
\sqrt{m_{\perp}^4 - 2(p_{\perp1}^2+p_{\perp2}^2)m_{\perp}^2 + (p_{\perp1}^2-p_{\perp2}^2)^2} >0, \label{real}
\end{eqnarray}
is fulfilled.
The upper limit for the integration over $p_{\perp1}$ we obtain when $\xi=0$ in Eq.~\ref{pp2}. Then
\begin{eqnarray}
p_{\perp1}=\frac{1}{2} (q_{\perp}+\sqrt{m+q_{\perp}^2}).
\end{eqnarray}
The Ceres acceptance function is
\begin{eqnarray}
&&Acc (M,y,q_{\perp }) ={\frac{\int d^{2}p_{1\perp }d^{2}p_{2\perp
}dy_{1}dy_{2}\,\,\,{\bf \phi }\,\,\,\delta (E_{q}-E_{p_{1}}-E_{p_{2}})\delta
^{(3)}({\bf q}-{\bf p}_{1}-{\bf p}_{2})}
{2\pi}},  \nonumber \\
&&  \label{acc2}
\end{eqnarray}
Using $\delta^{(3)} ({\bf q}-{\bf p}_{1}-{\bf p}_{2})=
\delta^{(2)}({\bf q_{\perp}}-{\bf p}_{\perp1}-{\bf p}_{\perp2}) 
\delta({\bf q}_{\parallel}-{\bf p}_{\parallel1}-{\bf p}_{\parallel2})$, where the two dimensional delta 
eliminates integration over $p_{\perp2}^2$, we are left with
\begin{eqnarray}
\delta (E_{q}-E_{p_{1}}-E_{p_{2}}) \delta ({\bf q}_{\parallel}-{\bf p}_{\parallel1}-{\bf p}_{\parallel2})= 
\frac{\sum_{i=1}^{2} \delta(y_{1}-y_{1}^{(i)})\delta(y_{2}-y_{2}^{(i)})}
{p_{\perp1}p_{\perp2} \hbox{cosh}(y_1 - y_2)}.
\end{eqnarray}
In Eq.~\ref{acc2}, $\phi$ is the product of step functions which includes the experimental cuts, such as  
$2.1<y_{1,2}<2.65,~p_{\perp1,2}>200~\rm{MeV},~\theta_{ee}>35~\rm{mrad}$. It has the explicit form
\begin{eqnarray}
\phi&=&\Theta(y_{1}-2.1)\Theta(2.65-y_{1})\Theta(y_{2}-2.1)(2.65-y_{2})\Theta(p_{\perp1}-0.2)
\Theta(p_{\perp2}-0.2) \times \nonumber \\
&\times&\Theta\left[\cos{(0.35)}-\frac{p_{\perp1}p_{\perp2}+p_{\perp1} \hbox{sinh}y_{1}p_{\perp2} \hbox{sinh}y_2}
{p_{\perp1} \hbox{cosh}y_{1} p_{\perp2} \hbox{cosh}y_2}\right]
\end{eqnarray}
As an example in Fig.~\ref{plotacc}, we show a plot of $Acc(M,y,q_{\perp})$ for different values of mass $M$. 
We calculate the CERES acceptance function using the Monte Carlo integration method in Fortran.

\section{Appendix D}
We are going to proof that the diagrams of Fig.~\ref{meddiag} with more than one $S_{D}$ propagator 
vanish for kinematic 
reasons. We write the two nucleon density propagators, $S_{D1}$ and $S_{D2}$ in the 
following way
\begin{eqnarray}
S_{D1}=i(\gamma^{\mu}k_{\mu}+m_{N})\frac{i\pi }{E_{k}}\delta (k_{0}-E_{k})\theta (k_{F}-|\bf{k}|)],
\label{sd1}  
\end{eqnarray}
\begin{eqnarray}
S_{D2}=i(\gamma^{\mu}k_{\mu}+ \gamma^{\mu}p_{\mu}+m_{N})\frac{i\pi }{E_{k+p}}\delta (k_{0}+p_{0}
-E_{k+p})\theta (k_{F}-|\bf{k}+\bf{p}|)].
\label{sd2}  
\end{eqnarray}
From above expressions \ref{sd1} and \ref{sd2} we have
\begin{eqnarray}
k_{0}=E_{k}= \sqrt{{\bf{k}}^2+m_{N}^2}~~~~{\rm{and}}~~~~k_{0}+p_{0}=E_{k+p}=\sqrt{({\bf{k}+\bf{p}})^2+m_{N}^{2}},
\end{eqnarray}
thus, after simply calculations, we obtain 
\begin{eqnarray}
m_{\rho}^2+2 p_{0}~k_{0}=2~{\bf{k}}~{\bf{p}}~ \hbox{cos}~\theta,
\end{eqnarray}
with $p_{0}=E_{p}= \sqrt{{\bf{p}}^2+m_{\rho}^2}$. We have that
\begin{eqnarray}
\hbox{cos}~\theta=\frac{m_{\rho}^2+2 \sqrt{{\bf{p}}^2+m_{\rho}^2}~\sqrt{{\bf{k}}^2+m_{N}^2}}{2~\bf{k}~\bf{p}}>1,
\end{eqnarray}
which ends the proof.

\bibliographystyle{npsty}
\bibliography{delta}

\begin{thebibliography}{100}

\bibitem{brscale}
G.~E. Brown and M. Rho, Phys. Rev. Lett. {\bf 66} (1991) 2720

\bibitem{chin}
S.~A. Chin, Ann. Phys. (NY) {\bf 108} (1977) 301

\bibitem{hatsuda}
T. Hatsuda, H. Shiomi, and H. Kuwabara, Prog. Theor. Phys. {\bf 95} (1996) 1009

\bibitem{ceres}
{CERES Collab., G. Agakichiev {\it et al.}}, Phys. Rev. Lett. {\bf 75} (1995)
  1272

\bibitem{helios}
{HELIOS/3 Collab., M. Masera {\it et al.}}, Nucl. Phys. {\bf A590} (1995) 93c

\bibitem{newceres}
{A. Martin for the CERES collaboration}, nucl-ex/0406007  (2004)

\bibitem{celenza}
L.~S. Celenza, A. Pantziris, C.~M. Shakin, and W.-D. Sun, Phys. Rev. {\bf C45}
  (1992) 2015

\bibitem{herrmann1}
M. Herrmann, B.~L. Friman, and W. Noerenberg, Nucl. Phys. {\bf A560} (1993) 411

\bibitem{herrmann2}
M. Herrmann, B.~L. Friman, and W. Noerenberg, Nucl. Phys. {\bf A545} (1992)
  267C

\bibitem{Mishra}
{A. Mishra, J. Reinhardt, H. Stoker, and W. Greiner}, Phys.Rev. {\bf C66}
  (2002) 064902

\bibitem{renkMish}
{T. Renk and A. Mishra}, Phys.Rev. {\bf C69} (2004) 054905

\bibitem{ab1}
{A. Bieniek, A. Baran, and W. Broniowski}, Phys.Lett. {\bf B526} (2002) 329

\bibitem{ceres2}
{G.~Agakichiev {\emph et.al}}, Nucl. Phys. {\bf A610} (1996) 317c

\bibitem{ceres3}
{G.~Agakichiev {\emph et.al}}, Phys. Lett. {\bf B422} (1998) 405

\bibitem{na38a}
{NA38 collaboration, S.~Ramos}, Nucl. Phys. {\bf A590} (1995) 93c

\bibitem{na38b}
{NA30 collaboration, M. C.~Abreu {\emph et al.}}, Phys. Lett. {\bf B368} (1996)
  230

\bibitem{na50}
{NA50 collaboration, M. C.~Abreu {\emph et al.}}, Nucl. Phys. {\bf A610} (1996)
  331

\bibitem{na60}
{H. Satz}, hep-ph/0405051  (2004)

\bibitem{bevalac1}
{DLS collaboration, R. J. Porter {\emph et al.}}, Phys. Rev. Lett {\bf 79}
  (1997) 1229

\bibitem{bevalac2}
{DLS collaboration, W. K. Wilson {\emph et al.}}, Phys. Rev. {\bf C57} (1997)
  1865

\bibitem{li}
G.~Q. Li, C.~M. Ko, and G.~E. Brown, Nucl. Phys. {\bf A606} (1996) 568

\bibitem{LiKoBrown1}
G.~Q. Li, C.~M. Ko, and G.~E. Brown, Phys. Rev. Lett. {\bf 75} (1995) 4007

\bibitem{CassingDil}
W. Cassing, W. Ehehalt, and C.~M. Ko, Phys. Lett. {\bf B363} (1995) 35

\bibitem{hatlee}
T. Hatsuda and S.~H. Lee, Phys. Rev. {\bf C46} (1992) R34

\bibitem{leupold}
S. Leupold, W. Peters, and U. Mosel, Nucl. Phys. {\bf A628} (1998) 311

\bibitem{Rappevol}
R. Rapp and J. Wambach, Eur. Phys. J. {\bf A6} (1999) 415

\bibitem{RappShur}
R. Rapp and E. Shuryak, Phys. Lett. {\bf B473} (2000) 13

\bibitem{hadrons}
{\em Hadrons in Nuclear Matter}, edited by {H. Feldmaier and W. {Noerenberg}}
  (GSI, Darmstadt, 1995), proc. Int. Workshop XXIII on Gross Properties of
  Nuclei and Nuclear Excitations, Hirschegg, Austria, 1995

\bibitem{tsuk}
{Quark matter '97. Ultra-relativistic nucleus nucleus collisions. Proceedings,
  48th Yamada Conference, 13th International Conference, Tsukuba, Japan,
  December 1- 5, 1997, edited by T. Hatsuda, Y. Miake, K. Yagi, and and S.
  Nagamiya}, Nucl. Phys. {\bf A638} (1998) 1

\bibitem{torino}
{Quark matter '99. Proceedings, 14th International Conference on
  ultra-relativistic nucleus nucleus collisions, Torino, Italy, May 10-15,
  1999, edited by L. Riccati, M. Masera, and E. Vercellin}, Nucl. Phys. {\bf
  A661} (1999) 1

\bibitem{jean}
H.-C. Jean, J. Piekarewicz, and A.~G. Williams, Phys. Rev. {\bf C49} (1994)
  1981

\bibitem{herrmann3}
M. Herrmann, B.~L. Friman, and W. Noerenberg, Z. Phys. {\bf A343} (1992) 119

\bibitem{pirner}
B. Friman and H.~J. Pirner, Nucl. Phys. {\bf A617} (1997) 496

\bibitem{Urban0}
M. Urban, M. Buballa, R. Rapp, and J. Wambach, Nucl. Phys. {\bf A641} (1998)
  433

\bibitem{urban}
M. Urban, M. Buballa, R. Rapp, and J. Wambach, Nucl. Phys. {\bf A673} (2000)
  357

\bibitem{rapp}
R. Rapp, G. Chanfray, and J. Wambach, Nucl. Phys. {\bf A617} (1997) 472

\bibitem{Post}
M. Post, S. Leupold, and U. Mosel,  preprint (2000), nucl-th/0008027

\bibitem{Peters}
{W. Peters, M. Post, H. Lenske, S. Leupold, and U. Mosel}, Nucl. Phys. {\bf
  A632} (1998) 109

\bibitem{LeupoldRev}
S. Leupold and U. Mosel, Prog. Part. Nucl. Phys. {\bf 42} (1999) 221

\bibitem{Gao}
{S. Gao, C. Gale, C. Ernst, H. Stocker, and W. Greiner}, Nucl. Phys. {\bf A661}
  (1999) 518

\bibitem{serot}
B.~D. Serot and J.~D. Walecka, Advances in Nuclear Physics {\bf 16} (1986) 1

\bibitem{lee2000}
S.~H. Lee, Nucl. Phys. {\bf A670} (2000) 119

\bibitem{fachini}
P. Fachini, J. Phys. {\bf G30} (2003) S565

\bibitem{eletsky}
V.~L. Eletsky, B.~L. Ioffe, and J.~I. Kapusta, Eur. J. Phys. {\bf A3} (1998)
  381

\bibitem{friman2}
B. Friman, Acta Phys. Pol. {\bf B29} (1998) 3195

\bibitem{Lutz99}
M. Lutz, B. Friman, and G. Wolf, Nucl. Phys. {\bf A661} (1999) 526

\bibitem{walecka}
J.~D. Walecka, Ann. of Phys. {\bf 83} (1974) 491

\bibitem{serot2}
C.~J. Horowitz and B.~D. Serot, Nuclear Physics {\bf A368} (1981) 503

\bibitem{serot3}
C.~J. Horowitz and B.~D. Serot, Phys. Lett. {\bf B140} (1984) 181

\bibitem{griegel}
T.~D. Cohen, R.~J. Furnstahl, and D.~K. Griegel, Phys. Rev. {\bf C45} (1992)
  1881

\bibitem{Jin}
X. Jin and D. Leinweber, Phys. Rev. {\bf C52} (1995) 3344

\bibitem{asakawa}
M. Asakawa and C. Ko, Nucl. Phys. {\bf A560} (1993) 399

\bibitem{eletsky2}
V.~L. Eletsky and B.~L. Ioffe, Phys. Rev. Lett. {\bf 78} (1997) 1010

\bibitem{galeLich}
C. Gale and B. Lichard, Phys. Rev. Lett. {\bf D49} (1994) 338

\bibitem{Song2}
C. Song and V. Koch, Phys. Rev. {\bf C54} (1996) 3218

\bibitem{Krippa}
B. Krippa, Nucl. Phys. {\bf A672} (2000) 270

\bibitem{bfh}
W. Broniowski, W. Florkowski, and B. Hiller, Acta Phys. Pol. {\bf B30} (1999)
  1079

\bibitem{BFH2}
W. Broniowski, W. Florkowski, and B. Hiller, Eur. Phys. J. {\bf A7} (2000) 287

\bibitem{BFH3}
W. Broniowski, W. Florkowski, and B. Hiller,  in {\em Hadron Physics: Effective
  theories of low-energy QCD, Coimbra, Portugal, September 1999, AIP Conference
  Proceedings}, edited by A.~H. Blin {\it et~al.} (AIP, Melville, New York,
  2000), Vol.~508, p.\ 218, nucl-th/9910057

\bibitem{rhopipi}
W. Broniowski, W. Florkowski, and B. Hiller, Nucl. Phys. {\bf A696} (2001) 870

\bibitem{rarita}
W. Rarita and J. Schwinger, Phys. Rev. {\bf 60} (1941) 61

\bibitem{Ben}
M. Benmerrouche, R.~M. Davidson, and N.~C. Mukhopadhyay, Phys. Rev. {\bf C39}
  (1989) 2339

\bibitem{EW}
T.~O.~E. Ericson and W. Weise, {\em Pions and nuclei} (Clarendon Press, Oxford,
  1988)

\bibitem{Gomez}
J.~A.~G. Tejedor and E. Oset, Nucl. Phys. {\bf A571} (1994) 667

\bibitem{ren}
C.-Y. Ren and M. Banerjee, Phys. Rev. {\bf C41} (1990) 2370

\bibitem{hohler}
G. Hohler and E. Pietarinen, Nucl. Phys. {\bf B95} (1976) 1

\bibitem{Pascal}
V. Pascalutsa and R. Timmermans, Phys. Rev. {\bf C60} (1999) 042201

\bibitem{Hemmert}
T.~R. Hemmert, B.~R. Holstein, and J. Kambor, J. Phys. G {\bf G24} (1998) 1831

\bibitem{Haber}
H. Haberzettl,  preprint (1998), nucl-th/9812043

\bibitem{feyncalc}
{R. Merting and F. Orellana}, http://www.feyncalc.org

\bibitem{BFH1}
W. Broniowski, W. Florkowski, and B. Hiller, Acta Phys. Polon. {\bf B30} (1999)
  1079

\bibitem{disp}
W. Florkowski and W. Broniowski, Nucl. Phys. {\bf A651} (1999) 397

\bibitem{RappRev}
R. Rapp and J. Wambach, Adv. Nucl. Phys {\bf 25} (2000) 1

\bibitem{Dom}
C.~A. Dominguez and M. Loewe, Z. Phys. {\bf C49} (1991) 423

\bibitem{bratko0}
E.~L. Bratkovskaya and C.~M. Ko, Phys. Lett. {\bf B445} (1999) 265

\bibitem{RappDil}
R. Rapp, G. Chanfray, and J. Wambach, Phys. Rev. Lett. {\bf 76} (1996) 368

\bibitem{sQGPshur}
{E. Shuryak}, hep-ph/0405066  (2004)

\bibitem{sQGPgul}
{M. Gyulassy and L. McLerran}, nucl-th/0405013  (2004)

\bibitem{satz}
T. Matsui and H. Satz, Phys. Lett. {\bf B178} (1986) 416

\bibitem{gavin}
S. Gavin and R. Vogt, Nucl. Phys. {\bf A610} (1996) 442c

\bibitem{kharzeev}
D. Kharzeev, Nucl. Phys. {\bf A610} (1996) 418c

\bibitem{jpblaizot}
J.~P. Blaizot, Nucl. Phys. {\bf A610} (1996) 452c

\bibitem{braunmunz}
{P. Braun-Munzinger, I. Heppe, and J. Stachel}, Phys. Lett. {\bf B465} (1999)
  15

\bibitem{bourg}
{M. Bourguin and J.M. Gaillard}, Nucl.Phys. {\bf B114} (1976) 334

\bibitem{alber}
{T. Alber \emph{et al.}}, Nucl.Phys. {\bf A566} (1994) 35c

\bibitem{KW}
{N. Kroll and W. Wada}, Phys.Rev {\bf 98} (1955) 1355

\bibitem{Land}
{L.S. Landsberg}, Phys.Rep {\bf 128} (1985) 301

\bibitem{sakurai}
{G.J. Gounaris and J.J. Sakurai}, Phys.Rev {\bf 21} (1968) 244

\bibitem{PhD}
P.~M. A.~G. Hering, nucl-ex/0203004  (2002)

\bibitem{40gevlow}
{D. Adamova, G. Agakichiev, {\emph et al.}}, nucl-exp/0209024.  (2004)

\bibitem{helios-1}
{HELIOS-1 collaboration, T. Akesson {\emph et al.}}, Z. Phys. {\bf C68} (1995)
  47

\bibitem{sorge}
{G. Q.~Li and C. M.~Ko and G. E.~Brown and H.~Sorge}, Nucl. Phys. {\bf A611}
  (1996) 539

\bibitem{helios-2}
{HELIOS-2 collaboration, T. Akesson {\emph et al.}}, Z. Phys. {\bf C51} (1990)
  369

\bibitem{kek}
{K. Ozawa {\emph et al.}}, Phys.Rev.Lett. {\bf 86} (2001) 5019

\bibitem{gsi}
{J. Friese, for HADES collaboration}, Nucl. Phys. {\bf A654} (1999) 1017c

\bibitem{CassingKralik}
{W. Cassing and W. Ehehalt and I. Kralik}, Phys. Lett. {\bf B377} (1996) 5

\bibitem{KapustaMc}
J. Kapusta, D. Kharzeev, and L. McLerran, Phys. Rev. {\bf D53} (1996) 5028

\bibitem{Huang}
Z. Huang and X.~N. Wang, Phys. Rev. {\bf D53} (1996) 5041

\bibitem{GaleKap}
C. Gale and J. Kapusta, Phys. Rev. {\bf C35} (1987) 2107

\bibitem{Song}
C. Song, V. Koch, S.~H. Lee, and C.~M. Ko, Phys. Lett. {\bf B366} (1996) 379

\bibitem{haglin}
K. Haglin, Phys. Rev. {\bf C53} (1996) R2606

\bibitem{murray}
J. Murray, W. Bauer, and K. Haglin, Phys. Rev. {\bf C57} (1998) 492

\bibitem{HungDil}
C.~M. Hung and E.~V. Shuryak, Phys. Rev. {\bf C56} (1997) 453

\bibitem{kluger}
{Y.~Kluger and V.~Koch and J.~Randrup and X. N.~Wang}, Phys. Rev. {\bf C57}
  (1998) 208

\bibitem{pdbook}
{Particle Data Group}, Phys. Rev. {\bf D66} (2002) 010001

\bibitem{PKoch}
P. Koch, Z. Phys. {\bf C57} (1993) 283

\bibitem{weise}
R.~A. Schneider and W. Weise, Eur. Phys. J. {\bf A617} (1997) 472

\bibitem{BaierDil}
R. Baier, M. Dirks, and K. Redlich, Phys. Rev. {\bf D55} (1997) 4344

\bibitem{ab2}
{A. Bieniek}, nucl-th/0401022  (2004)

\end{thebibliography}
\end{document}